%% file: v1.tex
\documentclass[sigplan,authorversion]{acmart}

\setcopyright{none}
\acmPrice{15.00}
\acmDOI{10.1145/3453483.3454111}
\acmYear{2021}
\copyrightyear{2021}
\acmSubmissionID{pldi21main-p853-p}
\acmISBN{978-1-4503-8391-2/21/06}
\acmConference[PLDI '21]{Proceedings of the 42nd ACM SIGPLAN International Conference on Programming Language Design and Implementation}{June 20--25, 2021}{Virtual, Canada}
\acmBooktitle{Proceedings of the 42nd ACM SIGPLAN International Conference on Programming Language Design and Implementation (PLDI '21), June 20--25, 2021, Virtual, Canada}

\author{Raven Beutner}
\affiliation{
	\institution{University of Oxford, and}
	\country{}                    %
}
\affiliation{
	\institution{Saarland University}
	\country{}                    %
}

\author{Luke Ong}
\affiliation{
	\institution{University of Oxford}
	\country{}                    %
}

\usepackage{microtype}

\usepackage{amsmath}
\usepackage{amssymb}
\usepackage{stmaryrd}
\usepackage{tikz}
\usetikzlibrary{decorations.pathmorphing}
\usetikzlibrary{patterns}
\usepackage{tcolorbox}
\tcbuselibrary{breakable}

\usepackage{subcaption}

\usepackage{bussproofs}
\usepackage{multirow}

\usepackage{imakeidx}

\usepackage{enumitem}

\usepackage{stackengine,trimclip,scalerel}

\usepackage{tasks}
\usepackage{mathtools}
\usepackage{soul}

\input{makrox.tex}

\usepackage{todonotes}

\renewcommand{\blist}[1]{\big\{ #1 \big\}}
\renewcommand{\Blist}[1]{\Big\{ #1 \Big\}}

\usepackage{multicol}
\makeatletter
	{\end{multicols}\if@restonecol\onecolumn\else\clearpage\fi}
\makeatother

\def\proofsnamefont{\itshape}
\def\proofsindent{\noindent}

\newenvironment{proofs}[1][Proof Sketch]{\par
	\pushQED{\qed}%
	\normalfont %
	\trivlist
	\item[\proofsindent\hskip\labelsep
	{\proofsnamefont #1.}]\ignorespaces
}{%
	\popQED\endtrivlist%
}

\newenvironment{theoremRE}[1]
{\noindent\textbf{Restatement of Thm.~#1.} \it}%
{}

\newenvironment{lemmaRE}[1]
{\noindent\textbf{Restatement of Lem.~#1.} \it}%
{}

\makeindex

\begin{document}

\title{On Probabilistic Termination of Functional Programs with Continuous Distributions}

\begin{abstract}
	We study termination of higher-order probabilistic functional programs with recursion, stochastic conditioning and sampling from continuous distributions. 
	
	Reasoning about the termination probability of programs with continuous distributions is hard, because the enumeration of terminating executions cannot provide any non-trivial bounds. 
	We present a new operational semantics based on \emph{traces of intervals}, which is sound and complete with respect to the standard sampling-based semantics, in which (countable) enumeration can provide arbitrarily tight lower bounds.
	Consequently we obtain the first proof that deciding almost-sure termination (AST) for programs with continuous distributions is $\Pi^0_2$-complete.
	We also provide a compositional representation of our semantics in terms of an intersection type system. 
	
	In the second part, we present a method of proving AST for \emph{non-affine programs}, i.e., recursive programs that can, during the evaluation of the recursive body, make \emph{multiple} recursive calls (of a first-order function) from \emph{distinct} call sites.
 	Unlike in a deterministic language, the number of recursion call sites has direct consequences on the termination probability. 
 	Our framework supports a proof system that can verify AST for programs that are well beyond the scope of existing methods.
 	
 	We have constructed prototype implementations of our method of computing lower bounds of termination probability, and AST verification.  
\end{abstract}

\ccsdesc[500]{Theory of computation~Operational semantics}
\ccsdesc[500]{Theory of computation~Program analysis}
\ccsdesc[500]{Theory of computation~Program verification}

\keywords{almost-sure termination, probabilistic programs, sampling-style operational semantics, intersection types, random walk}

\maketitle

\section{Introduction}
\label{sec:intro}

{Probabilistic (or randomised) programs have long been recognised as essential to the efficient solution of many algorithmic problems \cite{Rabin1976,MotwaniR1995,DBLP:books/daglib/0012859}.}
Recently, in probabilistic programming \cite{DBLP:conf/icse/GordonHNR14,rainforth2017Automating,vandemeent2018Introduction}, probabilistic programs, augmented with stochastic conditioning constructs, have been used as a means of expressing generative models whose posterior probability can be computed by general-purpose inference engines.
Though sampling from discrete distributions (such as binary probabilistic branching) can be considered algorithmically adequate\footnote{in the sense that they are enough to make any Turing complete programming language universal for probabilistic Turing machine \cite{ProbTuringMasch,DBLP:journals/ita/LagoZ12}} for probabilistic computation, the generation of real-world data---a basic capability expected of generative models---requires expressivity of the whole gamut of continuous distributions.
For this reason, sampling from continuous distributions is an essential feature of probabilistic programming languages. (See e.g.~Church \cite{DBLP:conf/uai/GoodmanMRBT08}, Stan \cite{carpenter2017stan}, Anglican \cite{DBLP:conf/pkdd/TolpinMW15}, Gen \cite{cusumano-towner2019Gen}, Pyro \cite{bingham2019Pyro}, Edward \citep{tran2016edward} and Turing \cite{ge2018Turing}.)

In this work we study a central property of probabilistic programs: \emph{termination}. 
In non-probabilistic (possibly nondeterministic) computation, termination is a purely qualitative, boolean property.
However, with randomness in the control flow, termination is characterised by a scalar quantity: the \emph{probability of termination}.
We say that a program is \emph{almost-surely terminating} (AST) if a run of it terminates with probability $1$.

Guarantees and bounds on the probability of termination are important both when viewing probabilistic programs as algorithmic solutions but also in the emerging field of probabilistic programming.
When a probabilistic program implements a solution to an algorithmic problem, one naturally requires the computation to terminate with a high (lower bounded) probability, usually $1$.
In probabilistic programming, lower bounds and guarantees of AST are equally important. 
Indeed, it is standard for designers and implementors of probabilistic programming systems to regard non-AST programs as defining invalid models, and hence inadmissible (see e.g.~\citep[\S 4.3.2]{rainforth2017Automating} and \cite{DBLP:conf/uai/GoodmanMRBT08}).
Moreover \cite{MakOPW20} have recently shown that AST programs have density (a.k.a.~weight) functions that are differentiable almost everywhere.
This is significant, because the latter property is a precondition for the correctness of some of the most scalable inference algorithms, such as Hamiltonian Monte Carlo \cite{DBLP:conf/aistats/ZhouGKRYW19,Nishimura2020b} and reparameterised gradient variational inference \cite{DBLP:conf/nips/LeeYY18}.
AST is thus a precondition for the correctness of inference algorithms and important both in theory and practice.

In this paper we tackle two key questions: computation of lower bounds on the probability of termination, and AST verification. %
While there has been much progress in the termination analysis of probabilistic programs with discrete distributions \cite{DBLP:conf/mfcs/KaminskiK15, DBLP:conf/lics/KobayashiLG19, DBLP:journals/fmsd/BrazdilEKK13}, programs with continuous distributions have received comparatively little attention.
Many methods and proofs hinge on the countable nature inherent to discrete distributions \cite{DBLP:conf/lics/KobayashiLG19,DBLP:journals/toplas/LagoG19,DBLP:journals/pacmpl/McIverMKK18,DBLP:series/mcs/McIverM05,DBLP:conf/ppdp/BreuvartL18,DBLP:journals/jacm/KaminskiKMO18,DBLP:conf/lics/OlmedoKKM16}.
It is not at all obvious if they can be extended to systems with continuous distributions. 

Using an idealised functional language with continuous samples and stochastic conditioning, %
we provide partial answers to these questions.
On the one hand, we give a definitive answer to the lower bound problem, and precisely determine the complexity of various termination problems in the arithmetic hierarchy. 
On the other hand, we provide a sound (but incomplete) proof method for AST which can be seen as orthogonal to \cite{DBLP:journals/toplas/LagoG19}.
\vspace{-0.01cm}%

\subsection{High Level Overview}\label{sec:overview}

\subsubsection*{Lower Bound Computation}

In languages with discrete distributions, evaluation can be seen as a step-indexed probability mass on terms \cite{DBLP:journals/toplas/LagoG19, DBLP:conf/popl/EhrhardTP14,DBLP:conf/lics/KobayashiLG19}.
By enumerating terminating executions, we can iteratively compute arbitrarily tight lower bounds on the probability of termination.
As a direct consequence, AST is a decision problem in $\Pi^0_2$ \cite{DBLP:conf/mfcs/KaminskiK15}, the second level of the arithmetic hierarchy \cite{kleene1955}\footnote{The class $\Pi^0_n$ in the arithmetic hierarchy contains a language $\mathcal{L}$ iff there exists a decidable relation $R(x, y_1, \cdots, y_n)$ such that $x \in \mathcal{L} \Leftrightarrow \forall y_1. \exists y_2. \forall y_3 \cdots . R(x, y_1, \cdots, y_n)$. $\Sigma^0_n$ is defined analogously starting with an existential instead of universal quantifier.
$\Sigma^0_1$ is thus the class of recursive enumerable languages. 
Almost-sure termination means that for \emph{all} (rational) termination probability $\delta$ strictly smaller than $1$, there \emph{exists} some finite set of terminating execution $T$ whose weight is at least $\delta$, making it a problem contained in $\Pi^0_2$.}.
In languages that admit continuous distributions, we cannot assign probability mass to terms directly. 
Rather, by viewing a probabilistic program as a deterministic program parameterised by an \emph{execution trace} (or simply, trace) (i.e.~the sequence of random draws made during the execution), we can organise such traces into a measure space \cite{DBLP:journals/jcss/Kozen81,DBLP:conf/icfp/BorgstromLGS16}.
The probability of termination can then be defined as the measure of all traces on which the program terminates \cite{MakOPW20}.
However, in general, a single terminating execution (or even a countable set thereof) cannot be assigned any positive probability measure. 
This leaves open problems such as sound computation of lower bounds, and the exact complexity of deciding AST. 

We approach these problems by introducing a novel operational semantics based on \emph{interval traces}, which are a summarisation of the relevant traces.  
We show soundness and completeness w.r.t.~the sampling-style semantics \cite{DBLP:conf/icfp/BorgstromLGS16}.
Instead of analysing a program using uncountably many traces, we work with interval traces, where only countably many such traces suffice. 
This yields an effective procedure to compute lower bounds on termination probability, enabling the first proof that deciding AST in the presence of continuous distributions is $\Pi^0_2$-complete (under mild assumptions on the primitive functions). 
Further, we show that positive almost sure termination (PAST) (i.e., finite expected time to termination) is $\Sigma^0_2$-complete, assuming the program is AST. 
For general PAST, we can only infer a (possibly non-tight) upper bound of $\Delta^0_3$.
This does \emph{not} match the $\Sigma^0_2$ bounds known for discrete distributions as a proof of this bound hinges on a countable set of executions \cite{DBLP:conf/mfcs/KaminskiK15}. 
See \refSection{ibs}.

In addition we give an alternative presentation of our semantics as an intersection type system in \refSection{intersection}.
Our system %
extends \cite{DBLP:conf/ppdp/BreuvartL18} and \cite{DBLP:conf/popl/EhrhardTP14} to languages with continuous distributions; moreover, both the probability of termination \emph{and} the expected time to termination can be obtained as the least upper bound of all derivations.
This gives a type-based, compositional method for lower bound computation.

\subsubsection*{AST Verification}

While our computation of lower bounds gives a $\Pi^0_2$ decision procedure for AST, it is not really effective for AST verification. 
Many of the recent advances in the development of AST verification methods \cite{DBLP:conf/pldi/ChenH20,DBLP:conf/cav/ChakarovS13,DBLP:conf/popl/FioritiH15,DBLP:journals/pacmpl/McIverMKK18,DBLP:conf/aplas/HuangFC18,DBLP:conf/popl/ChatterjeeNZ17,DBLP:journals/pacmpl/AgrawalC018,DBLP:conf/cav/ChatterjeeFG16,DBLP:conf/lics/OlmedoKKM16,DBLP:journals/pacmpl/Huang0CG19} are concerned with loop-based programs.
{We can view such loops as tail-recursive programs that, in particular, are \emph{affine recursive}, i.e., in each evaluation (or run) of the body of the recursion, recursive calls are made from at most one call site} \cite[\S 4.1]{DBLP:journals/toplas/LagoG19}.
By contrast, many probabilistic programming languages allow for richer recursive structures \cite{DBLP:conf/pkdd/TolpinMW15,DBLP:conf/uai/GoodmanMRBT08,DBLP:journals/corr/MansinghkaSP14}.
We propose a new verification method for probabilistic programs that are defined by {\emph{non-affine recursion}, i.e., in the evaluation of the body of the recursion, multiple recursive calls can be made from \emph{distinct} call sites.
(Note that whether a program is affine recursive cannot be checked by just counting textual occurrences of variables.)}

\begin{example}[Running Example]\label{ex:3dprinting}
Consider an unreliable 3d printing company.
Unfortunately, for every printing, the outcome is acceptable with only probability $p$; if it is unacceptable, 
{reprinting must take place on the following day, and thus, the process is repeated.}
We can model this scenario, starting with a single job, as the following program%
{\small\begin{align}\label{prog:intro1}
	\Big(\mu^\varphi_x. \myif \,\sample \leq p \mythen x \myelse \varphi (x+\num{1})\Big) \, \num{1}\tag*{\textbf{(1)}}
\end{align}}%
{where $\mu^\varphi_x.(\cdot)$ is a fixpoint constructor (that binds the variable $\varphi$ to the fixpoint)}, 
and $\sample$ evaluates to a random draw from the uniform distribution on $[0, 1]$. 
The value returned by the program is the number of days needed to complete the job. 
Luckily, as the program is AST for all success probabilities $p \in (0, 1]$, the company can assure its customers that it will finish the job eventually. 
However, in a bid to drum up business, a new quality policy is introduced.
The manager advises their customers: ``Each day our print attempt fails, we will print an additional copy for you.''
We model the situation as follows:
{\small\begin{align}\label{prog:intro2}
	\Big(\mu^\varphi_x. \myif \,\sample \leq p \mythen x \myelse \varphi \big(\varphi (x+\num{1}) \big) \Big) \num{1} \tag*{\textbf{(2)}}
\end{align}}%
Soon after implementing the new policy, it was noticed that some of the print jobs could never be completed.
Phrased differently: Program \ref{prog:intro2} is no longer AST for every $p \in (0, 1]$.

This example illustrates that non-affine recursion, 
as exhibited in program \ref{prog:intro2},
can complicate the analysis of termination. 
While the affine program \ref{prog:intro1} is clearly AST for every $p > 0$, program \ref{prog:intro2} is not.
It turns out that \ref{prog:intro2} is AST if and only if $p \geq \tfrac{1}{2}$; %
and in case $p = \tfrac{1}{2}$, while the process is AST, the expected time to termination is infinite.  
It is unsurprising that termination depends on the number of recursive calls, as termination itself is a quantitative property.
\end{example}

Termination analysis of non-affine recursive probabilistic programs does not seem to have received much attention.
Methods such as those presented in \cite{DBLP:journals/toplas/LagoG19} explicitly restrict to affine programs and are unsound otherwise. 
Our method for the analysis of non-affine recursive programs can be viewed as orthogonal to \cite{DBLP:journals/toplas/LagoG19}: {while they restrict to affine programs and investigate the recursive function argument for size information, we accept the function argument without examination, and admit non-affine programs.} 
We call our methods \emph{counting-based}, as we over-approximate the recursive behaviour by counting recursive calls from distinct call sites,
thus reducing AST analysis to the analysis of a random walk for which we show linear decidability.
See \refSection{nonaffine}.
Our method is the basis of an AST proof system that can verify programs (including the simple example above) {well beyond} the reach of existing methods (\refSection{proofsystem}). 
As a simple corollary, we obtain a functional \emph{generalisation} of the \emph{zero-one law} for termination of while-programs \cite[\S 2.6]{DBLP:series/mcs/McIverM05}\footnote{\label{fnote:0-1-law}{The \emph{zero-one law} states that a while-loop is almost-surely terminating if there is a positive lower bound on the probability of exiting it.}}

\subsubsection*{Contributions}

Our main contributions are as follows:
\begin{itemize}
	\item We propose a new sound and complete interval-based semantics that enables lower bound computation.
	We obtain a first proof that the (CbN) AST (resp.~PAST) decision problem is, under mild assumptions on primitive functions, $\Pi^0_2$-complete (resp. $\Sigma^0_2$-complete) even in the presence of continuous distributions. 
	
	\item We give a local representation of our semantics as an intersection type system where both the probability of termination and expected time to termination are characterised as the least upper bound over all derivations. 
	
	\item We provide a new proof method for AST verification of non-affine recursive programs. 
  {We show how our proof system can be automated.} 
\end{itemize}

Our theoretical results give rise to practical algorithms. We provide prototype implementations for both lower bound computation and AST verification based on our novel semantics and proof system respectively\footnote{\label{fnote:source}Both tools are available at \url{https://github.com/ravenbeutner/astnar}} (see \refSection{implementation}). 
{Missing proofs and further discussions can be found in the appendix.}

\section{Statistical PCF (SPCF)}
\label{sec:spcf}

We begin by introducing some basics of probability theory and presenting our language of study.

\subsection{Basic Probability Theory}

A \emph{$\sigma$-algebra} on a set $\Omega$, typically written $\Sigma_\Omega$, is a collection of subsets of $\Omega$ such that $\Omega \in \Sigma_\Omega$, and $\Sigma_\Omega$ is closed under complementation and countable unions (and hence countable intersections).
A \emph{measurable space} is a pair $(\Omega, \Sigma_\Omega)$ where $\Omega$ is a set (of outcomes) and $\Sigma_\Omega$ is a $\sigma$-algebra on $\Omega$.  
A function $f: \Omega_1 \to \Omega_2$ between measurable spaces, $(\Omega_1, \Sigma_{\Omega_1})$ and $(\Omega_2, \Sigma_{\Omega_2})$, is called \emph{measurable} if for every $A \in \Sigma_{\Omega_2}$, $f^{-1}(A) \in \Sigma_{\Omega_1}$. 
A \emph{measure} on $(\Omega, \Sigma_\Omega)$ is a function $\mu : \Sigma_\Omega \to \overline{\posreal}$ that satisfies $\mu(\emptyset) = 0$ and is $\sigma$-additive: if $\{A_i\}_{i \in \natnum}$ is a countable family of pairwise disjoint sets from $\Sigma_\Omega$ then $\mu(\cup_i A_i) = \sum_i \mu(A_i)$. 
If $\mu(\Omega) \leq 1$ we call $\mu$ a subprobability measure and if $\mu(\Omega) = 1$ we call it a probability measure (or distribution). 
For the $n$-dimensional Euclidean space $\real^n$ we write $\Sigma_{\real^n}$ for the Borel $\sigma$-algebra over $\real^n$, which is the smallest $\sigma$-algebra that contains all open and closed $n$-dimensional boxes.
In the special case of $n = 1$, this is the set generated by all open (and closed) intervals. 
The $n$-dimensional Lebesgue measure, denoted $\lambda_n$, is the unique measure on $(\real^n, \Sigma_{\real^n})$ that satisfies $\lambda_n(\myint{a_1, b_1} \times \cdots \times \myint{a_n, b_n}) = \prod_{i=1}^n (b_i - a_i)$.

\subsubsection*{Discrete Sample Space}

In case $\Omega$ is countable, we often work with the powerset $2^\Omega$ as the trivial $\sigma$-algebra. 
Every probability measure is then uniquely determined by a \emph{probability mass function (pmf)}, a function $p : \Omega \to \intreal$ with $\sum_{x \in \Omega} p(x) = 1$.
Every pmf $p$ gives rise to a probability measure by defining $\mu(A) \defined \sum_{x \in A} p(x)$; conversely, for every probability measure $\mu$ on the powerset we can recover a generating pmf by defining $p(x) \defined \mu(\{x\})$. 
A subprobability mass function is defined analogously.

\subsection{SPCF}

Statistical PCF (SPCF) is an extension of PCF \cite{DBLP:journals/tcs/Plotkin77} with support to sample\footnote{{Sampling from other real-valued distributions can be obtained from $\sample$ by applying the inverse of the distribution’s cumulative distribution function; see e.g.~\cite[\S 2.3.1]{RubinsteinK2017}.}} from the uniform distribution on $[0, 1]$ and condition executions (see \cite{DBLP:conf/icse/GordonHNR14}).
Terms in SPCF are implicitly parametrised over a set $\mathbb{F}$ of \emph{measurable} functions $f : \real^n \to \real$ that model primitive operations.
Each function $f \in \mathbb{F}$ has an arity $|f| \geq 0$.
The sets of terms and values are defined by the following grammar where $x$ and $\varphi$ are distinct variables (from a fixed denumerable set of symbols), $r \in \real$ and $f \in \mathbb{F}$:
\begin{align*}
	V &\defined x \mid \num{r} \mid \lambda x. M \mid \mu^\varphi_x. M\\
	M, N, P &\defined V \mid M N \mid \myif(M, N, P) \mid f(M_1, \cdots, M_{|f|})\\
	 &\quad\quad\mid \sample \mid \score(M) 
\end{align*}%
As usual, we identify terms modulo $\alpha$-conversion.\index{$M, N, P$, standard SPCF terms} \index{$V$, SPCF value}
The fixpoint constructor, $\mu^\varphi_x.(\cdot)$, binds the recursively defined function $\varphi$ and its argument $x$.
We abbreviate\footnote{{Our conditional statement, $\myif(P, M, N)$, branches on whether $P \leq 0$.}} $$M \oplus_P N \defined \myif~(\sample - P, M, N)$$ in the style of \cite{DBLP:series/mcs/McIverM05} and write $M \oplus N$ for $M \oplus_{\num{.5}} N$.
We type terms using a standard simple type system with types defined by $\alpha, \beta \defined \typeReal \mid \alpha \to \beta$.
A selection of typing rules is given in \refFig{simpleTypes} (see appendix). %
We denote the set of typable SPCF terms by $\Lambda$ and its subset of closed terms by $\Lambda_0$.

{In this paper we consider both call-by-name (CbN) and call-by-value (CbV) evaluation strategies.
We use CbN for the first part of this paper, as the results (especially those about intersection types) are cleaner this way \cite{DBLP:conf/popl/EhrhardTP14,DBLP:conf/ppdp/BreuvartL18,DBLP:conf/lics/KobayashiLG19}. 
(Our CbN SPCF can express CbV computation at base types, 
giving it a suitable algorithmic expressiveness; c.f.~\cite{DBLP:journals/jacm/EhrhardPT18}.)
We switch to CbV SPCF when presenting our AST proof system, thereby enabling a more straightforward comparison to related approaches such as \cite{DBLP:journals/toplas/LagoG19}.}

\begin{figure}
	\begin{tcolorbox}[colback=white, colframe=black, arc=0mm, boxrule=1pt,bottom=0pt, top=0pt,after skip=0pt,before skip=0pt]
		{\small
			
			\begin{minipage}{0.5\textwidth}
				\begin{prooftree}
					\AxiomC{}
					\UnaryInfC{$\Gamma \vdash \sample : \typeReal $}
				\end{prooftree}
			\end{minipage}%
			\begin{minipage}{0.5\textwidth}
				\begin{prooftree}
					\AxiomC{$\Gamma \vdash M : \typeReal$}
					\UnaryInfC{$\Gamma \vdash \score(M) : \typeReal $}
				\end{prooftree}
			\end{minipage}
			
			\begin{minipage}{0.5\textwidth}
				\begin{prooftree}
					\AxiomC{$\Gamma, \varphi: \alpha \to \beta, x:\alpha \vdash M : \beta$}
					\UnaryInfC{$\Gamma \vdash \mu^\varphi_x. M : \alpha \to \beta $}
				\end{prooftree}
			\end{minipage}%
			\begin{minipage}{0.5\textwidth}
				\begin{prooftree}
					\AxiomC{$\{\Gamma \vdash M_i : \typeReal\}_{i=1}^{|f|}$}
					\UnaryInfC{$\Gamma \vdash f(M_1, \cdots, M_{|f|}) : \typeReal $}
				\end{prooftree}
		\end{minipage}}
	\end{tcolorbox}
	
	\caption{Selection of SPCF typing Rules} \label{fig:simpleTypes}
\end{figure}

\subsection{Operational Semantics}

We give a sampling-style operational semantics for SPCF.
The idea (going back to Kozen \cite{DBLP:journals/jcss/Kozen81}) is to evaluate a term $M$ together with a sequence of (fixed) probabilistic outcomes for each \sample\ statement~\cite{DBLP:conf/icfp/BorgstromLGS16,MakOPW20}.
We then generate a probabilistic interpretation of programs by endowing the set of traces with a measure.

\subsubsection*{CbN SPCF}

\begin{figure*}
	\begin{tcolorbox}[colback=white, colframe=black, arc=0mm, boxrule=1pt,bottom=0pt, top=0pt,after skip=0pt,before skip=0pt]
		
		\begin{minipage}{0.33\textwidth}
			\begin{prooftree}
				\AxiomC{}
				\UnaryInfC{$\langle (\lambda x. M) N, \tr \rangle \to \langle M[N/x], \tr \rangle$}
			\end{prooftree}
		\end{minipage}
		\begin{minipage}{0.33\textwidth}
			\begin{prooftree}
				\AxiomC{}
				\UnaryInfC{$\langle (\mu^\varphi_x. M) N, \tr \rangle \to \langle M[N/x, (\mu^\varphi_x. M)/\varphi], \tr \rangle $}
			\end{prooftree}
		\end{minipage}
		\begin{minipage}{0.33\textwidth}
			\begin{prooftree}
				\AxiomC{}
				\UnaryInfC{$\langle \sample, r \, \tr \rangle \to \langle \num{r}, \tr \rangle $}
			\end{prooftree}
		\end{minipage}

		\begin{minipage}{0.33\textwidth}
			\begin{prooftree}
				\AxiomC{$r \leq 0$}
				\UnaryInfC{$\langle \myif(\num{r}, N, P), \tr \rangle \to \langle N, \tr \rangle $}
			\end{prooftree}
		\end{minipage}
		\begin{minipage}{0.33\textwidth}
			\begin{prooftree}
				\AxiomC{$r > 0$}
				\UnaryInfC{$\langle \myif(\num{r}, N, P), \tr \rangle \to \langle P, \tr \rangle $}
			\end{prooftree}
		\end{minipage}
		\begin{minipage}{0.33\textwidth}
			\begin{prooftree}
				\AxiomC{$r \geq 0$}
				\UnaryInfC{$\langle \score(\num{r}), \tr \rangle \to \langle \num{r}, \tr \rangle $}
			\end{prooftree}
		\end{minipage}
		
		\begin{minipage}{0.5\textwidth}
			\begin{prooftree}
				\AxiomC{}
				\UnaryInfC{$\langle f(\num{r_1}, \cdots, \num{r_{|f|}}), \tr \rangle \to \langle \num{f(r_1, \cdots, r_{|f|})}, \tr \rangle $}
			\end{prooftree}
		\end{minipage}
		\begin{minipage}{0.5\textwidth}
			\begin{prooftree}
				\AxiomC{$\langle R, \tr \rangle \to \langle M, \tr' \rangle$}
				\UnaryInfC{$\langle E[R], \tr \rangle \to \langle E[M], \tr' \rangle $}
			\end{prooftree}
		\end{minipage}
		
	\end{tcolorbox}
	\caption{Call by Name small-step reduction for SPCF.} \label{fig:cbnRules}
\end{figure*}

We define the set of traces $\stdtrset$\index{$\stdtrset$, the set of standard traces} as all finite sequences of real numbers from $\intreal := \{r \in \real \mid 0 \leq r \leq 1\}$), i.e., $\stdtrset \defined \intreal^* = \bigcup_{n \in \natnum} \intreal^n$. 
We let $\tr$ range over elements in $\stdtrset$, denote the empty trace with $\epsilon$; for $r \in \intreal$ write $r$ for the one element trace; and $\tr_1, \tr_2$ for concatenation.\index{$\tr$, a standard trace}
The set of CbN redexes and evaluation contexts is defined by:
\begin{align*}
	R &\defined (\lambda x. M) N \mid (\mu^\varphi_x. M) N \mid \myif(\num{r}, N, P) \\
	&\quad\quad\mid f(\num{r_1}, \cdots, \num{r_{|f|}}) \mid \sample \mid \score(\num{r})\\
	E &\defined [\cdot] \mid E M \mid \myif(E, N, P) \mid \score(E)  \\
	&\quad\quad\mid f(\num{r_1}, \cdots, \num{r_{k-1}}, E, M_{k+1}, \cdots, M_{|f|}) 
\end{align*}
Given a context $E$ and a term $M$ the (capture-permitting) substitution $E[M]$ is defined in the obvious way. 
An easy induction establishes that every $M \in \Lambda_0$ is either a value or there are \emph{unique} $E$ and $R$, s.t., $M = E[R]$ (see e.g.~\cite{DBLP:conf/icfp/BorgstromLGS16}). 
The small-step reduction relation has the from $\langle M, \tr \rangle \to \langle M', \tr'\rangle$ where $M, M'$ are terms and $\tr, \tr'$ are traces.
It is defined inductively by the rules given in \refFig{cbnRules} where $M[N_i/x_i]_{i}$ denotes standard capture-avoiding substitution \cite{DBLP:books/daglib/0067558}. 
Note that our reduction does not enjoy \emph{progress}, as e.g.~redex $\score(\num{r})$ cannot reduce if $r < 0$.

The score constructs is used to stochastic condition of executions (see e.g. \cite{DBLP:conf/icse/GordonHNR14}) by weighting each execution \cite{DBLP:conf/icfp/BorgstromLGS16}. 
As this work is a study of termination properties, we elide the weight parameter used for stochastic conditioning as the weight of a execution is irrelevant for the termination behaviour\footnote{The weight function can be seen as a function mapping terminating traces to weights (i.e., $\real$). The denotation of a program is then the Lebesgue integral of this weight functions over the set of terminating traces (\cite[\S 3.4]{DBLP:conf/icfp/BorgstromLGS16}). To get e.g., the almost-everywhere differentiability of the weight function (needed for correct inference), it is sufficient to show that the measure of the set of termination traces is $1$ (irrespective of the weight on these traces) \cite[\S 4.3]{MakOPW20}. The \score-construct has, nevertheless, a subtle effect on  termination as we require the conditioned value to be positive.}.

\subsubsection*{A Measure on Traces}

{To interpret probabilistic programs using traces, we first need to endow the set of traces with a measure.
We cannot assign probability mass to individual traces directly, as there are uncountably many traces.}
Instead we define a suitable measurable space of program traces following \cite{DBLP:conf/icfp/BorgstromLGS16}.
Let $\Sigma_{\intreal^n}$ be the Borel $\sigma$-algebra on $\intreal^n$  (We set $\Sigma_{\intreal^0} \defined \big\{ \emptyset, \{\epsilon\} \big\}$).
We can then define a $\sigma$-algebra on traces ($\Sigma_{\stdtrset})$ and a measure ($\mu_{\stdtrset}$) by:
\begin{align*}
	\textstyle\Sigma_{\stdtrset} &\textstyle\defined \{ \biguplus_{n \in \natnum} B_n \mid B_n \in \Sigma_{\intreal^n} \}\\
	\textstyle\mu_{\stdtrset}\big(\biguplus_{n \in \natnum} B_n\big) &\textstyle\defined \sum_{n \in \natnum} \lambda_n(B_n)
\end{align*}
As shown in \cite[Lem.~7 \& 8]{DBLP:conf/icfp/BorgstromLGS16}, $(\stdtrset, \Sigma_{\stdtrset})$ is a measurable space and $\mu_{\stdtrset}$ a ($\sigma$-finite) measure on $(\stdtrset, \Sigma_{\stdtrset})$.

\subsection{Probabilistic Termination}\label{sec:AST}

With $\to^n$ we denote the $n$-fold self-composition, and with $\to^*$ the reflexive-transitive closure, of $\to$.
We define 
$$\termTr{M} \defined \big\{\tr \in \stdtrset \mid \exists V: \; \langle M, \tr \rangle \to^* \langle V, \epsilon \rangle\big\}$$%
as the set of traces on which a term $M$ terminates, which is measurable (similar to \cite[Lem.~9]{DBLP:conf/icfp/BorgstromLGS16}).
\index{$\termTr{M}$, the set of traces on which $M$ terminates}
As shown in \cite[Lem.~7]{MakOPW20}, $\mu_{\stdtrset}\big( \termTr{M} \big) \leq 1$; we are therefore justified in calling the interpretation $\mu_{\stdtrset}\big( \termTr{M} \big)$ a ``probability''.

\begin{definition}
  The \emph{probability of termination} of $M \in \Lambda_0$ is defined by $\termProb{M} \defined \mu_{\stdtrset}(\termTr{M})$. \index{$\termProb{M}$, the probability of termination of $M$}
	$M$ is called \emph{almost-surely terminating} (AST) if $\termProb{M} = 1$.
\end{definition}

\subsubsection*{Positive Almost-Sure Termination}

An even stronger property than AST is {finiteness of the expected time to termination}. 
For any trace $\tr \in \termTr{M}$ we define $\numberSteps{\tr}{M} \in \natnum$ as the unique number $n$ such that %
$\langle M, \tr \rangle \to^n \langle V, \epsilon \rangle$ for some value $V$. \index{$\numberSteps{\tr}{M}$, number of reduction steps of trace $\tr$ on $M$}
For any $n \in \natnum$ we define
$$\termTr{M}^{\leq n} \defined \big\{\tr \in \termTr{M} \mid \numberSteps{\tr}{M} \leq n\big\}$$
as the set of traces on which termination occurs within $n$ steps, which is measurable. {We define $\termTr{M}^{n}$ analogously.}

\begin{definition}\label{def:past}
	For $M \in \Lambda_0$ we define the \emph{expected time to termination}, $\expectedTermSteps{M} \in \overline{\posreal}$, by
	\index{$\expectedTermSteps{M}$, expected termination time of $M$}
	$$\textstyle\expectedTermSteps{M} \defined \sum_{n=0}^{\infty} \Big(1- \mu_{\stdtrset} \big(\termTr{M}^{\leq n}\big) \Big)$$
	$M$ is \emph{positive almost-surely terminating} if $\expectedTermSteps{M} < \infty$.
\end{definition}

It is easy to see that any program that is PAST is also AST.
Following \cite{DBLP:conf/mfcs/KaminskiK15}, $\expectedTermSteps{M}$ can be phrased as $\sum_{n=0}^\infty \mathbb{P} \big($``$M$ runs for more than $n$ steps''$\big) = \sum_{n=0}^\infty \big(1 - \mathbb{P} ($``$M$ terminates within $n$ steps''$)\big)$, with the latter expressed in Def.~\ref{def:past}. 
We can show that, provided $M$ is AST, the expected time to termination is the expected value of the random variable that gives the number of reduction steps:

\begin{lemma}\label{lem:altExp}
	If $M$ is AST, $\expectedTermSteps{M} = \sum\limits_{n = 0}^{\infty} \mu_{\stdtrset} \Big( \termTr{M}^{n} \Big) \cdot n$
\end{lemma}

\subsubsection*{CbV SPCF}

{Our CbV SPCF is essentially the system of ~\cite{MakOPW20}, except that we use a simpler CbV fixpoint reduction rule.}

\section{Interval-based Semantics}
\label{sec:ibs}

{It is impractical to use the standard trace-based (or sampling-style) semantics to reason about termination properties of SPCF programs, because the trace measure $\mu_{\stdtrset}$ is continuous.
Suppose we are interested in the decidability of the \emph{lower bound question}: does a term terminate with probability \emph{strictly} greater than $p$?}
For discrete distributions, this problem is r.e.~(in $\Sigma^0_1$) as we can enumerate terminating paths until the sum of the weight of those paths exceeds $p$ \cite{DBLP:conf/lics/KobayashiLG19,DBLP:conf/mfcs/KaminskiK15}. 
In the presence of continuous distributions, this is no longer possible. 
A well-known property of the Lebesgue measure on $\intreal^n$ (inherited by the trace measure $\mu_{\stdtrset}$) is that every countable set of elements is a null set. 
So even if we can identify a countably infinite set of traces $A \subseteq \termTr{M}$, we cannot obtain any non-trivial lower bound on $\termProb{M}$. 
{Thus the semantics itself cannot be used to settle such complexity questions as whether the lower bound problem for SPCF is in $\Sigma^0_1$, or whether the AST problem is in $\Pi^0_2$, or whether the PAST problem in $\Sigma^0_2$.}
In this section, we introduce a novel operational semantics for SPCF by executing terms parameterised by a trace of \emph{intervals}. 
{We demonstrate that this semantics, which is complete w.r.t.~the trace-based semantics, is well-suited to the derivation of lower bounds.}
The completeness hinges on the observation that, under mild restrictions on primitive functions, interval-based reasoning can effectively abstract actual traces. 
{This is the basis of our positive answer to the questions above.}

\subsubsection*{Syntax of Interval Terms}

{We adjust the syntax of terms slightly and treat intervals as constant symbols of type $\typeReal$.}
We define interval values and interval terms as follows where $a \leq b \in \real$. 
\begin{align*} 
	\calV &\defined x \mid \num{\myint{a, b}} \mid \lambda x. \calM \mid \mu^\varphi_x. \calM\\
	\calM, \calN, \calP &\defined \calV \mid \calM \calN \mid \myif(\calM, \calN , \calP) \mid f(\calM_1, \cdots, \calM_{|f|})\\
	&\quad\quad\quad \mid \sample \mid \score(\calM)
	\end{align*}
Our simple type system extends naturally. \index{$\calM, \calN, \calP$, interval terms} \index{$\calV$, interval term value}
We denote the set of (closed, bounded) intervals by $\interval$, and write $\interval_{0, 1} \defined \{\myint{a, b} \mid a, b \in \real, 0 \leq  a \leq b \leq 1\}$ as the set of intervals with endpoints between $0$ and $1$.
With $\interval^\mathbb{Q}$ and $\interval_{0, 1}^\mathbb{Q}$ we denote the sets $\interval$ and $\interval_{0, 1}$ respectively, restricted to \emph{rational endpoints}.

\begin{definition}
	We call $f: \real^n \to \real$ \emph{interval preserving} (resp.~$\ration$-\emph{interval preserving}) if there is a function $\hat{f} : \real^{2n} \to \interval$ (resp.~$\hat{f} : \ration^{2n} \to \interval^\ration$) such that for every sequence of intervals $\myint{a_1, b_1}, \cdots, \myint{a_n, b_n} \in \interval$ (resp.~$\in \interval^\ration$) we have $f\big(\myint{a_1, b_1} \times \cdots \times \myint{a_n, b_n}\big) = \hat{f}(a_1, b_1, \cdots, a_n, b_n)$, i.e., the image of every $n$-dimensional box (resp.~with rational endpoints) is an interval (resp.~with rational endpoints). 
\end{definition}

We restrict the primitive functions to those that are interval preserving to ensure that interval-based reasoning is compatible with primitive operations.  
As the following shows, most interesting functions (including e.g.~$+, \cdot, -, \exp, |\cdot|, \cdots$) are interval preserving.

\begin{lemma}\label{lem:suf1}
	If $f: \real^n \to \real$ is continuous then $f$ is interval preserving.
\end{lemma}

\begin{figure}[!t]
	\begin{tcolorbox}[colback=white, colframe=black, arc=0mm, boxrule=1pt,left=0pt, right=0pt,bottom=0pt, top=0pt,after skip=0pt,before skip=0pt]
		\footnotesize
		\begin{minipage}{0.5\textwidth}
			\begin{prooftree}
				\def\ScoreOverhang{0pt}
				\AxiomC{$b \leq 0$}
				\UnaryInfC{$\langle \myif(\num{\myint{a, b}}, \calN, \calP), \wpi \rangle \too \langle \calN, \wpi \rangle $}
			\end{prooftree}
		\end{minipage}
		\begin{minipage}{0.49\textwidth}
			\begin{prooftree}
				\def\ScoreOverhang{0pt}
				\AxiomC{$a > 0$}
				\UnaryInfC{$\langle \myif(\num{\myint{a, b}}, \calN, \calP), \wpi \rangle \too \langle \calP, \wpi \rangle $}
			\end{prooftree}
		\end{minipage}
	
		\begin{minipage}{0.5\textwidth}
			\begin{prooftree}
				\def\ScoreOverhang{0pt}
				\AxiomC{}
				\UnaryInfC{$\langle \sample, \myint{a, b} :: \wpi \rangle \too \langle \num{\myint{a, b}}, \wpi \rangle $}
			\end{prooftree}
		\end{minipage}
		\begin{minipage}{0.49\textwidth}
			\begin{prooftree}
				\def\ScoreOverhang{0pt}
				\AxiomC{$0 \leq a$}
				\UnaryInfC{$\langle \score(\num{\myint{a, b}}),\wpi \rangle \too \langle \num{\myint{a, b}}, \wpi \rangle $}
			\end{prooftree}
		\end{minipage}
	
		\begin{minipage}{1\textwidth}
			\begin{prooftree}
				\def\ScoreOverhang{0pt}
				\AxiomC{}
				\UnaryInfC{$\langle f\big(\num{\myint{a_1, b_1}}, \cdots, \num{\myint{a_{|f|}, b_{|f|}}} \big), \wpi \rangle \too \langle \num{\hat{f}(a_1, b_1, \cdots, a_{|f|}, b_{|f|})}, \wpi \rangle $}
			\end{prooftree}
		\end{minipage}
		
	\end{tcolorbox}
	
	\caption{Selection of interval-based reduction rules} \label{fig:intervalSemantics}
\end{figure}

\subsection{Interval-based Syntax and Semantics}

We define the set of interval traces by $\inttrset \defined \bigcup_{n \in \natnum} \interval_{0, 1}^n$\index{$\inttrset$, the set of interval traces}, i.e., finite sequences of intervals with endpoints between 0 and 1 (inclusive). 
We let $\wpi$ range over elements in $\inttrset$.  \index{$\wpi$, a interval trace}
To avoid confusion, we shall refer to elements of $\inttrset$ as \emph{interval traces}, and elements of $\stdtrset$ as \emph{standard traces}.

Redexes and evaluation contexts of interval terms are defined as expected.
As we only replace real-valued numerals with interval-valued, our standard small-step semantics (\refFig{cbnRules}) mostly extends to interval terms.
The specific reduction rules concerning the control flow and primitive functions are given in \refFig{intervalSemantics}.
As a useful intuition, it is helpful to view an interval numeral $\num{\myint{a, b}}$ as an unknown value within that interval. 
As in the standard semantics,  we are interested in the interval traces that lead to a normal form.
$$\termInterval{\calM} \defined \{\wpi \in \inttrset \mid \exists \calV: \langle \calM, \wpi \rangle \, \too^* \langle \calV, \epsilon \rangle\}$$ \index{$\termInterval{\calM}$, the set of terminating interval traces for $\calM$}%
For any $\wpi \in \termInterval{\calM}$, we define $\numberSteps{\wpi}{\calM}$\index{$\numberSteps{\wpi}{\calM}$, number of reduction steps of interval trace $\wpi$ on $\calM$} as the number of reduction steps to termination.

\subsubsection*{Embedding Into Intervals}

While we want to analyse the termination probability of standard terms, our interval-based semantics builds on interval terms.
We define a natural embedding $\toIntervalTerm{(\cdot)}$ that maps every standard term $M$ to the interval term $\toIntervalTerm{M}$ obtained by replacing every numeral $\num{r}$ by the interval numeral $\num{\myint{r, r}}$.
Our soundness and completeness results are now based on the operational behavior of $\toIntervalTerm{M}$ (in the interval semantics), {and they allow us to draw conclusions about the behavior of $M$ (in the standard semantics)}.\index{$\toIntervalTerm{M}$, the natrual embedding of stanrd terms $M$ as a interval term}

\subsection{Soundness}

We now show that the interval-based semantics gives %
lower bounds on the probability of termination in the standard semantics. 
We define the \emph{weight of an interval trace} $\wpi$, denoted by $\omega(\wpi)$, in the obvious way:
$$\textstyle\omega(\myint{a_1, b_1}, \cdots, \myint{a_n, b_n}) \defined \prod_{i=1}^{n} (b_i - a_i)$$

To combine the weight of multiple terminating interval traces we need to ensure that the interval traces are disjoint, i.e., we do not account twice for the same standard trace. 
\begin{definition}
	Two interval traces $\wpi = \myint{a_1, b_1}, \cdots, \myint{a_n, b_n}$ and $\wpi' = \myint{a'_1, b'_1}, \cdots, \myint{a'_m, b'_m}$ are \emph{compatible} if $n \neq m$ or there exists $i$ such that
	$b_i \leq a_i'$ or $b_i' \leq a_i$. 
\end{definition}
 
For example, the four interval traces, $\myint{0, 1}\myint{0, \tfrac{1}{3}}$, $\myint{0, 1}\myint{\tfrac{1}{3}, \tfrac{1}{2}}$, $\myint{0, 1}\myint{\tfrac{3}{4}, 1}$ and $\myint{0, 1}$, are  pairwise compatible. 
For a \emph{countable} set of interval traces $A$ we define $\omega(A) \defined \sum_{\wpi \in A} \omega(\wpi)$;
\index{$\omega(A)$, the cumulative weight of a countable set of standard traces}
if $A \subseteq \termInterval{\calM}$ we also define the expected value of $A$, denoted $\expVal(\calM, A)$, by\index{$\expVal(\calM, A)$, the expected number of steps on a countable set of interval traces $A$}
$$\textstyle\expVal(\calM, A) \defined \sum_{\wpi \in A} \omega(\wpi) \cdot \numberSteps{\wpi}{\calM}$$
We can now state soundness as follows:
\begin{theorem}\label{theo:intSound}
	For every countable set of \emph{pairwise compatible} traces $A \subseteq \termInterval{\toIntervalTerm{M}}$ the following holds:
	\begin{tasks}[style=itemize](2)
		\task $\omega(A) \leq \termProb{M}$ 
		\task $\expVal(\toIntervalTerm{M}, A) \leq \expectedTermSteps{M}$
	\end{tasks}
\end{theorem}

This (perhaps unsurprising) soundness result is the basis of an effective tool to verify lower bounds on $\termProb{M}$ and $\expectedTermSteps{M}$. 
The real force of the interval-based semantics lies in its completeness. 

\subsection{Completeness}

We show that, under mild assumptions on the primitive functions, a countable number of traces for $\toIntervalTerm{M}$ already gives the exact probability of termination $\termProb{M}$. 
Consequently, by an incremental search of terminating interval-traces, we can compute arbitrarily tight lower bounds on $\termProb{M}$.

\begin{example} 
Consider the term 
\[
M = \big(\mu^\varphi_x. \myif \, \sample + \sample - \num{1} \myelse x \myelse \varphi \, x\big) \, \num{0}.
\] 
For the moment we focus on the set of traces on which this term terminates \emph{without} making a single recursive call which is $T = \{r_1 \, r_2 \in \intreal^2 \mid r_1 + r_2 \leq 1\}$.
This set cannot be described by a countable union of interval traces, i.e., there are no interval traces $\{\wpi_i\}_{i \in \natnum}$ such that $\tr \in T \Leftrightarrow \exists i \in \natnum: \tr \triangleleft \wpi_i$, where $\tr \triangleleft \wpi_i$ means that 
$\tr$ refines $\wpi_i$ (see appendix).
Nevertheless, as $M$ is AST, our completeness result states that we can find a countable family of (pairwise compatible) interval traces, $A \subseteq \termInterval{\toIntervalTerm{M}}$, 
{whose cumulative weight (i.e.~$\omega(\calA)$) equals $\termProb{M} = 1$.}
\end{example}

To achieve completeness we need the concept of \emph{interval separable} primitive functions.
For measurable $A, B \subseteq \real^n$ we write $A \almostSub B$ if $A \subseteq B$ and $\lambda_n(B \setminus A) = 0$, i.e., 
{$A$ is contained in, and, up to a null set, equal to, $B$.}
Interval separability now states that the preimage of every interval can be written, up to a null set, as a countable union of boxes. Precisely:
\begin{definition}
	A function $f: \real^n \to \real$ is called \emph{interval separable} if for every interval $\myint{a, b} \in \interval$, there exists a family of \emph{boxes} $\{B_i\}_{i \in \natnum}$ with $B_i \subseteq \real^n$ such that {$\cup_i B_i \almostSub f^{-1}([a, b])$}. %
\end{definition}
Most interesting functions such as $+$, $\cdot$, $\exp$, etc. are interval separable. 

\begin{lemma}\label{lem:suf2}
	\label{lem:sufficient cond interval sep}
	If $f: \real^n \to \real$ is continuous, and for all $y \in \real$, $f^{-1}(\{y\})$ is a Lebesgue null set, then $f$ is interval separable. 
\end{lemma}

\begin{theorem}\label{theo:intComp}
	If every $f \in \mathbb{F}$ is interval separable, then for every $M \in \Lambda_0$ there exists a countable set of pairwise-compatible interval traces $A \subseteq \termInterval{\toIntervalTerm{M}}$ such that $\omega(A) = \termProb{M}$; and if $M$ is AST then $\expVal(\toIntervalTerm{M}, A) = \expectedTermSteps{M}$.
\end{theorem}
\begin{proofs}
	We first partition $\termTr{M}$ according to the branching behaviour, i.e., sequences in $\{0, 1\}^*$ indicating if the left or the right branch of conditionals was taken. 
	We then fix a branching behaviour (notice that $\{0, 1\}^*$ is countable) and {employ \emph{stochastic symbolic execution} (in the sense of \citep{MakOPW20})} by executing a term on a trace of variables, while collecting symbolic constraints along the way. 
	As primitive functions are interval separable, we show that the corresponding constraints can be exhausted via interval traces.
\end{proofs}

\subsubsection*{Incompleteness}

While the collection of primitive functions with respect to which our semantics is complete is very broad (c.f.~\refLemma{suf2}), interval-based reasoning is incomplete in the presence of arbitrary continuous functions.  

\begin{example}
Let $C \subseteq \real$ be any Smith-Volterra-Cantor set, i.e., $C$ has positive Lebesgue measure but is nowhere dense, i.e., there are no $a < b$ with $\myint{a, b} \in C$. 
Now construct function $f_C : \real \to \real$ by $f_C(x) := d(x, C)$, %
the distance of $x$ to $C$. As $C$ is a closed set, the function is well-defined and obviously continuous; and the roots of $f_C$ coincide with $C$.
Then $M \defined \myif~f_c(\sample) \mythen \num{0} \myelse \num{1}$ is clearly AST. 
However, in the interval-based semantics, we can never derive a termination probability of more than $1-\lambda_1(C) < 1$ as there is no non-trivial interval trace taking the left branch.
\end{example}

\subsection{AST and PAST in the Arithmetic Hierarchy}

If we only consider functions that are $\mathbb{Q}$-interval preserving we can restrict the previous reasoning to intervals and boxes with rational endpoints.
This has direct recursion-theoretic consequences.
\begin{theorem}\label{theo:compl}
	Assume that every $f \in \mathbb{F}$ is $\mathbb{Q}$-interval preserving and interval separable, and $\hat{f}$ is computable and we consider CbN evaluation.
	For any term $M$ (containing only rational numerals), deciding AST is in $\Pi^0_2$. 
	If $M$ is AST, deciding PAST is in $\Sigma^0_2$. 
	In general, deciding PAST is in $\Delta^0_3$.
\end{theorem}
\begin{proof}
	Thanks to \refTheo{intSound} and \refTheo{intComp} we can express ``$M$ is AST'' by the following $\forall\exists$-formula:
	{\small\begin{align*}
		\forall \epsilon > 0 \in \ration. \; \exists A. \; A \subseteq \termInterval{\toIntervalTerm{M}} \land \omega(A) \geq 1 - \epsilon
	\end{align*}}%
	where $A$ ranges over (encodings of) \emph{finite}, \emph{pairwise compatible} sets of interval traces with \emph{rational} endpoints. 
	If $M$ is AST we can express PAST ($\expectedTermSteps{M} < \infty$) as this $\exists\forall$-formula:
	{\small\begin{align*}
			\exists c \in \ration. \; \forall A. \; A \subseteq \termInterval{\toIntervalTerm{M}} \Rightarrow \expVal(\toIntervalTerm{M}, A) \leq c 
	\end{align*}}%
	{In general, $M$-is-PAST $\Leftrightarrow$ $M$-is-AST $\land$  $\expectedTermSteps{M} < \infty$,
	so the general PAST decision problem is in $\Delta^0_3$.}
\end{proof}

{If addition is {definable}, then---thanks to the hardness results in \cite{DBLP:conf/mfcs/KaminskiK15}---deciding AST in the presence of continuous distributions (and suitable primitive functions) is $\Pi^0_2$-complete; and deciding PAST (assuming AST) is $\Sigma^0_2$-complete.\footnote{The reduction from the complement of the of the universal halting problem used to establish $\Sigma^0_2$-hardness of deciding PAST \cite[Thm. 8]{DBLP:conf/mfcs/KaminskiK15} always yields programs that are AST. 
		It is therefore $\Sigma^0_2$-hard to decide PAST even if the program in question is already assumed AST. }}
We remark that the $\Delta^0_3$ upper bound for the general PAST problem does not match the corresponding bound for discrete distributions \cite{DBLP:conf/mfcs/KaminskiK15}.
The approach in \cite{DBLP:conf/mfcs/KaminskiK15} uses the fact that 
{there are finitely many traces of a given length}; 
this property obviously does not hold in the presence of continuous distributions.

\section{Intersection Type System}\label{sec:intersection}

Intersection types have long been studied in termination analysis as they can give a complete characterisation of termination:
A $\lambda$-term is typable in a (suitable) intersection type system iff it is strongly normalising.  
A first study of the quantitative notion of AST, and whether the intriguing completeness of intersection types can be extended to a probabilistic language, was conducted in \cite{DBLP:conf/ppdp/BreuvartL18}.
Owing to the intrinsic $\Pi^0_2$-hardness of AST \cite{DBLP:conf/mfcs/KaminskiK15}, we cannot hope for a semi-decidable type system in which a term is typable iff it is AST. 
Instead \cite{DBLP:conf/ppdp/BreuvartL18} presented two approaches to termination analysis, where the probability of termination is either a sum over all (countably many) typing derivation (called the oracle system) or the least upper bound (lub) thereof.
We show that completeness of intersection types w.r.t.~termination can also be established for a language with continuous samples %
(where a program admits uncountably many distinct runs),
thereby giving a \emph{local representation} of our interval-based semantics.
In our system the lub over countably many derivations gives the probability of termination \emph{and} the expected number of computation steps. 
Thus we obtain a complete, %
compositional and recursion-theoretically optimal method for computing lower bounds on the probability of termination and the expected time to termination.

\begin{figure*}
	\begin{tcolorbox}[colback=white, colframe=black, arc=0mm, boxsep=1pt,left=0pt, right=0pt,after skip=0pt,before skip=0pt,top=0pt, bottom=0pt]
		\small
		
		\begin{minipage}{0.2\textwidth}
			\begin{prooftree}
				\AxiomC{$\calA \in \sigma$}
				\RightLabel{\rulename{var}}
				\UnaryInfC{$\Gamma, x:\sigma \vdash x : \calA$}
			\end{prooftree}
		\end{minipage}
		\begin{minipage}{0.3\textwidth}
			\begin{prooftree}
				\AxiomC{}
				\RightLabel{\rulename{num}}
				\UnaryInfC{$\Gamma\vdash \num{\myint{a, b}} : \Blist{ (\myint{a, b}, \epsilon, 0)}$}
			\end{prooftree}
		\end{minipage}
		\begin{minipage}{0.5\textwidth}
			\begin{prooftree}
				\AxiomC{$\{\myint{a_i, b_i}\}_{i \in [n]} \text{ are almost disjoint}$}
				\RightLabel{\rulename{sample}}
				\UnaryInfC{$\Gamma\vdash \sample : \Blist{ (\myint{a_i, b_i}, \myint{a_i, b_i}, 1 ) \mid i \in [n]}$}
			\end{prooftree}
		\end{minipage}

		\vspace{0.0cm}
		
		\begin{minipage}{0.5\textwidth}
			\begin{prooftree}
				\AxiomC{$\Gamma, x:\sigma, \varphi:\gamma \vdash \calM : \calA$}
				\AxiomC{$\Big\{\Gamma \vdash \mu^\varphi_x. \calM : \calB \mid \forall \calB \in \gamma \Big\}$}
				\RightLabel{\rulename{fix}}
				\BinaryInfC{$\Gamma\vdash \mu^\varphi_x. \calM : \Blist{(\sigma \to \calA, \epsilon, 0)}$}
			\end{prooftree}
		\end{minipage}
		\begin{minipage}{0.5\textwidth}
			\begin{prooftree}
				\AxiomC{$\Gamma \vdash \calM : \calA$}
				\AxiomC{$  \{\Gamma \vdash \calN : \calC \mid (\sigma \to \calB, \wpi, \tau) \in \calA, \calC \in \sigma \}$}
				\RightLabel{\rulename{app}}
				\BinaryInfC{$\Gamma\vdash \calM \calN : \bigcup\limits_{(\sigma \to \calB, \wpi, \tau) \in \calA} \calB^{(\uparrow \wpi, \tau+1)} $}
			\end{prooftree}
		\end{minipage}
		
		\vspace{0.0cm}
		
		\begin{minipage}{0.15\textwidth}
			\begin{prooftree}
				\AxiomC{}
				\RightLabel{\rulename{$\lBrace race$}}
				\UnaryInfC{$\Gamma\vdash \calM : \blist{}$}
			\end{prooftree}
		\end{minipage}
		\begin{minipage}{0.25\textwidth}
			\begin{prooftree}
				\AxiomC{$\Gamma, x:\sigma \vdash \calM : \calA$}
				\RightLabel{\rulename{abs}}
				\UnaryInfC{$\Gamma\vdash \lambda x. \calM : \Blist{(\sigma \to \calA, \epsilon, 0)}$}
			\end{prooftree}
		\end{minipage}
		\begin{minipage}{0.6\textwidth}
			\begin{prooftree}
				\AxiomC{$\Gamma \vdash \calM : \calA$}
				\RightLabel{\rulename{score}}
				\UnaryInfC{$\Gamma\vdash \score(\calM) : \Blist{(\myint{a, b}, \wpi, \tau + 1) \mid (\myint{a, b}, \wpi, \tau) \in \calA,  a \geq 0} $}
			\end{prooftree}
		\end{minipage}
		
		\vspace{0.0cm}

		\begin{prooftree}
			\AxiomC{$\Gamma \vdash \calM : \calA$}
			\AxiomC{$\{\Gamma \vdash \calN : \calB_{(\myint{a, b}, \wpi, \tau)} \mid (\myint{a, b}, \wpi, \tau) \in \calA, b \leq 0\}$}
			\AxiomC{$\{\Gamma \vdash \calP : \calC_{(\myint{a, b}, \wpi, \tau)} \mid (\myint{a, b}, \wpi, \tau) \in \calA, a > 0 \}$}
			\RightLabel{\rulename{\myif}}
			\TrinaryInfC{$\Gamma\vdash \myif(\calM, \calN, \calP) : \bigcup\limits_{(\myint{a, b}, \wpi, \tau) \in \calA \mid b \leq 0} \calB_{(\myint{a, b}, \wpi, \tau)}^{(\uparrow \wpi, \tau+1)} \cup \bigcup\limits_{(\myint{a, b}, \wpi, \tau) \in \calA \mid a > 0} \calC_{(\myint{a, b}, \wpi, \tau)}^{(\uparrow \wpi, \tau+1)} $}
		\end{prooftree}
		
		\vspace{0.0cm}
		
		\begin{prooftree}
			\AxiomC{$\Gamma \vdash \calM : \calA$}
			\AxiomC{$\{\Gamma \vdash \calN : \calB_{(\myint{a, b}, \wpi, \tau)} \mid (\myint{a, b}, \wpi, \tau) \in \calA \}$}
			\RightLabel{\rulename{$f_2$}}
			\BinaryInfC{$\Gamma\vdash f(\calM, \calN) : \bigcup\limits_{(\myint{a, b}, \wpi, \tau) \in \calA} \quad \bigcup\limits_{(\myint{c, d}, \wpi', \tau') \in \calB_{(\myint{a, b}, \wpi, \tau)} } \Blist{(\hat{f}(a, b, c, d), \wpi\wpi', \tau+\tau'+1)}$}
		\end{prooftree}
		
	\end{tcolorbox}
	\caption{Intersection Type System for SPCF.} \label{fig:multiType}
\end{figure*}

\subsection{Intersection Type System For SPCF}

Our system conceptually lies between the two approaches of \cite{DBLP:conf/ppdp/BreuvartL18} {(alluded to above):} we reason about the lub, and at the same type explicitly enumerate terminating (interval) traces 
{as in the oracle system of \cite{DBLP:conf/ppdp/BreuvartL18}}.
The system in \cite{DBLP:conf/ppdp/BreuvartL18} relies on the countable nature of the execution tree and can exhibit subject reduction by taking the weighted (finite) sum over the reduction relation. 
This approach does not work for SPCF because of the uncountable nature of the latter.
Instead, our proofs hinge on the soundness and completeness of the interval-based semantics (\refSection{ibs}).

\subsubsection*{Set Types}
We define \emph{set types} by the following grammar:
{\small\begin{align*}
	\alpha \defined &\myint{a, b} \mid \sigma \to \calA \quad  \quad \sigma \defined \bset{\calA_1, \cdots, \calA_n}\\
	\calA &\defined \Blist{(\alpha_1, \wpi_1, \tau_1), \cdots, (\alpha_m, \wpi_m, \tau_m)}
	\end{align*}}%
where each $\wpi_i$ is an interval trace, and $\tau_i$ a natural number. \index{$\calA, \calB$, a set type} \index{$\sigma, \gamma$, a intersection}
We refer to elements $\sigma$ as \emph{intersections} and $\calA$ as \emph{set types}. 
To effectively type conditionals we need to integrate first-order data, in our case intervals, in the types themselves.
This is similar to the type system in \cite{DBLP:conf/popl/EhrhardTP14}. 
For a set type $\calA = \blist{(\alpha_i, \wpi_i, \tau_i)}_i$ we write $\calA^{(\uparrow \wpi, \tau)}$ for the set type $\blist{(\alpha_i, \wpi \, \wpi_i, \tau_i+\tau)}_i$, i.e., the set obtained by {prepending} $\wpi$ to every trace and adding $\tau$ to every count. 
We call two interval \emph{almost disjoint} if their intersection contains at most one element.

\subsubsection*{Type System}

Typing judgments are of the form $\Gamma \vdash \calM : \calA$. Valid judgments are defined by induction over the rules in Fig.~\ref{fig:multiType}. 
Intuitively, if
\(
\vdash \calM : \blist{(\alpha_i, \wpi_i, \tau_i)}_i
\) 
then $\wpi_i$ are all terminating traces for $M$ on which exactly $\tau_i$ steps are made until a value is reached. 
We advise the reader to compare this system with the monadic system given in \cite[\S 6.1]{DBLP:conf/ppdp/BreuvartL18}. 
Note, in particular, that the type of an application is %
determined by the left argument, matching the CbN $\beta$-reduction where arguments are passed unevaluated.  
While the \rulename{if}-rule looks %
{complicated at first sight}, the subscript $(\myint{a, b}, \wpi, \tau)$ for each set type is merely used as an index, i.e., if $\Gamma \vdash \calM : \calA$ we can combine a different type derivation for every element in $\calA$. 
Although we restrict primitive functions to have arity 2, the rules can easily be extended to handle higher arities. 
We omitted the general rule as it gets chaotic. 
For set type $\calA =  \blist{(\alpha_i, \wpi_i, \tau_i) \mid i \in [n]}$ we define $\omega(\calA) := \sum_{i \in [n]} \omega(\wpi_i)$ and $\expVal(\calA) := \sum_{i \in [n]} \omega(\wpi_i) \cdot \tau_i$.\index{$expVal(\calA)$, the expectation of a set type $\calA$}
We can then show correctness.

\begin{theorem}\label{theo:soundCOmInt}
	For every term $M \in \Lambda_0$,
	\begin{enumerate}
		\item $\bigvee\limits_{\vdash \toIntervalTerm{M} : \calA} \omega(\calA) = \termProb{M}$, and
		\item If $M$ is AST, $\bigvee\limits_{\vdash \toIntervalTerm{M} : \calA} \expVal(\calA) = \expectedTermSteps{M}$
	\end{enumerate}
\end{theorem}
This gives a recursion-theoretically optimal characterisation of AST that is purely based on the type system (c.f.~\cite[\S 5.3]{DBLP:conf/ppdp/BreuvartL18}). 
{Thus we can computationally analyse termination, not just by evaluation (c.f.~\refSection{ibs}), but also via a local typing system.}
By incrementally searching for typing derivations, we can compute arbitrarily tight bounds. 
Compared to \cite{DBLP:conf/ppdp/BreuvartL18}, a novel feature of our work lies in the fact that we can explicitly reason about execution time, thus enabling a type-based characterisation of PAST (for terms that are AST). 
{While our system as a whole may look a little intimidating, each rule is actually simple by itself, requiring no complex operations.}
The idea of annotating types by a step count is also applicable to the setting of \cite{DBLP:conf/ppdp/BreuvartL18}, giving a strict generalisation of their system.
Our system can also be easily generalised to the untyped $\lambda$-calculus considered in \cite{DBLP:conf/icfp/BorgstromLGS16}.
Lastly, while our correctness proof hinges on the completeness of the interval-based semantics, we can present the system without referring to interval traces directly, and instead consider probability mass functions on types (as done in e.g.~\cite{DBLP:conf/ppdp/BreuvartL18}).

\section{Counting-based Recursion Analysis}\label{sec:nonaffine}

We now turn our attention to devising a method that, unlike the preceding approach, can prove AST efficiently.
We focus on programs that can make multiple recursive calls from distinct call sites (during evaluation of the recursive body); we call such recursion \emph{non-affine}.
As evident from Ex.~\ref{ex:3dprinting}, non-affine recursion complicates AST analysis considerably.
Intuitive results such as the \emph{zero-one law of termination}\textsuperscript{\ref{fnote:0-1-law}} \cite{DBLP:series/mcs/McIverM05} 
are only valid for affine recursion.

Our framework builds on the idea of counting. 
We show that analysis of the resulting distributions on natural numbers suffices for proving AST of the program.
As a corollary we obtain a functional generalisation of the \emph{zero-one law}, which specialises to the original law in case the recursion is affine.
Our approach to proving non-affine recursion can be viewed as orthogonal to \cite{DBLP:journals/toplas/LagoG19} (which is restricted to affine recursion):
rather than using the size-related information of the recursive function argument, we count the number of recursive calls from distinct call sites in the evaluation of the body of the recursion. 
Moreover, compared to \cite{DBLP:journals/toplas/LagoG19}, our approach supports continuous distributions, and it is not restricted to binary probabilistic choice.
Compared to techniques based on ranking functions -- a dominant approach to AST verification \cite{DBLP:conf/popl/FioritiH15,DBLP:journals/pacmpl/McIverMKK18,DBLP:conf/pldi/ChenH20,DBLP:journals/pacmpl/AgrawalC018,DBLP:conf/cav/ChakarovS13,DBLP:conf/popl/ChatterjeeNZ17,DBLP:series/mcs/McIverM05,AKROng}, our method is fully automatic and easy to implement, as we show in \refSection{proofsystem}.

\begin{example}\label{ex:runningExample}
	Let's revisit the 3d printing company (Ex.~\ref{ex:3dprinting}). 
  The new situation is that the staff gets tired over time and prints an incorrect number of copies. 
	In case the print is faulty, there is a probability $\mathit{sig}(x)$ of the operator becoming tired and making mistakes, where $\mathit{sig}$ is the sigmoid function. 
	(Thus with increasing time ($x$), the probability of making mistakes approaches $1$.) 
	The operator's mistake takes the form of printing $3$ instead of the intended $2$ copies with probability $.5$.
We model this scenario by the following term: 
	\begin{align*}
		\mu^\varphi_x. x \oplus_p \Big( \big( \varphi^3(x+1) \oplus \varphi^2(x+1) \big) \oplus_{\mathit{sig}(x)} \varphi^2(x+1) \Big).
	\end{align*}%
	The question now becomes: for which $p$ is this term AST? 

\end{example}

We keep track of the number of calls by extracting a \emph{counting distribution}, a (sub) pmf on $\natnum$, that models the distribution on new calls made. 
To account for the fact that a recursive function that is called $n$ times (inclusive of the original call) contributes $n-1$ to the total number of \emph{pending calls}\footnote{equivalently, the maximum number of stack frames on the function's call stack}, we shift the counting pattern by $-1$, obtaining a (sub) pmf on $\intnum$. 
The counting distribution is analysed via a random walk whose current value can be seen as the number of {pending} calls. 
In this section, we first introduce the necessary tools to analyse a random walk (\refSection{MarkovNat}), then present the extraction of the counting distribution from programs (\refSection{extraction}), {and the soundness theorem that relates termination behavior of the shifted random walk with that of the non-affine recursive program in question (\refSection{term})}.

\subsection{Random Walk on $\natnum$}\label{sec:MarkovNat}

Assume a countable state space $X$. 
A \emph{stochastic matrix on $X$} is a function $\mathfrak{P} : X \times X \to \intreal$ such that $\sum_{y \in X} \mathfrak{P}(x, y) = 1$ for every $x \in X$ (see e.g.~\cite[\S 10.1]{PMC} or \cite{MeynT09}).
$\mathfrak{P}(x, y)$ gives the probability of transitioning from $x$ to $y$.
Given stochastic matrices $\mathfrak{P}$ and $\mathfrak{P}'$, we write $\mathfrak{P} \, \mathfrak{P}'$ for their product, which is a stochastic matrix; and $\mathfrak{P}^n$ for the $n$-fold product of $\mathfrak{P}$.

We consider Markov chains whose step behaviour is definable in terms of relative change, independently of the current state.
The relative change in each step is given by a \emph{step distribution} which is a (sub)probability mass function $s : \mathbb{Z} \to \intreal$.
We call $s$ \emph{finite} if it has finite support.
We interpret the ``missing probability'', $1 - \sum_{i \in \mathbb{Z}} s(i)$, as failure.

\begin{definition}\label{def:transitionMatrix}
	Given a {step distribution} $s : \mathbb{Z} \to \intreal$ we define a stochastic matrix $\mathfrak{P}_s$ on $\natnum_\bot \defined \natnum \cup \{\bot\}$ by:\index{step distribution, a sub-pmf $s, t: \intnum \to \intreal$}\index{$\mathfrak{P}_s$, the transition matrix for step distribution $s$}
	\begin{center}
		\begin{tabular}{c|c|c|c|}
			\cline{2-4}
			& $\bot$ & $0$ & $m > 0$ \\ \hline
			\multicolumn{1}{ |c|  }{$\bot$} & $1$ & $0$ & $0$ \\ \hline
			\multicolumn{1}{ |c|  }{$0$} & $0$ & $1$ & $0$ \\ \hline
			\multicolumn{1}{ |c|  }{$n > 0$} & {$1 - \sum_{i \in \mathbb{Z}} s(i)$} & $\sum_{i \leq -n} s(i)$ & $s(m-n)$\\ \hline
		\end{tabular}
	\end{center}
\end{definition}

Note that $s$ gives the relative change in each step, the walk is truncated (trapped) at $0$, %
{and the probability mass deficit in $s$ (if any) is balanced by the probability of transitioning (from a good state) to the failure state $\bot$}.
We call $s$ AST if the associated walk reaches state $0$ a.s. 

\begin{definition}
	A {step distribution} $s$ is called \emph{AST} if for every $m \in \natnum$, $\lim\limits_{n \to \infty} \mathfrak{P}_s^n(m, 0) = 1$.
\end{definition}

Note that the limit in the definition above always exists (and lies between 0 and 1) as the sequence $(\mathfrak{P}_s^n(m, 0))_n$ is monotone \emph{increasing} and bounded.
A {step distribution} can be shown AST by reduction to a one-counter Markov decision process (MDP) (following \cite{DBLP:journals/toplas/LagoG19}), for which a.s.~termination can be decided in polynomial time \cite{DBLP:conf/soda/BrazdilBEKW10}.
We present a new proof that avoids the detour to MDPs, giving a tighter (in fact optimal) complexity upper bound than that in \cite{DBLP:journals/toplas/LagoG19}.
{The crux lies in a simple, and decidable---if $s$ is finite and rational valued---characterisation of AST which directly gives \emph{linear-time decidability}.} 
\begin{theorem}\label{theo:conditionForAST}
	A finite {step distribution} $s$ is AST if and only if all of the following hold
	\begin{tasks}(3)
		\task \label{item:linear1}$\sum\limits_{i \in \mathbb{Z}} s(i) = 1$
		\task \label{item:linear2}$s \neq \delta_0$
		\task \label{item:linear3}$\sum\limits_{i \in \mathbb{Z}} i \cdot s(i) \leq 0$
	\end{tasks}
\end{theorem}

\subsubsection*{Uniform AST}

We also model the case where in each time step, a different distribution can be chosen from an available set of step distributions, similar to a MDP \cite{PMC}.

\begin{definition}\label{def:uniAST}
	A family of {step distributions} $\{s_i\}_{i \in \indx}$ is \emph{uniform AST} if for every $m \in \natnum$
	$$\lim_{n \to \infty} \big( \inf_{i_1, \cdots, i_n} \mathfrak{P}_{s_{i_1}} \cdots  \mathfrak{P}_{s_{i_n}} (m, 0) \big) = 1$$
\end{definition}

Informally it reads that as step count $n$ tends to $\infty$, no matter which step distribution from $\{s_i\}_{i \in \indx}$ is chosen at each step, the walk eventually reaches $0$ almost surely.
Obviously, uniform AST implies AST for each of the $s_i$ but, in general, not conversely.
However we can show:

\begin{lemma}\label{lem:uniformAST}
	If $\{s_i\}_{i \in \indx}$ is a \emph{finite} family of {step distributions} and each $s_i$ is AST then $\{s_i\}_{i \in \indx}$ is uniform AST.
\end{lemma}

\subsection{Counting-based Extraction of Random Walks}\label{sec:extraction}

\begin{figure*}
	
	\begin{tcolorbox}[colback=white, colframe=black, arc=0mm, boxrule=1pt,after skip=0pt,before skip=0pt]
		\begin{minipage}{0.4\textwidth}
			\begin{prooftree}
				\AxiomC{}
				\UnaryInfC{$\langle (\lambda x. M) V, \tr, n \rangle \starto \langle M[V/x], \tr, n \rangle $}
			\end{prooftree}
		\end{minipage}
		\begin{minipage}{0.3\textwidth}
			\begin{prooftree}
				\AxiomC{}
				\UnaryInfC{$\langle \boxed{\mu} \, V, \tr, n \rangle \starto \langle \star, \tr, n+1 \rangle $}
			\end{prooftree}
		\end{minipage}
		\begin{minipage}{0.3\textwidth}
			\begin{prooftree}
				\AxiomC{}
				\UnaryInfC{$\langle \sample, r::\tr, n \rangle \starto \langle \num{r}, \tr, n \rangle $}
			\end{prooftree}
		\end{minipage}

		\vspace{0.1cm}
		
		\begin{minipage}{0.3\textwidth}
			\begin{prooftree}
				\AxiomC{$r \leq 0$}
				\UnaryInfC{$\langle \myif(\num{r}, N, P), \tr, n \rangle \starto \langle N, \tr, n \rangle $}
			\end{prooftree}
		\end{minipage}
		\begin{minipage}{0.3\textwidth}
			\begin{prooftree}
				\AxiomC{$r > 0$}
				\UnaryInfC{$\langle \myif(\num{r}, N, P), \tr, n \rangle \starto \langle P, \tr, n \rangle $}
			\end{prooftree}
		\end{minipage}
		\begin{minipage}{0.4\textwidth}
			\begin{prooftree}
				\AxiomC{}
				\UnaryInfC{$\langle f(\num{r_1}, \cdots, \num{r_{|f|}}), \tr,n \rangle \starto \langle \num{f(r_1, \cdots, r_{|f|})}, \tr, n \rangle$}
			\end{prooftree}
		\end{minipage}

		\vspace{0.1cm}

		\begin{minipage}{0.4\textwidth}
			\begin{prooftree}
				\AxiomC{}
				\UnaryInfC{$\langle f(V_1, \cdots, \star, \cdots,  V_{|f|}), \tr, n \rangle \starto \langle \star, \tr, n \rangle $}
			\end{prooftree}
		\end{minipage}
		\begin{minipage}{0.3\textwidth}
			\begin{prooftree}
				\AxiomC{$r \geq 0$}
				\UnaryInfC{$\langle \score(\num{r}), \tr, n \rangle \starto \langle \num{r}, \tr,n \rangle $}
			\end{prooftree}
		\end{minipage}
		\begin{minipage}{0.3\textwidth}
			\begin{prooftree}
				\AxiomC{$\langle R, \tr, n \rangle \starto \langle M, \tr', n' \rangle$}
				\UnaryInfC{$\langle E[R], \tr, n \rangle \starto \langle E[M], \tr', n' \rangle$}
			\end{prooftree}
		\end{minipage}

	\end{tcolorbox}
	
	\caption{Small-step reduction rules for $\starto$.}\label{fig:countRecCalls}
\end{figure*}

{Let's fix a \emph{1st-order} program $\mu^\varphi_x. M$ with \emph{no} nested recursion.
To extract the counting pattern of $\mu^\varphi_x. M$, we  instrument a counting-based reduction relation $\starto$, and use it to analyse a related term $\repTerm{r} \defined M[\num{r}/x, \boxed{\mu}/\varphi]$, i.e., the body of the program $\mu^\varphi_x. M$, with $x$ instantiated to a fixed actual argument $\num r$, and a special symbol $\boxed{\mu}$ in place of all recursive calls.} 
The counting-based reduction relation $\starto$ is presented in \refFig{countRecCalls} and acts on configuration of the form $\langle N, \tr, n \rangle$ where $n \in \natnum$ counts recursive calls. 
The main idea is to replace outcomes of recursive calls by a distinguished value $\star$ of type $\typeReal$ which stands for an unknown numeral.
Note that the unknown numeral $\star$ can end up in the guard of a conditional if recursive outcomes affect the control flow of the program.
This is, however, unavoidable if we want to count recursive calls (without reference to the program denotation) as the number of function call sites can depend on the (probabilistic) outcome of a prior call.
We define 
$$\termTrCount{N}{n} \defined \{\tr \in \stdtrset \mid \exists V:  \langle N, \tr, 0 \rangle \starto^* \langle V, \epsilon, n \rangle\}$$ \index{$\termTrCount{N}{n}$}%
{the set of traces on which recursive calls from exactly $n$ \emph{distinct} call sites are made.} \index{$\termTrCount{N}{n}$, the set of terminating traces w.r.t.~$\starto$ on which exactly $n$ recursive calls are made}
As the reduction relation is deterministic we get that $\{\termTrCount{N}{n}\}_{n \in \natnum}$ are pairwise disjoint. 
Using similar arguments in \cite{DBLP:conf/icfp/BorgstromLGS16}, it is easy to see that $\termTrCount{N}{n}$ is a measurable set of traces. 

\begin{definition}
	Given a term $\mu^\varphi_x. M$ we define the $r$-indexed family $\{\talloblong \mu^\varphi_x. M \mid r \talloblong : \natnum \to \intreal\}_{r \in \real}$, called the \emph{counting pattern of $\mu^\varphi_x. M$}, whereby 
	$$\talloblong \mu^\varphi_x. M \mid r \talloblong(n) \defined \mu_{\stdtrset} \big(
	\termTrCount{\repTerm{r}}{n}\big)$$ \index{$\talloblong \mu^\varphi_x. M \mid r \talloblong(n)$, counting pattern of $\mu^\varphi_x. M$}
\end{definition}

\vspace{-0.5cm}

{In words, $\talloblong \mu^\varphi_x. M \mid r\talloblong(n) $ gives the probability of \emph{a run of $\mu^\varphi_x.M$}, on the actual argument $r$, making recursive calls from $n$ distinct call sites.}
It is straightforward to see that for every $r$ we have $\sum_n \talloblong \mu^\varphi_x. M \mid r \talloblong(n) \leq 1$, by the same argument as in \cite[Lem.~7]{MakOPW20}.

\begin{example} \label{ex:cp}
	Consider the term $\mu^\varphi_x. M$ from Ex.~\ref{ex:runningExample}.
	We get $\talloblong \mu^\varphi_x. M \mid r \talloblong(0) = p$, $\talloblong \mu^\varphi_x. M \mid r \talloblong(1) = 0$, 
  $\talloblong \mu^\varphi_x. M \mid r \talloblong(2) = (1-p) \cdot \tfrac{1}{2} \cdot (2-\mathit{sig}(r))$, 
  $\talloblong \mu^\varphi_x. M \mid r \talloblong(3) = (1-p) \cdot \tfrac{1}{2} \cdot \mathit{sig}(r)$ and $\talloblong \mu^\varphi_x. M \mid r \talloblong(n) = 0$ for all other $n$. 
\end{example}

\subsection{Termination via Counting Patterns}\label{sec:term}

Our main result of this section is that we can use the counting pattern of a program to soundly reason about its termination property.
For any \emph{counting distribution}, i.e., (sub) pmf $s: \natnum \to \intreal$, we define the shifted step distribution $\overline{s} : \intnum \to \intreal$ by $\overline{s}(z) = s(z+1)$ for $z \geq -1$ and $\overline{s}(z) = 0$ otherwise
\footnote{ The shifting of the distribution accounts for the fact that resolving a recursive call by making $n$ recursive calls changes the number of pending calls by $n-1$. In the extreme case, making no recursive call, decreases the number of pending calls by $1$. }
.\index{counting distribution, a sub-pmf $\natnum \to \intreal$}
\index{$\overline{s}$, the step distribution obtained by shifting $s$ by $-1$}

\begin{theorem}\label{theo:ASTCount}
	If $\{\overline{\talloblong \mu^\varphi_x. M \mid r\talloblong}\}_{r \in \real}$ {(qua family of step distributions)} is uniform AST then $\mu^\varphi_x. M$ is AST on every actual argument.
\end{theorem}
\begin{proofs}
	We decompose the set of terminating traces on a fixed argument according to the arguments of recursive calls arranged in a tree. We can lower bound the probability of each partition in terms of $\{\overline{\talloblong \mu^\varphi_x. M \mid r\talloblong}\}_{r \in \real}$ and show that uniform AST implies that the cumulative weight over every decomposed part equals $1$, i.e., the program is AST.
\end{proofs}

\subsubsection*{A Partial Order For Counting Distributions}

We can equip the set of {counting distributions} (i.e.~(sub)pmfs $s, t: \natnum \to \intreal$) with a partial order that is compatible with the termination behavior.
We define
$$\textstyle s \sqsubseteq t \Leftrightarrow \forall n \in \natnum. \; \sum_{m \leq n} s(m) \leq \sum_{m \leq n} t(m)$$
i.e., $s \sqsubseteq t$ if the cumulative weight of $s$ is no greater than that of $t$ at any point.
It is easy to see that $\sqsubseteq$ is a partial order.\index{$\sqsubseteq$, the terminating preserving partial order defined on counting distribution}
Furthermore, we can show compatibility w.r.t.~AST (using \refTheo{conditionForAST}):

\begin{lemma}\label{lem:order}
  If $s$, $\{t_i\}_{i \in \indx}$ are counting distributions and for all $i \in \indx$, $s \sqsubseteq t_i$ and $\overline{s}$ is AST then $\{\overline{t_i}\}_{i \in \indx}$ is uniform AST.
\end{lemma}

\begin{example}\label{ex:simplepBound}
	The counting pattern presented in Ex.~\ref{ex:cp} for the term from Ex.~\ref{ex:runningExample} %
  {satisfies the preconditions} of \refLemma{order} for $s \defined p \delta_0 + (1-p)\tfrac{1}{2} \delta_2  + (1-p)\tfrac{1}{2} \delta_3$ (where $\delta_i$ denotes the Dirac-distribution).
	For $p \geq \tfrac{3}{5}$, we can deduce that the counting pattern is uniform AST (via \refLemma{order} and \refTheo{conditionForAST}) and thus the example is AST on every input (via \refTheo{ASTCount}). 
\end{example}

\subsection{$\epsilon$-Recursion Avoiding Fixpoint Terms}\label{sec:pra}

{An interesting quantity that arises from analysing \refTheo{ASTCount} is $\talloblong \mu^\varphi_x. M \mid r \talloblong(0)$, i.e., the probability of a run of $\mu^\varphi_x. M$ (on argument $\num r$) making no further recursive calls. 
Let's consider programs where this probability has a positive lower bound.}

\begin{definition}
  A recursive program $\mu^\varphi_x. M$ is \emph{$\epsilon$-recursion avoiding ($\epsilon$-RA) } if for all $r \in \real$, $\talloblong \mu^\varphi_x. M \mid r \talloblong(0) \geq \epsilon$.
\end{definition}

Lets assume $\sum_n \talloblong \mu^\varphi_x. M \mid r \talloblong(n) = 1$, i.e., the $\starto$-reduction is never stuck. 
In the appendix %
we show how this can be statically ensured via a type system.
Note that a program may be $\epsilon$-RA for a positive $\epsilon$, and yet not AST (as evident from Ex.~\ref{ex:3dprinting}). 
To ensure AST, the positive probability $\epsilon$ must be ``large enough'', in relation to the number of recursive calls.  
We define the \emph{recursive rank} of $\mu^\varphi_x. M$ to be the minimal $m$ such that for all $n > m$, and $r$, $\talloblong \mu^\varphi_x. M \mid r \talloblong(n) = 0$
{(or, equivalently, the maximal number of call sites from which recursive calls are made in a run of $(\mu^\varphi_x. M) \, \num r$, for any $r$).}
(In the appendix, we show that the recursive rank can be upper bounded via a decidable non-idempotent intersection type system.)
Now, using \refTheo{conditionForAST} combined with \refTheo{ASTCount}, we can get an easy corollary:

\begin{corollary}\label{corr:era}
	If $\mu^\varphi_x. M$ has recursive rank $m$ and is $\epsilon$-RA for some $\epsilon > 0$ that satisfies $m(1-\epsilon) \leq 1$ then $\mu^\varphi_x. M$ is AST on every argument.
\end{corollary}

\begin{example}
	The program \ref{prog:intro2} in Ex.~\ref{ex:3dprinting} has recursive rank $2$, and is $p$-RA. So Cor.~\ref{corr:era} is applicable whenever $2(1-p)  \leq 1 \Leftrightarrow p \geq \tfrac{1}{2}$.
	Note that Cor.~\ref{corr:era} is weaker than Thm.~\ref{theo:ASTCount}; for example, Cor.~\ref{corr:era} on Ex.~\ref{ex:runningExample} is only applicable for $p \geq \tfrac{2}{3}$ whereas \refTheo{ASTCount} is applicable for $p \geq \tfrac{3}{5}$ (\refExample{simplepBound}).
\end{example}

\begin{example}\label{ex:complexExample}
	As a further example, consider yet another variation to our 3d-printing program from \refExample{runningExample}: 
  \begin{align*}
		\mu^\varphi_x. &\mylet e = \sample \myin \myif \, e \leq p \mythen x \myelse \\
		&\,\Big(\big( \varphi^3(x+1) \oplus_e \varphi^2(x+1) \big) \oplus_{\mathit{sig}(x)} \varphi^2(x+1) \Big)
	\end{align*}%
	We sample the error value $e$ (the higher $e$ is, the more damaged the print) and accept the print whenever $e \leq p$. 
	If the print is unacceptable, we replace the binary choice in \refExample{runningExample} with one that depends on the sampled value of $e$.
	In the appendix we show how  \refTheo{ASTCount} and \refLemma{order} can be used to prove this program AST whenever $p \geq \sqrt{7} - 2 \approx 0.646$. 
  As this example illustrates well, termination analysis of terms that use continuous random samples as first-class values can become very intricate.
	Such examples are not expressible in PHORS \cite{DBLP:conf/lics/KobayashiLG19} or with binary probabilistic choice \cite{DBLP:journals/toplas/LagoG19,DBLP:journals/jacm/KaminskiKMO18,DBLP:conf/lics/OlmedoKKM16}.
	Our framework can analyse such examples efficiently, even automatically. 
\end{example}

\subsubsection*{Special Case: Affine Recursion}

Every affine-recursive program \cite[\S 4.1]{DBLP:journals/toplas/LagoG19} has recursive rank at most $1$, so by Cor.~\ref{corr:era}, $\epsilon$-RA for \emph{any} $\epsilon > 0$ implies AST. 
This can be seen as the functional equivalent of the \emph{zero-one-law for termination} (c.f.~\cite[Sec.~2.6]{DBLP:series/mcs/McIverM05}).
However, the real novelty of our result lies in the fact that sophisticated methods are necessary to deal with the case of non-affine recursion. 
Our proof rules (\refTheo{ASTCount} and Cor.~\ref{corr:era}) give a powerful tool to verify AST for non-affine programs where the standard zero-one law fails. 
Similarly to the language studied e.g. in \cite{DBLP:journals/pacmpl/LewCSCM20}, we can use this to design languages that are AST-by-construction. 
In particular, in any probabilistic programming system we can {(safely) add a special fixpoint operator that comes with the guarantee of $\epsilon$-RA for a sufficiently large $\epsilon$}, whose size can be determined statically via the recursive rank.
This corresponds to a generalization of the stochastic while-loop in \cite{DBLP:journals/pacmpl/LewCSCM20}.  
As demonstrated in \cite{DBLP:journals/pacmpl/LewCSCM20}, probabilistic programming languages with this seemingly severe restriction can still describe complex models with arbitrary precision {and convergence guarantee}, supporting correct inference of (AST) programs.

\section{A Proof System For Non-affine Recursion}\label{sec:proofsystem}

The framework of Sec.~\ref{sec:nonaffine} (\refTheo{ASTCount}) relies on the \emph{counting pattern}, $\{\talloblong \mu^\varphi_x. M \mid r\talloblong\}_{r \in \real}$, of the program $\mu^\varphi_x. M$. 
This family can contain uncountably many different counting distributions, making it impractical for analysis. 
As we saw in Ex.~\ref{ex:cp}, for the counting pattern $\{\talloblong \mu^\varphi_x. M \mid r\talloblong\}_{r \in \real}$ of Ex.~\ref{ex:runningExample}, we have $\talloblong \mu^\varphi_x. M \mid r\talloblong \neq \talloblong \mu^\varphi_x. M \mid r'\talloblong$ for every $r \neq r'$. 
{So how can we automate analysis so that \refTheo{ASTCount} can be applied without explicitly computing the counting pattern?}
{In this section we use a simple \emph{game-playing} perspective to solve the problem.}
We show that we can replace probabilistic branching that depends on the {actual argument}, by nondeterministic branching and thus obtain a sound method to apply \refTheo{ASTCount}.
As a rule of thumb, our system can verify all programs that exhibit an AST counting pattern which is independent of the (exact values of the) actual arguments (of the recursive function in question).
In this section we give an overview of our approach, and direct readers to the Appendix for a full account, including more complex examples.

\subsection{Stochastic symbolic Execution}

\begin{figure}
	\begin{tcolorbox}[colback=white, colframe=black, arc=0mm, boxrule=1pt,left=0pt, right=0pt,after skip=0pt,before skip=0pt]
		\begin{subfigure}{0.35\textwidth}
			\begin{center}
				\begin{tikzpicture}[scale=0.7]
				\node[circle, inner sep=1.7pt, draw, fill=black!10, label={right:\footnotesize$\alpha_1 - p$}] at (0,0) (n1){};
				
				\node[circle, draw=blue, fill=white, inner sep=1.7pt, label=below:{}] at (-1,-0.8) (n12){};

				\draw[-, thick] (n1) to (n12);

				\node[circle, inner sep=1.7pt, draw, fill=red!70, label=right:{}] at (1,-0.8) (n21){};
				
				\node[rectangle, draw=blue] at (1.7,-1.6) (n22){\tiny$\mu$};
				
				\node[rectangle, draw=blue] at (1.7,-2.4) (n23){\tiny$\mu$};
				
				\node[circle, draw=blue, fill=white, inner sep=1.7pt, label=below:{}] at (1.7,-3.2) (n24){};

				\node[circle, inner sep=1.7pt, draw, fill=black!10, label={left:\footnotesize$\alpha_3 - \tfrac{1}{2}$}] at (0.3,-1.6) (n25){};

				\node[rectangle, draw=blue] at (-0.2,-2.4) (n26){\tiny$\mu$};
				
				\node[rectangle, draw=blue] at (-0.2,-3.2) (n27){\tiny$\mu$};
				
				\node[circle, draw=blue, fill=white, inner sep=1.7pt, label=below:{}] at (-0.2,-4) (n28){};
				
				\node[rectangle, draw=blue] at (0.8,-2.4) (n29){\tiny$\mu$};
				\node[rectangle, draw=blue] at (0.8,-3.2) (n210){\tiny$\mu$};
				
				\node[rectangle, draw=blue] at (0.8,-4) (n211){\tiny$\mu$};
				
				\node[circle, draw=blue, fill=white, inner sep=1.7pt, label=below:{}] at (0.8,-4.8) (n212){};

				\draw[-, thick] (n21) to (n22);
				\draw[-, thick] (n22) to (n23);
				\draw[-, thick] (n23) to (n24);
				
				\draw[-, thick] (n21) to (n25);
				
				\draw[-, thick] (n25) to (n26);
				\draw[-, thick] (n26) to (n27);
				\draw[-, thick] (n27) to (n28);
				
				\draw[-, thick] (n25) to (n29);
				\draw[-, thick] (n29) to (n210);
				\draw[-, thick] (n210) to (n211);
				\draw[-, thick] (n211) to (n212);

				\draw[-, thick] (n1) to (n21);
				\end{tikzpicture}
			\end{center}
			\vspace{-0.5cm}
			\subcaption{}\label{fig:exTreea}
		\end{subfigure}
		\begin{subfigure}{0.65\textwidth}
			
			\begin{minipage}{0.45\textwidth}
				\begin{center}
					\begin{tikzpicture}[scale=0.7]
						\node[circle, inner sep=1.7pt, draw, fill=black!10, label={right:\footnotesize$\alpha_1 - p$}] at (0,0) (n1){};
						
						\node[circle, draw=blue, fill=white, inner sep=1.7pt, label=below:{}] at (-1,-0.8) (n12){};

						\draw[-, thick] (n1) to (n12);

						\node[circle, inner sep=1.7pt, draw, fill=red!70, label=right:{}] at (1,-0.8) (n21){};

						\node[circle, inner sep=1.7pt, draw, fill=black!10, label={left:\footnotesize$\alpha_3 - \tfrac{1}{2}$}] at (0.3,-1.6) (n25){};

						\node[rectangle, draw=blue] at (-0.2,-2.4) (n26){\tiny$\mu$};
						
						\node[rectangle, draw=blue] at (-0.2,-3.2) (n27){\tiny$\mu$};
						
						\node[circle, draw=blue, fill=white, inner sep=1.7pt, label=below:{}] at (-0.2,-4) (n28){};
						
						\node[rectangle, draw=blue] at (0.8,-2.4) (n29){\tiny$\mu$};
						\node[rectangle, draw=blue] at (0.8,-3.2) (n210){\tiny$\mu$};
						
						\node[rectangle, draw=blue] at (0.8,-4) (n211){\tiny$\mu$};
						
						\node[circle, draw=blue, fill=white, inner sep=1.7pt, label=below:{}] at (0.8,-4.8) (n212){};

						\draw[-, thick] (n21) to (n25);
						
						\draw[-, thick] (n25) to (n26);
						\draw[-, thick] (n26) to (n27);
						\draw[-, thick] (n27) to (n28);
						
						\draw[-, thick] (n25) to (n29);
						\draw[-, thick] (n29) to (n210);
						\draw[-, thick] (n210) to (n211);
						\draw[-, thick] (n211) to (n212);

						\draw[-, thick] (n1) to (n21);
					\end{tikzpicture}
				\end{center}
			\end{minipage}
			\begin{minipage}{0.5\textwidth}
				\begin{center}
					\begin{tikzpicture}[scale=0.7]
						\node[circle, inner sep=1.7pt, draw, fill=black!10, label={right:\footnotesize$\alpha_1 - p$}] at (0,0) (n1){};
						
						\node[circle, draw=blue, fill=white, inner sep=1.7pt, label=below:{}] at (-1,-0.8) (n12){};

						\draw[-, thick] (n1) to (n12);

						\node[circle, inner sep=1.7pt, draw, fill=red!70, label=right:{}] at (1,-0.8) (n21){};
						
						\node[rectangle, draw=blue] at (1.7,-1.6) (n22){\tiny$\mu$};
						
						\node[rectangle, draw=blue] at (1.7,-2.4) (n23){\tiny$\mu$};
						
						\node[circle, draw=blue, fill=white, inner sep=1.7pt, label=below:{}] at (1.7,-3.2) (n24){};

						\draw[-, thick] (n21) to (n22);
						\draw[-, thick] (n22) to (n23);
						\draw[-, thick] (n23) to (n24);

						\draw[-, thick] (n1) to (n21);
					\end{tikzpicture}
				\end{center}
			\end{minipage}
			
			\vspace{-0.5cm}
			\subcaption{}\label{fig:exTreeb}
		\end{subfigure}
	\end{tcolorbox}
	
	\caption{Symbolic execution trees for the running example and all possible strategies \textbf{\small(b)}.}\label{fig:exTree}
	\vspace{-0.5cm}
\end{figure}
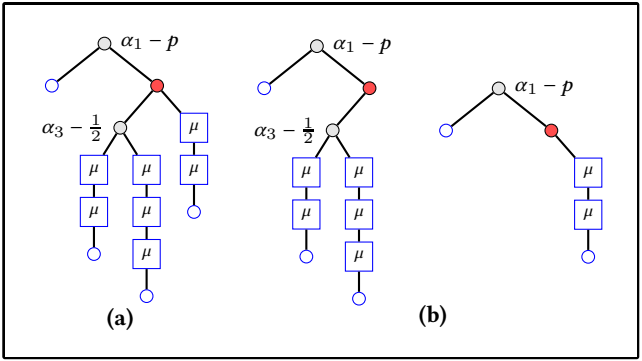

The first idea we use for our system is \emph{stochastic symbolic execution}.
Instead of executing a program on a fixed trace (as done in $\starto$, Fig.~\ref{fig:countRecCalls}) we evaluate on a trace of \emph{sample variables} ($\alpha_0, \alpha_1, \cdots$) whose values can be {instantiated} later. 
We organise execution in the form a binary tree where each branching represents a conditional which is annotated with the value {on which control flow branches}. 
We also record \score-statements as well as recursive calls. 
We now replace the {actual} argument with an unknown value $\circledast$ {(corresponding to the analysis of $\repTerm{\circledast}$ in Sec.~\ref{sec:extraction})}.
In {\refFig{exTreea}} the execution tree that corresponds to the running example, Ex.~\ref{ex:runningExample}, is depicted, we have coloured each branching that relies on the concrete argument $\circledast$ in red. 

\subsection{Strategies on Trees}

As some of the branching (coloured red) depends on the actual argument, we cannot analyse it probabilistically. 
{This can be overcome by an intuitive 2-player game reading of the execution tree: 
for every such node the Environment (player) can resolve its branching by an explicit strategy that indicates a left/right choice for each coloured node.} 
For example, all possible strategies for the tree in \refFig{exTreea} are depicted in \refFig{exTreeb}.
As a strategy no longer relies on branching at nodes that contain $\circledast$, we can recover a probabilistic interpretation of paths, which is just the Lebesgue-measure of the possible assignment to the sample variables $\alpha_0, \alpha_1, \cdots$, such that a path is indeed followed. 
Given a strategy $\trees$, we denote with $\mathbb{P}(\trees, n)$ the probability of taking a path such that \emph{at most} $n$ recursive calls are made (i.e., a fixpoint node is traversed at most $n$ times). \index{$\trees$, a strategy on an execution tree}
Depending on the set of primitive functions, this value can be computed effectively. 
We now define the counting distribution $\mathbb{P}_{\mathit{approx}}$ by (where $n > 0$)
{\small\begin{align*}
	\mathbb{P}_{\mathit{approx}}(0) &\defined \min_{\trees \in \mathit{Strat}(\treex)} \mathbb{P}(\trees, 0)\\
	\mathbb{P}_{\mathit{approx}}(n) &\defined \Big(\min_{\trees \in \mathit{Strat}(\treex)} \mathbb{P}(\trees, n)\Big) - \Big(\min_{\trees \in \mathit{Strat}(\treex)} \mathbb{P}(\trees, n-1)\Big)
\end{align*}}%
We can understand $\mathbb{P}_{\mathit{approx}}(n)$ as the least probability that $n$ calls are made even if the Environment chooses in the worst (meaning maximal no.~of recursive calls) possible way. 

\begin{example}\label{ex:ex2}
	Consider all strategies for the running example listed in \refFig{exTreeb}.
	We can compute  $\mathbb{P}_{\mathrm{approx}}(0) = p$; $\mathbb{P}_{\mathrm{approx}}(2) = \mathbb{P}_{\mathrm{approx}}(3) = (1-p) \cdot \tfrac{1}{2}$; and $\mathbb{P}_{\mathrm{approx}}(n) = 0$ for all other $n$.
\end{example}

We can show that replacing probabilistic branching with nondeterministic one gives a lower bound (w.r.t.~to the order $\sqsubseteq$) on the counting pattern.

\begin{theorem}\label{theo:soundPS}
	For every $r \in \real$, $\mathbb{P}_{\mathrm{approx}} \sqsubseteq \talloblong \mu^\varphi_x. M \mid r \talloblong$
\end{theorem}

Thus, if $\overline{\mathbb{P}_{\mathrm{approx}}}$ is AST (which is checkable via \refTheo{conditionForAST}) we get that $\{\overline{\talloblong \mu^\varphi_x. M \mid r \talloblong}\}_{r\in \real}$ is uniform AST; and via \refTheo{ASTCount} $\mu^\varphi_x. M$ is AST on every actual argument.
\refTheo{soundPS} and {the values computed in \refExample{ex2}} allow us to deduce (\emph{automatically}) that Ex.~\ref{ex:runningExample} is AST on every argument if $p \geq \tfrac{3}{5}$.
Similarly, our tool can verify AST of \refExample{complexExample} if $p \geq \sqrt{7}- 2$.

\begin{table}[!t]
	\caption{Experimental results for lower bound computations. We give the actual probability of termination (if known), the lower bound computed (LB), the depth ($d$) at which we stopped the exploration and the time ($t$) in milliseconds. 
	$\mathit{geo}_p$ describes a geometric distribution with parameter $p$, $\mathit{1dRW}_{p, s}$ a 1-dimensional $p$-biased random walk starting at $s$ \cite{DBLP:journals/pacmpl/McIverMKK18}, $\mathit{gr}$ a term analysed in \cite{DBLP:conf/lics/OlmedoKKM16} terminating with a probability given as the inverse golden ratio, $\mathit{3print}_p$ the natural extension of \refExample{3dprinting} \ref{prog:intro2} to $3$ recursive calls, $\mathit{bin}_{p, s}$ a 1-dimensional random walk in one direction \cite{DBLP:journals/pacmpl/McIverMKK18} and $\mathit{pedestrian}$ a stochastic program modelling a pedestrian taken inspired by \cite{MakOPW20}. 
	See the appendix for a detailed description of the example terms.  
    }
	\label{tab:resLB}
	\begin{tabular}{|c||c|c|c|c|}
		\hline
		Term $M$ & $\termProb{M}$ & LB & d & t \\
		\hline
		\hline
		$\mathit{geo}_{\tfrac{1}{2}}$ & $1$ & $0.9999990463$ & $100$ & $78$ \\
		\hline
		$\mathit{geo}_{\tfrac{1}{5}}$ & $1$ & $0.9995620416$ & $200$ & $192$ \\
		\hline
		$\mathit{1dRW}_{\tfrac{1}{2}, 1}$ & $1$ & $0.8036193847$ & $200$ & $28223$ \\
		\hline
		$\mathit{1dRW}_{\tfrac{7}{10}, 1}$ & $1$ & $0.9720964250$ & $150$ & $10224$ \\
		\hline
		$\mathit{gr}$ & $\tfrac{\sqrt{5}-1}{2}$ & $0.6112594604$ & $80$ & $4389$ \\
		\hline
		\refExample{3dprinting}, $p = \tfrac{1}{2}$ & $1$  & $0.8318119049$ & $90$ & $15749$ \\
		\hline
		\refExample{3dprinting}, $p = \tfrac{1}{4}$ & $? (< 1)$ & $0.3328795089$ & $90$ & $15749$ \\
		\hline
		$\mathit{3print}_{\tfrac{3}{4}}$ & $1$ & $0.9606655982$ & $80$ & $4622$\\
		\hline
		$\mathit{bin}_{\tfrac{1}{2}, 2}$ & $1$ & $0.9998493194$ & $100$ & $2265$\\
		\hline
		$\mathit{pedestrian}$ & $1$ & $0.6002376673$ & $40$ & $4493$\\
		\hline
	\end{tabular}
\end{table}

\begin{table}[!t]
	\caption{Experimental results for AST verification. For each term (all of which our tool can verify to be AST) we give the counting distribution $\mathbb{P}_{\mathit{approx}}$ computed by our tool (which is analysed via \refTheo{conditionForAST}) and the total time $t$ in milliseconds. }
	\label{tab:resLAST}
	\begin{tabular}{|c||c|c|c|c|}
		\hline
		Term $M$ & $\mathbb{P}_{\mathit{approx}}$  & $t$\\
		\hline
		\hline
		\refExample{3dprinting}, \ref{prog:intro1}, $p = \tfrac{1}{2}$ & $\tfrac{1}{2}\delta_0 + \tfrac{1}{2}\delta_1$ & 239\\
		\hline
		\refExample{3dprinting}, \ref{prog:intro2}, $p = \tfrac{1}{2}$ & $\tfrac{1}{2}\delta_0 + \tfrac{1}{2}\delta_2$ & 237\\
		\hline
		$\mathit{3print}_{\tfrac{2}{3}}$ & $\tfrac{2}{3}\delta_0 + \tfrac{1}{3}\delta_3$& 297\\
		\hline
		\refExample{runningExample}, $p = 0.6$ & $ 0.6\delta_0 + 0.2\delta_2 + 0.2 \delta_3$  & 396\\
		\hline
		\refExample{complexExample}, $p = 0.65$ & $ 0.65\delta_0 + 0.06125\delta_2 + 0.28875\delta_3$ & 373\\
		\hline
	\end{tabular}
\end{table}

\section{Implementation}
\label{sec:implementation}

We provide prototype implementations for computing lower bound of the termination probability, and for AST verification, building on \refSection{ibs} and \refSection{proofsystem} respectively.
This is the first prototype to compute lower bounds in the presence of continous distributions and one of the first that can automatically proof AST for non-affine recursive programs\textsuperscript{\ref{fnote:source}}. 
In this section we present the results of our experiments. 
In the appendix we give a more detailed description of the algorithm and example terms.
The experimental results were obtained on a Intel(R) Core i5-6200U process with $8$GB of memory.

\subsection{Lower Bounds of Termination Probability}

Our prototype for lower bound computation exploits the completeness of the interval-traces semantics. %
Our tool combines the symbolic exploration of terms with a simple sweep algorithm to search for terminating interval traces.
In the stochastic symbolic execution, we execute terms while substituting sample-variable for random outcomes. Each trace leading to a symbolic value is then analysed by iteratively searching for terminating interval traces by splitting the unit box $[0, 1]^m$ (where $m$ is the number of sample-variables along a path). 
As lower bound computation is an intrinsically non-terminating process the user must specify a target depth (or timeout) at which the computation is stopped. 
Even in its current, unoptimized, form our tool is able to compute meaningful lower bounds in a reasonable time. 
We evaluate our tool on various examples taken from \cite{DBLP:conf/lics/KobayashiLG19, DBLP:journals/pacmpl/McIverMKK18, DBLP:conf/lics/OlmedoKKM16} and \cite{MakOPW20} (possibly modified to match the CbN evaluation).
The computed lower bounds can be found in table \ref{tab:resLB}. 
Our tool computes rational lower-bounds to avoid rounding errors. For presentation we only gave the first $10$ digits of the decimal representation. 

\subsection{AST Verification}

The challenge in implementing the ideas from \refSection{proofsystem} lies in the computation of branching probabilities, i.e., given a fixed strategy for the environment what is the probability of traversing at most $n$ fixpoint nodes. 
We adopt a geometric interpretation of probability and make use of various results and implementation techniques for volume computation.
For simplicity we restrict primitive operations to addition, and multiplication by a constant (and thus subtraction), as the probability of branching can be seen as the volume of a convex polytope \cite{DBLP:journals/siamcomp/DyerF88} (a subset of $\real^d$ of the from $\{\vec{x} \mid A \vec{x} \leq b \}$).
We make use of the analytic formula for this volume in \cite{lasserre1983analytical} and its subsequent implementation in \cite{Bueler2000}, and appeal to \refTheo{conditionForAST}.
Our implementation then performs the basic-tree operations outlined in \refSection{proofsystem} (see the appendix for a fuller account) and uses \cite{Bueler2000} as a volume-computation oracle. 
Our prototype implementation can verify many examples, including those from \refSection{overview}, \refSection{nonaffine}, and \refSection{proofsystem} (see table \ref{tab:resLB}), thereby illustrating that our approach is well-suited to implementation.
All of those terms can be verified to be AST in less than a second.

\section{Related Work and Conclusion}
\label{sec:related}

Our interval-traces approach can be seen as a probabilistic interpretation of interval analysis, a standard approach to infer bounds on program variables \cite{DBLP:books/daglib/0022249,DBLP:journals/toplas/Burke90}.
The attractive feature of intervals in our work is its completeness w.r.t.~the Lebesgue measure for a broad range of primitive operations. 
 
The only comparable lower bound computation we are aware of is presented in \cite{DBLP:conf/lics/KobayashiLG19}.
Kobayashi et al.~show that the termination probability of CbN order-$n$ probabilistic recursion schemes ($n$-PHORS) can be obtained as the least fixpoint of suitable order-$(n-1)$ fixpoint equations, which can be solved using standard Kleene fixpoint iteration.  
By contrast, our approach works on programs directly, and can handle continuous distributions. 
{It is worth noting that (order-$n$) PHORS is readily encodable as (order-$n$) CbN SPCF, but the former is strictly less expressive (because the underlying recursion schemes are not Turing complete). 
Some interesting SPCF terms such as \refExample{complexExample} cannot be expressed as PHORS.
Since recursive Markov chains \cite{DBLP:journals/jacm/EtessamiY09} (equivalently, probabilistic pushdown automata \cite{DBLP:journals/fmsd/BrazdilEKK13}) are essentially equivalent to 1-PHORS \cite{DBLP:conf/lics/KobayashiLG19}, it follows that order-1 SPCF (which contains the term in Ex.~\ref{ex:complexExample}) is strictly more expressive than recursive Markov chains.}

Our intersection type system is inspired by, and builds upon, the ideas of \cite{DBLP:conf/ppdp/BreuvartL18,DBLP:conf/popl/EhrhardTP14}. 
However, unlike \cite{DBLP:conf/ppdp/BreuvartL18}, we cannot prove correctness directly (because of continuous samples), rather we need to appeal to the completeness of our interval-based semantics. 
The step annotation of types enables us to reason about expected termination time. We conjecture these ideas to also be applicable to the system of \cite{DBLP:conf/ppdp/BreuvartL18}.
Independent of us, \cite{DBLP:journals/pacmpl/LagoFR21} designed an intersection type system that is also able to reason about the expected time to termination. Their approach is, however restricted to a language with discrete samples and it is not obvious whether the approach extends to continuous samples.

Our AST verification method is closely related to \cite{DBLP:conf/lics/OlmedoKKM16}, in that they also study recursive programs and allow for non-affine behaviour.
Our work differs nonetheless in several key aspects:
While they study an imperative language with discrete distributions, we work with a purely functional language with continuous distributions. 
Though their proposed rules can produce lower bounds on the probability of termination, they seem cumbersome to use.
Their rule informally reads: if, for all $n$, we assume that each recursive call terminates with probability $l_n$ after $n$ fixpoint unfoldings, and we can prove that it terminates with probability at least $l_{n+1}$ after $n+1$ unfoldings, then the program terminates with probability at least $\sup_n l_n$ (c.f.~\cite[Thm.~4.2]{DBLP:conf/lics/OlmedoKKM16}).
In order to apply this rule, the user must manually find an explicit (and often non-trivial) sequence $(l_n)_{n\in \natnum}$.
By contrast, our system provides a sound reduction to a random walk which can be analysed efficiently in linear time. 

As already mentioned, our method can be seen as orthogonal to that in \cite{DBLP:journals/toplas/LagoG19}. 
It is not at all obvious if their techniques can be extended to our setting with sampling from the uniform distribution. %
An interesting future direction is to develop a unified framework that analyses both the size-related information of the recursive function argument and the number of recursion call sites.

\subsection*{Conclusion}

Recent advances in probabilistic programming systems and allied areas (such as \cite{MakOPW20}) provide strong impetus for the study of AST of programs with continuous distribution.
We have presented a first comprehensive study of the lower bound problem, and ascertained the recursion-theoretic complexity of several termination problems.
We have introduced a novel proof system for AST verification of non-affine programs which is easily implementable. 
While some of the existing AST proof methods support continuous distributions \cite{DBLP:conf/popl/FioritiH15,DBLP:conf/pldi/ChenH20,DBLP:journals/toplas/ChatterjeeFNH18} the majority do not. 
It would be interesting to investigate if they \cite{DBLP:journals/toplas/LagoG19,DBLP:journals/jacm/KaminskiKMO18,DBLP:conf/lics/OlmedoKKM16,DBLP:conf/lics/KobayashiLG19,DBLP:conf/mfcs/KaminskiK15,DBLP:series/mcs/McIverM05} can be so extended.

\bibliographystyle{ACM-Reference-Format}
\bibliography{references}

\newpage

\appendix

\section{Additional Material - Section~\ref{sec:spcf}}

\begin{figure*}
	\begin{tcolorbox}[colback=white, colframe=black, arc=0mm, boxrule=1pt]
		\begin{minipage}{0.2\textwidth}
			\begin{prooftree}
				\AxiomC{$x:\alpha \in \Gamma$}
				\UnaryInfC{$\Gamma \vdash x: \alpha $}
			\end{prooftree}
		\end{minipage}
		\begin{minipage}{0.25\textwidth}
			\begin{prooftree}
				\AxiomC{$\Gamma, x:\alpha \vdash M : \beta$}
				\UnaryInfC{$\Gamma \vdash \lambda x. M : \alpha \to \beta $}
			\end{prooftree}
		\end{minipage}
		\begin{minipage}{0.35\textwidth}
			\begin{prooftree}
				\AxiomC{$\Gamma, \varphi: \alpha \to \beta, x:\alpha \vdash M : \beta$}
				\UnaryInfC{$\Gamma \vdash \mu^\varphi_x. M : \alpha \to \beta $}
			\end{prooftree}
		\end{minipage}
		\begin{minipage}{0.18\textwidth}
			\vspace{3mm}
			\begin{prooftree}
				\AxiomC{}
				\UnaryInfC{$\Gamma \vdash \num{r} : \typeReal $}
			\end{prooftree}
		\end{minipage}
		
		\begin{minipage}{0.5\textwidth}
			\begin{prooftree}
				\AxiomC{$\Gamma\vdash M : \beta \to \alpha$}
				\AxiomC{$\Gamma \vdash N : \beta$}
				\BinaryInfC{$\Gamma \vdash M N : \alpha$ }
			\end{prooftree}
		\end{minipage}
		\begin{minipage}{0.49\textwidth}
			\vspace{2mm}
			\begin{prooftree}
				\AxiomC{$\Gamma \vdash M : \typeReal$}
				\AxiomC{$\Gamma \vdash N : \alpha$}
				\AxiomC{$\Gamma \vdash P : \alpha$}
				\TrinaryInfC{$\Gamma \vdash \myif(M, N, P) : \alpha $}
			\end{prooftree}
		\end{minipage}
		
		\begin{minipage}{0.2\textwidth}
			\vspace{3.2mm}
			\begin{prooftree}
				\AxiomC{}
				\UnaryInfC{$\Gamma \vdash \sample : \typeReal $}
			\end{prooftree}
		\end{minipage}
		\begin{minipage}{0.5\textwidth}
			\begin{prooftree}
				\AxiomC{$\Gamma \vdash M_1 : \typeReal$}
				\AxiomC{$\cdots$}
				\AxiomC{$\Gamma \vdash M_{|f|} : \typeReal$}
				\TrinaryInfC{$\Gamma \vdash f(M_1, \cdots, M_{|f|}) : \typeReal $}
			\end{prooftree}
		\end{minipage}
		\begin{minipage}{0.27\textwidth}
			\vspace{0.5mm}
			\begin{prooftree}
				\AxiomC{$\Gamma \vdash M : \typeReal$}
				\UnaryInfC{$\Gamma \vdash \score(M) : \typeReal $}
			\end{prooftree}
		\end{minipage}
	\end{tcolorbox}
	\caption{Full SPCF Typing rules} \label{fig:simpleTypesFull}
\end{figure*}

\subsection{Typing Rules for SPCF}

The full typing system of SPCF is given in \refFig{simpleTypesFull}.

\subsection{Additional Proofs}

\begin{lemmaRE}{\ref{lem:altExp}}
	If $M$ is AST then
	$$\expectedTermSteps{M} = \sum_{n = 0}^{\infty} \mu_{\stdtrset} \Big( \termTr{M}^{n} \Big) \cdot n$$
\end{lemmaRE}
\begin{proof}
	We have $\biguplus_n \termTr{M}^{n} = \termTr{M}$.
	Now define 
	$$\termTr{M}^{> n} \triangleq \biguplus_{i > n} \termTr{M}^{i}$$
	It is easy to see that $\mu_{\stdtrset} \big( \termTr{M}^{> n} \big) + \mu_{\stdtrset} \big( \termTr{M}^{\leq n} \big) = 1$ as by assumption $\mu_{\stdtrset} \big(\termTr{M}\big) = 1$.
	Now
	\begin{align*}
			\expectedTermSteps{M} &= \sum\limits_{n=0}^{\infty} \Big(1- \mu_{\stdtrset} \big(\termTr{M}^{\leq n}\big) \Big)\\
			&\myeq{\textbf{(1)}} \sum\limits_{n=0}^{\infty} \mu_{\stdtrset} \big(\termTr{M}^{> n}\big)\\
			&\myeq{\textbf{(2)}} \sum\limits_{n=0}^{\infty} \sum_{j > n}  \mu_{\stdtrset} \big(\termTr{M}^{i}\big)\\
			&\myeq{\textbf{(3)}} \sum\limits_{n=0}^{\infty} \mu_{\stdtrset} \big(\termTr{M}^{n}\big) \cdot n
	\end{align*}
	where \textbf{(1)} holds as $\mu_{\stdtrset} \big( \termTr{M}^{> n} \big) + \mu_{\stdtrset} \big( \termTr{M}^{\leq n} \big) = 1$, \textbf{(2)} follows as the union in the definition of $\termTr{M}^{> n}$ is disjoint and \textbf{(3)} is an easy combinatorial argument. 
\end{proof}

\begin{lemma}[PAST implies AST]
	If $M$ is PAST then $M$ is AST
\end{lemma}
\begin{proof}
	Assume $M$ is PAST so by definition 
	$$\sum_{n=0}^{\infty} \big(1- \mu_{\stdtrset} \big(\termTr{M}^{\leq n}\big) \big)$$
	is a finite sum.
	As this sum converges to a finite value the sequence $\big(\mu_{\stdtrset} \big(\termTr{M}^{\leq n}\big)\big)_{n \in \natnum}$ must converge to $1$.
	And as $\termTr{M}^{\leq n} \subseteq \termTr{M}$	for every $n$ and $\mu_{\stdtrset}$ is a measure (in particular monotone w.r.t.~to $\subseteq$ and $\leq$) we get $\mu_{\stdtrset} (\termTr{M}) = 1$, so $M$ is AST.
\end{proof}

\subsection{Call by Value}

\begin{figure*}[!t]
	\begin{tcolorbox}[colback=white, colframe=black, arc=0mm, boxrule=1pt]
		\begin{minipage}{0.5\textwidth}
			\begin{prooftree}
				\AxiomC{}
				\UnaryInfC{$\langle (\lambda x. M) V, \tr \rangle \cbvto \langle M[V/x], \tr \rangle$}
			\end{prooftree}
		\end{minipage}
		\begin{minipage}{0.5\textwidth}
			\begin{prooftree}
				\AxiomC{}
				\UnaryInfC{$\langle (\mu^\varphi_x. M) V, \tr \rangle \cbvto \langle M[V/x, (\mu^\varphi_x. M)/\varphi], \tr \rangle $}
			\end{prooftree}
		\end{minipage}
		
		\begin{minipage}{0.33\textwidth}
			\begin{prooftree}
				\AxiomC{$r \leq 0$}
				\UnaryInfC{$\langle \myif(\num{r}, N, P), \tr \rangle \cbvto \langle N, \tr \rangle $}
			\end{prooftree}
		\end{minipage}
		\begin{minipage}{0.33\textwidth}
			\vspace{0.5mm}
			\begin{prooftree}
				\AxiomC{$r > 0$}
				\UnaryInfC{$\langle \myif(\num{r}, N, P), \tr \rangle \cbvto \langle P, \tr \rangle $}
			\end{prooftree}
		\end{minipage}
		\begin{minipage}{0.3\textwidth}
			\vspace{3mm}
			\begin{prooftree}
				\AxiomC{}
				\UnaryInfC{$\langle \sample, r::\tr \rangle \cbvto \langle \num{r}, \tr \rangle $}
			\end{prooftree}
		\end{minipage}
		
		\begin{minipage}{0.45\textwidth}
			\vspace{5.3mm}
			\begin{prooftree}
				\AxiomC{}
				\UnaryInfC{$\langle f(\num{r_1}, \cdots, \num{r_{|f|}}), \tr \rangle \cbvto \langle \num{f(r_1, \cdots, r_{|f|})}, \tr \rangle $}
			\end{prooftree}
		\end{minipage}
		\begin{minipage}{0.25\textwidth}
			\vspace{1.7mm}
			\begin{prooftree}
				\AxiomC{$r \geq 0$}
				\UnaryInfC{$\langle \score(\num{r}), \tr \rangle \cbvto \langle \num{r}, \tr \rangle $}
			\end{prooftree}
		\end{minipage}
		\begin{minipage}{0.25\textwidth}
			\begin{prooftree}
				\AxiomC{$\langle R, \tr \rangle \cbvto \langle M, \tr' \rangle$}
				\UnaryInfC{$\langle E[R], \tr \rangle \cbvto \langle E[M], \tr' \rangle $}
			\end{prooftree}
		\end{minipage}
		
	\end{tcolorbox}
	\caption{Call by Value small-step reduction $\cbvto$ for SPCF. If it is clear from the context that we work in a CbV strategy we drop the annotation and simply write $\to$.\index{$\cbvto$, the CbV SPCF reduction relation}} \label{fig:cbvRules}
\end{figure*}

We now introduce a Call by Value evaluation strategy for SPCF. 
We define call by value redexes and evaluation contexts by:

\begin{align*}
	R &\defined (\lambda x. M) V \mid (\mu^\varphi_x. M) V \mid \myif(\num{r}, N, P) \\
	&\quad\quad\mid f(\num{r_1}, \cdots, \num{r_{|f|}}) \mid \sample \mid \score(\num{r})\\
	E &\defined [\cdot] \mid E M \mid (\lambda y. M) E \mid (\mu^\varphi_x. M) E \mid \myif(E, N, P) \\
	&\quad \quad \mid f(\num{r_1}, \cdots, \num{r_{k-1}}, E, M_{k+1}, \cdots, M_{|f|}) \mid \score(E)
\end{align*}
Note that for a $\beta$-redex to reduce, the argument must be a value and we conversely reduce the left hand side of applications. 
We define the CbV reduction relation by the rules in \refFig{cbvRules}.
The definitions in \refSection{AST} regarding AST and PAST extend naturally to CbV\footnote{While the concepts extend naturally, they obviously are not identical. E.g., the probability of a termination in CbV may very well differ from the one in CbN.}.  
Throughout this paper we always make clear what evaluation strategy we are using, so the notation never clashes. 

\section{Additional Material - Section~\ref{sec:ibs}}

\subsection{Interval-Based Semantics}

\begin{lemmaRE}{\ref{lem:suf1}}
	If $f: \real^n \to \real$ is continuous then $f$ is interval preserving
\end{lemmaRE}
\begin{proof}
	Let $A \triangleq \myint{a_1, b_1} \times \cdots \times \myint{a_n, b_n}$ as in the definition of interval perseverance.
	We need a higher-dimensional version of the intermediate value theorem (IVT): if $\mathbf{x}, \mathbf{y} \in A$ and $c \in \real$ be such that $f(\mathbf{x}) \leq c \leq f(\mathbf{y})$ then there is a $\mathbf{z} \in A$ such that $f(\mathbf{z}) = c$ \textbf{(1)}.
	The IVT implies that $f(A)$ is a connected set. 
	A standard property of continuous functions is that the images of compact sets are compact sets.
	Due to the Heine–Borel theorem, compact euclidean sets are exactly those that are bounded and closed.
	As $A$ is obviously compact, we get that $f(A)$ is compact and thus bounded and closed.
	As $f(A)$ is also connected (by the IVT), it is a closed (bounded) interval as required. 
	
	It remains to show \textbf{(1)}:
	Let $\gamma : \myint{0, 1} \to A$ be defined by $\gamma(t) \triangleq t \mathbf{x} + (1-t) \mathbf{y}$ which is obviously continuous.
	In particular, note that since $A$ is a box (and thus convex) $\gamma(t) \in A$ for every $t \in \myint{0, 1}$. 
	Now define $\phi : \myint{0, 1} \to \real$ by $\phi \triangleq f \circ \gamma$ which is the composition of continuous functions and thus also continuous. 
	Now $\phi(0) = f(\mathbf{x}) \leq c \leq f(\mathbf{y}) = \phi(1)$ so by the intermediate value theorem in the 1d case there exists a $t \in \myint{0, 1}$ with $\phi(t) = c$. 
	We can define $\mathbf{z} \triangleq \gamma(t)$ which satisfies the requirement by definition of $\phi$.
\end{proof}

\begin{figure*}[!t]
	\begin{tcolorbox}[colback=white, colframe=black, arc=0mm, boxrule=1pt]
		\begin{minipage}{0.5\textwidth}
			\begin{prooftree}
				\AxiomC{}
				\UnaryInfC{$\langle (\lambda x. \calM) \calN, \wpi \rangle \too \langle \calM[\calN/x], \wpi \rangle $}
			\end{prooftree}
		\end{minipage}
		\begin{minipage}{0.5\textwidth}
			\begin{prooftree}
				\AxiomC{}
				\UnaryInfC{$\langle (\mu^\varphi_x. \calM) \calN, \wpi \rangle \too \langle \calM[\calN/x, (\mu^\varphi_x. \calM)/\varphi], \wpi \rangle $}
			\end{prooftree}
		\end{minipage}
		
		\begin{minipage}{0.5\textwidth}
			\begin{prooftree}
				\AxiomC{$b \leq 0$}
				\UnaryInfC{$\langle \myif(\num{\myint{a, b}}, \calN, \calP), \wpi \rangle \too \langle \calN, \wpi \rangle $}
			\end{prooftree}
		\end{minipage}
		\begin{minipage}{0.5\textwidth}
			\begin{prooftree}
				\AxiomC{$a > 0$}
				\UnaryInfC{$\langle \myif(\num{\myint{a, b}}, \calN, \calP), \wpi \rangle \too \langle \calP, \wpi \rangle $}
			\end{prooftree}
		\end{minipage}
		
		\begin{minipage}{0.5\textwidth}
			\vspace{2.5mm}
			\begin{prooftree}
				\AxiomC{}
				\UnaryInfC{$\langle \sample, \myint{a, b} :: \wpi \rangle \too \langle \num{\myint{a, b}}, \wpi \rangle $}
			\end{prooftree}
		\end{minipage}
		\begin{minipage}{0.5\textwidth}
			\begin{prooftree}
				\AxiomC{$0 \leq a$}
				\UnaryInfC{$\langle \score(\num{\myint{a, b}}),\wpi \rangle \too \langle \num{\myint{a, b}}, \wpi \rangle $}
			\end{prooftree}
		\end{minipage}
		
		\begin{minipage}{0.65\textwidth}
			\vspace{5mm}
			\begin{prooftree}
				\AxiomC{}
				\UnaryInfC{$\langle f\big(\num{\myint{a_1, b_1}}, \cdots, \num{\myint{a_{|f|}, b_{|f|}}} \big), \wpi \rangle \too \langle \num{\hat{f}(a_1, b_1, \cdots, a_{|f|}, b_{|f|})}, \wpi \rangle $}
			\end{prooftree}
		\end{minipage}
		\begin{minipage}{0.34\textwidth}
			\begin{prooftree}
				\AxiomC{$\langle \calR, \wpi \rangle \too \langle \calM, \wpi' \rangle$}
				\UnaryInfC{$\langle \calE[\calR], \wpi \rangle \too \langle \calE[\calM], \wpi' \rangle$}
			\end{prooftree}
		\end{minipage}

	\end{tcolorbox}
	
	\caption{Internal-based (CbN) small-step reduction.\index{$\too$, the interval-based (CbN) reduction relation}} \label{fig:intervalSemantics}
\end{figure*}

\subsection{Interval-Based Reduction}

The full (CbN) reduction system for interval terms is given in \refFig{intervalSemantics}.

\begin{figure*}[!t]
	\begin{tcolorbox}[colback=white, colframe=black, arc=0mm]
		\begin{minipage}{0.15\textwidth}
			\vspace{1.3mm}
			\begin{prooftree}
				\AxiomC{}
				\UnaryInfC{$x \triangleleft x$}
			\end{prooftree}
		\end{minipage}
		\begin{minipage}{0.25\textwidth}
			\vspace{2.3mm}
			\begin{prooftree}
				\AxiomC{}
				\UnaryInfC{$\sample \triangleleft \sample$}
			\end{prooftree}
		\end{minipage}
		\begin{minipage}{0.25\textwidth}
			\begin{prooftree}
				\AxiomC{$r \in \myint{a, b}$}
				\UnaryInfC{$\num{r} \triangleleft \num{\myint{a, b}}$}
			\end{prooftree}
		\end{minipage}
		\begin{minipage}{0.24\textwidth}
			\begin{prooftree}
				\AxiomC{$M \triangleleft \calM$}
				\UnaryInfC{$\lambda x. M \triangleleft \lambda x. \calM$}
			\end{prooftree}
		\end{minipage}
		
		\begin{minipage}{0.33\textwidth}
			\vspace{0.1mm}
			\begin{prooftree}
				\AxiomC{$M \triangleleft \calM$}
				\AxiomC{$N \triangleleft \calN$}
				\BinaryInfC{$M N \triangleleft \calM \calN$}
			\end{prooftree}
		\end{minipage}
		\begin{minipage}{0.33\textwidth}
			\begin{prooftree}
				\AxiomC{$M \triangleleft \calM$}
				\UnaryInfC{$\mu^\varphi_x. M \triangleleft \mu^\varphi_x. \calM$}
			\end{prooftree}
		\end{minipage}
		\begin{minipage}{0.32\textwidth}
			\begin{prooftree}
				\AxiomC{$M \triangleleft \calM$}
				\UnaryInfC{$\score(M) \triangleleft \score(\calM)$}
			\end{prooftree}
		\end{minipage}
		
		\begin{minipage}{0.5\textwidth}
			\vspace{1.2mm}
			\begin{prooftree}
				\AxiomC{$M \triangleleft \calM$}
				\AxiomC{$N \triangleleft \calN$}
				\AxiomC{$P \triangleleft \calP$}
				\TrinaryInfC{$\myif(M, N, P) \triangleleft \myif(\calM, \calN, \calP)$}
			\end{prooftree}
		\end{minipage}
		\begin{minipage}{0.49\textwidth}
			\begin{prooftree}
				\AxiomC{$M_1 \triangleleft \calM_1$}
				\AxiomC{$\cdots$}
				\AxiomC{$M_{|f|} \triangleleft \calM_{|f|}$}
				\TrinaryInfC{$f(M_1, \cdots, M_{|f|}) \triangleleft f(\calM_1, \cdots, \calM_{|f|})$}
			\end{prooftree}
		\end{minipage}	
		
	\end{tcolorbox}
	\caption{Inductive definition of the refinement relation $\triangleleft$ between the set of well-typed terms $\Lambda$ and the set of well-typed interval terms $\LambdaInterval$.} \label{fig:subsumption}
\end{figure*}

\subsection{Soundness}
\label{apx:soundness}

This subsection is devoted to give a full proof of \refTheo{intSound}.

Assume $(\Omega, \Sigma_{\Omega})$ is a measurable space and $\mu$ is a measure on $(\Omega, \Sigma_{\Omega})$.
$A, B \in \Sigma_{\Omega}$ are called almost-disjoint if $\mu(A \cap B) = 0$.
In the case of $\Omega = \real$ we get that intervals $\myint{a, b}$ and $\myint{c, d}$ are almost disjoint iff $b \leq c$ or $d \leq a$. 

\subsubsection*{Embedding and Refinement}
To state soundness it is fruitful to investigate the embedding of standard terms in interval terms ($\toIntervalTerm{\cdot}$).
We define a relation $M \triangleleft \calM$ in \refFig{subsumption} which models the intuitive idea of viewing every interval numeral $\num{\myint{a, b}}$ as any value within $\myint{a, b}$. 
Then $M \triangleleft \calM$ is derivable if and only if $M$ and $\calM$ agree structurally and every standard numeral in $M$ is contained in the repressive interval numeral in $\calM$.
We can see that the canonical embedding is compatible with this refinement, i.e., for every standard term $M$, $M \triangleleft \toIntervalTerm{M}$.  \index{$\triangleleft$, blabla}
We can also define a refinement between standard traces and interval by
$$ r_0 \cdots r_{n-1} \triangleleft \myint{a_0, b_0} \cdots \myint{a_{n-1}, b_{n-1}} \Leftrightarrow \forall i: r_i \in \myint{a_i, b_i} $$
For an interval trace $\wpi$ we define $\containedTraces{\wpi} \defined \{\tr \mid \tr \triangleleft \wpi\}$, i.e., the set of all traces refining $\wpi$. \index{$\triangleleft$, the refinement relation between interval term and terms as well as interval traces and traces} \index{$\containedTraces{\wpi}$ the set of standard traces that refine $\wpi$}

\begin{lemma}\label{lem:correspondance}
	If $\langle \calM, \wpi \rangle \too^n \langle \calN, \wp' \rangle$ and $M \triangleleft \calM$ and $\tr \triangleleft \wp$ then there exists a $N \triangleleft \calN$ and $\tr' \triangleleft \wp'$ such that $\langle M, \tr \rangle \to^n \langle N, \tr' \rangle$.
\end{lemma}
\begin{proof}
	We first observe the following obvious result:
	If $M \triangleleft \calM$ and $N_i \triangleleft \calN_i$ for $i \in [n]$ then $M[N_i/x_i]_{i \in [n]} \triangleleft \calM[\calN_i/x_i]_{i \in [n]}$ which can be proved by induction on $M$ (or $\calM$). 
	Call this observation \textbf{(1)}.
	We now show the statement for $n = 1$. The case for $n = 0$ is trivial and for $n > 1$ follows by a simple induction.
	We do structural induction on $\calM$. 
	
	\begin{itemize}
		\item If $\calM = (\lambda x. \calP) \calQ$: then $\calN = \calP[\calQ/x]$ and as $M \triangleleft \calM$, $M = (\lambda x. P) V$ for some $P \triangleleft \calP, Q \triangleleft \calQ$ and $\wpi' = \wpi$. 
		Define $\tr' \triangleq \tr$ and $N \triangleq P[Q/x]$. Clearly $\langle M, \tr \rangle \to \langle N, \tr' \rangle$.
		Now $\tr' \triangleleft \wpi'$ is obvious and from \textbf{(1)} we also get $N \triangleleft \calN$.
		
		\item If $\calM = (\mu^\varphi_x. \calP) \calQ$: similar to the previous case.
		
		\item $\calM = f\big(\num{\myint{a_1, b_1}}, \cdots, \num{\myint{a_{|f|}, b_{|f|}}} \big)$. Then  
		$$\calN = \num{\hat{f}(a_1, b_1, \cdots, a_{|f|}, b_{|f|})}$$
		As $M \triangleleft \calM$ we get $M = f(r_1, \cdots, r_{|f|})$ and $r_i \in \myint{a_i, b_i}$.
		Define $\tr' \triangleq \tr$ and $N \triangleq \num{f(r_1, \cdots, r_n)}$. 
		Clearly $\langle M, \tr \rangle \to \langle N, \tr' \rangle$.
		As $f$ is interval preserving we also get that $f(r_1, \cdots, r_{|f|}) \in \num{\hat{f}(a_1, b_1, \cdots, a_{|f|}, b_{|f|})}$ so $N \triangleleft \calN$.
		
		\item $\calM = \myif\big( \myint{a, b}, \calP, \calQ \big)$. Assume $b \leq 0$, the case where $a > 0$ is analogous. 
		So $\calN = \calP$.  As $M \triangleleft \calM$, $M = \myif\big(r, P, Q \big)$ for some $r \in \myint{a, b}$ and $P \triangleleft \calP$. 
		So we can choose $N = P$.
		
		\item $\calM = \sample$ so $\wpi = \myint{a, b}::\wpi'$ and $\calN = \myint{a, b}$. Now $M = \sample$ and as $\tr \triangleleft \wpi$, $\tr = r::\tr'$ for $r \in \myint{a, b}$.
		Now define $N = \num{r}$.
		
		\item $\calM = \score(\num{\myint{a, b}})$: So $M = \score(\num{r})$ and $r \in \myint{a, b}$. So $N = \num{r}$.
		
		\item $\calM = \calE[\calR]$ for $\calE \neq [\cdot]$: Then $\langle \calR, \wpi \rangle \to \langle \calM, \wpi' \rangle$.
		As $M \triangleleft \calM$ we have $M = E[R]$ for $R \triangleleft \calR$. 
		Now by induction on $\calR$ we get a $M$ with $M \triangleleft \calM$ and $\tr'  \triangleleft \wpi'$ such that $\langle R, \tr \rangle \to \langle M, \tr' \rangle$. 
		Now $E[M] \triangleleft \calE[\calM]$ as $\triangleleft$ is obviously closed under evaluation contexts. 
		And $\langle E[R], \tr \rangle \to \langle E[M], \tr' \rangle$ as required. 
	\end{itemize}
\end{proof}

\begin{lemma}\label{lem:termTrCon}
	If $\wpi \in \termInterval{\calM}$ and $M \triangleleft \calM$ then $\containedTraces{\wpi} \subseteq \termTr{M}$ and for each $\tr \in \containedTraces{\wpi}, \numberSteps{\wpi}{\calM} = \numberSteps{\tr}{M}$
\end{lemma}
\begin{proof}
	Follows directly from \refLemma{correspondance}.
\end{proof}

\begin{proposition}\label{prop:sound}
	For every countable set of \emph{pairwise compatible} traces $A \subseteq \termInterval{\calM}$ and every $M \triangleleft \calM$ we have the following:
	\begin{tasks}[style=itemize](2)
		\task $\omega(A) \leq \termProb{M}$ 
		\task $\expVal^\calM(A) \leq \expectedTermSteps{M}$
	\end{tasks}
\end{proposition}
\begin{proof}
	We first note that for every interval trace $\wpi$, $\containedTraces{\wpi}$ is a measurable set of traces and furthermore $\mu_{\stdtrset}(\containedTraces{\wpi}) = \omega(\wpi)$ by definition of the Lebesgue measure. 
	
	Now as $A$ is by assumption pairwise compatible the family $(\containedTraces{\wpi})_{\wpi \in A}$ is pairwise almost disjoint.
	Thus
	\begin{align*}
		\omega(A) &= \sum_{\wpi \in A} \omega(\wpi) \myeq{\textbf{(1)}} \sum_{\wpi \in A} \mu_{\stdtrset} \big( \containedTraces{\wpi} \big) \\
		&\myeq{\textbf{(2)}} \mu_{\stdtrset} \Big( \bigcup_{\wpi \in A} \containedTraces{\wpi} \Big)
		\myleq{\textbf{(3)}} \mu_{\stdtrset} \big( \termTr{M}\big) = \termProb{M}
	\end{align*}%
	where \textbf{(1)} follows from the definition of the Lebesgue measure on boxes, \textbf{(2)} from the fact that family is pairwise almost disjoint and thus differs by a countable union of null sets.
	\textbf{(3)} follows from \refLemma{termTrCon}.
	
	For the second part we can observe that if $\wpi \in \termInterval{\calM}$, $M \triangleleft \calM$ and $\tr \triangleleft  \wpi$, then $\numberSteps{\tr}{M} = \numberSteps{\wpi}{\calM}$. 
	Now
	\begin{align*}
		\expVal^\calM(A) &= \sum_{\wpi \in A} \omega(\wpi) \cdot \numberSteps{\wpi}{\calM} \\
		&\myeq{\textbf{(1)}} \sum_{n=0}^\infty \omega(\{\wpi \in A \mid \numberSteps{\wpi}{\calM} = n\}) \cdot n \\
		&\myleq{\textbf{(2)}} \sum_{n = 0}^{\infty} \mu_{\stdtrset} \Big( \termTr{M}^{n} \Big) \cdot n \myleq{\textbf{(3)}} \expectedTermSteps{M}
	\end{align*}%
	where \textbf{(1)} follows from simple reordering, \textbf{(2)} from the fact that every interval trace in $\{\wpi \in A \mid \numberSteps{\wpi}{\calM} = n\}$ we get $\containedTraces{\wpi} \subseteq \termTr{M}^{n}$ and the same reasoning as above.  
	\textbf{(3)} is standard and can e.g.~be inferred from the proof of \refLemma{altExp}.
\end{proof}

\begin{theoremRE}{\ref{theo:intSound}}
	For every countable set of \emph{pairwise compatible} traces $A \subseteq \termInterval{\toIntervalTerm{M}}$ the following holds:
	\begin{tasks}[style=itemize](2)
		\task $\omega(A) \leq \termProb{M}$ 
		\task $\expVal(\toIntervalTerm{M}, A) \leq \expectedTermSteps{M}$
	\end{tasks}
\end{theoremRE}
\begin{proof}
	Follows directly from \refProp{sound} as $M \triangleleft \toIntervalTerm{M}$. 
\end{proof}

\subsection{Completeness}

\begin{lemmaRE}{\ref{lem:suf2}}
	If $f: \real^n \to \real$ is continuous and for every $y \in \real$, $f^{-1}(\{y\})$ is a Lebesgue Null-set then $f$ is $\mathbb{Q}$-interval separable. 
\end{lemmaRE}
\begin{proof}
	Let $I = \myint{a, b}$ be an interval as in the definition of interval separable. 
	We have $f^{-1}(\myint{a, b}) = f^{-1}((a, b)) \cup f^{-1}(\{a\}) \cup f^{-1}(\{b\})$.
	By assumption $f^{-1}(\{a\})$ and $f^{-1}(\{b\})$ are null sets.
	As $f$ is continuous and $(a, b)$ is an open set we get that $f^{-1}(a, b)$ is an open set. A well know result in $\real^n$ is that every open-set can be covered exactly by a countable number of boxes that have rational endpoints.
	So there are boxes $B_1, B_2, \cdots$ (with rational endpoints) such that $\cup_i B_i = f^{-1}(a, b)$. 
\end{proof}

The remaining pats of this section are devoted to give a proof of \refTheo{intComp}.
We assume that all primitive function $f \in \mathbb{F}$ are interval separable. 
As a simple example while this is not easy consider the following:

\begin{example}\label{ex:nonInterval}
	Consider the term $M \triangleq \myif (\sample - \num{0.5}, \num{0}, \num{1})$ which is clearly AST.
	In fact, we have $\termTr{M} = \{s_1 \mid s_1 \in \intreal \}$, so the set of terminating traces is itself an interval. 
	However, the interval trace $\wpi = \myint{0, 1}$ is not terminating for $\toIntervalTerm{M}$ (formally $\myint{0, 1} \not\in \termInterval{\toIntervalTerm{M}}$).
\end{example}

\begin{figure*}[!t]
	\begin{tcolorbox}[colback=white, colframe=black, arc=0mm, boxrule=1pt]
		\begin{minipage}[t]{0.3\textwidth}
			\begin{prooftree}
				\AxiomC{}
				\UnaryInfC{$\langle (\lambda x. M) N, \tr, \kappa \rangle \toConditional \langle M[N/x], \tr, \kappa \rangle $}
			\end{prooftree}
		\end{minipage}
		\begin{minipage}{0.37\textwidth}
			\begin{prooftree}
				\AxiomC{$r \leq 0$}
				\UnaryInfC{$\langle \myif(\num{r}, N, P), \tr, \leftP::\kappa \rangle \toConditional \langle N, \tr, \kappa \rangle$ }
			\end{prooftree}
		\end{minipage}
		\begin{minipage}{0.37\textwidth}
			\begin{prooftree}
				\AxiomC{$r > 0$}
				\UnaryInfC{$\langle \myif(\num{r}, N, P), \tr, \rightP::\kappa \rangle \toConditional \langle P, \tr, \kappa \rangle$ }
			\end{prooftree}
		\end{minipage}

		\begin{minipage}[t]{0.25\textwidth}
			\begin{prooftree}
				\AxiomC{}
				\UnaryInfC{$\langle \sample, r::\tr, \kappa \rangle \toConditional \langle \num{r}, \tr, \kappa \rangle $}
			\end{prooftree}
		\end{minipage}
		\begin{minipage}{0.49\textwidth}
			\vspace{3mm}
			\begin{prooftree}
				\AxiomC{}
				\UnaryInfC{$\langle (\mu^\varphi_x. M) N, \tr, \kappa \rangle \toConditional \langle M[N/x, (\mu^\varphi_x. M)/\varphi], \tr, \kappa \rangle $}
			\end{prooftree}
		\end{minipage}
		\begin{minipage}{0.25\textwidth}
			\begin{prooftree}
				\AxiomC{$r \geq 0$}
				\UnaryInfC{$\langle \score(\num{r}), \tr, \kappa \rangle \toConditional \langle \num{r}, \tr, \kappa \rangle $}
			\end{prooftree}
		\end{minipage}

		\begin{minipage}{0.5\textwidth}
			\vspace{4.5mm}
			\begin{prooftree}
				\AxiomC{}
				\UnaryInfC{$\langle f(\num{r_1}, \cdots, \num{r_{|f|}}), \tr, \kappa \rangle \toConditional \langle \num{f(r_1, \cdots, r_{|f|})}, \tr, \kappa \rangle $}
			\end{prooftree}
		\end{minipage}
		\begin{minipage}{0.5\textwidth}
			\begin{prooftree}
				\AxiomC{$\langle R, \tr, \kappa \rangle \toConditional \langle M, \tr', \kappa' \rangle$}
				\UnaryInfC{$\langle E[R], \tr, \kappa \rangle \toConditional \langle E[M], \tr', \kappa' \rangle $}
			\end{prooftree}
		\end{minipage}
		
	\end{tcolorbox}
	
	\caption{Small-step reduction relation with conditional oracles. } \label{fig:conditionalOracle}
\end{figure*}

The key step to constructing a countable set of interval traces is to focus on branching.
We, therefore, annotate the reduction relation with explicit information which branch of a conditional was taken. 
We define the set of \emph{directions} by $D = \{\leftP, \rightP\}$. 
A \emph{conditional oracle} is then a sequence $\kappa \in D^*$.\index{$\kappa$, a conditional oracle}
To define the meaning of a conditional oracle we use a modified reduction relation ${\toConditional} \subseteq {(\Lambda \times \stdtrset \times D^*)^2}$ via the rules in \refFig{conditionalOracle}. \index{$\toConditional$, the conditional oracle reduction relation}
We can easily see:
\begin{lemma}\label{lem:disjointMkappa}
	If $\tr \in \termTr{M}$ then there exists a \emph{unique} $\kappa \in D^*$ and value $V$ with $\langle M, \tr, \kappa \rangle \toConditional^* \langle V, \epsilon, \epsilon \rangle$.
\end{lemma}

We now partition the set of terminating traces according to their branching behaviour. 
For $\kappa \in D^*$ we define 
$$\termTrCO{M}{\kappa} \triangleq \{\tr \in \stdtrset \mid \exists V:  \langle M, \tr,  \kappa \rangle \toConditional^* \langle V, \epsilon, \epsilon \rangle \}$$
I.e., all traces that branch according to $\kappa$. \index{$\termTrCO{M}{\kappa}$, terminating traces of $M$ that branch according to $\kappa$}
By \refLemma{disjointMkappa} it is easy to see that that the family $\big\{\termTrCO{M}{\kappa}\big\}_{\kappa \in D^*}$ forms a partition of the set of terminating traces.
Note that by fixing the branching, we also fix the number of reduction steps and the number of samples: if $\tr_1, \tr_2 \in \termTrCO{M}{\kappa}$, $\numberSteps{\tr_1}{M} = \numberSteps{\tr_2}{M}$ and $|\tr_1| = |\tr_2|$. 

In \refExample{nonInterval}, we have seen that there exist interval traces $\wpi \in \inttrset$ with $\containedTraces{\wpi} \subseteq \termTr{M}$ that are not terminating for the canonical embedding $\toIntervalTerm{M}$ (i.e., $\wpi \not\in \termInterval{\toIntervalTerm{M}}$).
We can, however, show that if all traces in $\containedTraces{\wpi}$ follow the same branching $\wpi \in \termInterval{\toIntervalTerm{M}}$ does hold.
We first need the following:

\begin{lemma}\label{lem:techReduction}
	Let $\calM$ be any term not already a value, $\kappa, \kappa' \in D^*$ and $\wpi \in \inttrset$ and $n \in \natnum$.
	If for every pair $(M, \tr)$ with $M \triangleleft \calM$ and $\tr \triangleleft \wpi$
	$$\langle M, \tr, \kappa\rangle \toConditional^n \langle M_{(M, \tr)}, \tr_{(M, \tr)}, \kappa' \rangle$$
	(Note that $M_{(M, \tr)}$ and $\tr_{(M, \tr)}$ are uniquely determined).
	Then $\langle \calM, \wpi \rangle \too^n \langle \calM', \wpi' \rangle$ for some $\calM'$ and $\wpi'$ such that for every pair $(M, \tr)$, $M_{(M, \tr)} \triangleleft \calM'$ and $\tr_{(M, \tr)} \triangleleft \wpi'$. 
\end{lemma}
\begin{proof}
	We show the result for $n = 1$. The case for $n = 0$ is trivial and for $n > 1$ follows by easy induction using the case for $n = 1$.
	The proof goes by induction on $\calM$. 
	We only consider the case were $\calM$ is itself a redex. 
	The case where $\calM = \calE[\calR]$ follow by induction on $\calR$.
	The only interesting case is where $\calM$ is a conditional redex:
	So lets focus on the case where $\calM = \myif(\myint{a, b}, \calN, \calP)$: 
	Now any $M \triangleleft \calM$ must have the from $M = \myif(r_M, N_M, P_M)$ and there exist at least on such (as we work with closed, non, empty intervals).
	As any such $M$ can reduce via $\toConditional$ by assumption we get that $\kappa = \leftP::\kappa'$ or $\kappa = \rightP::\kappa'$ as otherwise no reduction can take place.
	W.l.o.g.~assume $\kappa = \leftP\kappa'$.
	We now claim that $b \leq 0$. Assume for contradiction that $b > 0$. 
	Then choose the term $\dot{M} \triangleleft \myif(b, N, P)$ where $N \triangleleft \calN$ and $P \triangleleft \calP$ are arbitrary (they always exist).
	Now $\langle \dot{M}, \tr, \leftP::\kappa'\rangle$ cannot make a reduction step via $\toConditional$ which contradicts the assumption.
	So $b \leq 0$. 
	
	This means that $\calM = \myif(\myint{a, b}, \calN, \calP) \too \calN$.
	Now any $M \triangleleft \calM$ of the form $M = \myif(r_M, N_M, P_M)$ and reduces to $N_M$.
	So $M_{(M, \tr)} = N_M \triangleleft \calN = \calM'$ as required and obviously $\tr_{(M, \tr)} = \tr \triangleleft \wpi = \wpi'$. 
\end{proof}

\begin{lemma}\label{lem:ensureTerm}
	If $\wpi \in \inttrset$, $\kappa \in D^*$ and $\containedTraces{\wpi} \subseteq \termTrCO{M}{\kappa}$ then $\wpi \in \termInterval{\toIntervalTerm{M}}$.
\end{lemma}
\begin{proof}
	We show the following stronger lemma which immediately implies the result as the only term refining $\toIntervalTerm{M}$ is $M$ itself:
	If $\calM$ is any interval term, $\wpi \in \inttrset$ and $\kappa \in D^*$ and for all $M \triangleleft \calM$, $\containedTraces{\wpi} \subseteq \termTrCO{M}{\kappa}$ then $\wpi \in \termInterval{\calM}$.
	
	We can proof this as follows: As for any $M \triangleleft \calM$, $\containedTraces{\wpi} \subseteq \termTrCO{M}{\kappa}$ we get that every $M \triangleleft \calM$ and $\tr \triangleleft \wpi$, $\langle M, \tr, \kappa\rangle \toConditional^n \langle V_{(M, \tr)}, \epsilon, \epsilon \rangle$ for a fixed $n$ (As soon as $\kappa$ is fixed each reduction takes the same number of steps).
	We can thus apply \refLemma{techReduction} and get that $\langle \calM, \wpi \rangle \too^n \langle \calV, \epsilon \rangle$, so $\wpi \in \termInterval{\calM}$ as required. 
\end{proof}

\subsection{Symbolic Terms and Symbolic Execution}\label{sec:symbolicTerms}

The second key ingredient is symbolic execution as this gives us a better understanding of the sets $\termTrCO{M}{\kappa}$.
The idea of symbolic terms is to not evaluate a term on a fixed trace of real numbers but instead on a generic trace consisting of variables. 
Whenever we resolve a \sample-statement we do not substitute in a real number but a variable.
This does prohibit us from evaluating primitive functions or resolve conditionals.
To circumvent the former we use symbolic values, which can be seen as partially evaluate primitive functions.
To resolve the latter we make use of  the conditional oracles. For an overview of a similar system of symbolic execution we refer the reader to \citep{MakOPW20}.

Let $\alpha_0, \alpha_1, \cdots$ be a denumerable set of \emph{sample-variables} indexed by natural numbers. 
We use them to postpone every sample statement by instead substitution a fresh variable. 
Symbolic values and terms are defined by:
\begin{align*}
		\symV &\triangleq x \mid \num{r} \mid \alpha_j \mid \lambda x. \symM \mid \mu^\varphi_x. \symM \mid \boxed{f}(\symV_1, \cdots, \symV_{|f|})\\
		\symM, \symN, \symP &\triangleq \symV \mid  \symM \symN \mid \myif(\symM, \symN, \symP)\\
		&\quad\quad\mid f(\symM_1, \cdots, \symM_{|f|}) \mid \sample \mid \score(\symM) 
\end{align*}%
Note that the only new syntactic additions, compared with standard SPCF, are the sample variables $\alpha_j$ and the symbolic primitive functions $\boxed{f}(\symM_1, \cdots, \symM_{|f|})$.\index{$\symM, \symN, \symP$, symbolic term} \index{$\symV$, symbolic values}
We again focus on typable terms.
The simple type system for standard SPCF (given in \refFig{simpleTypes}) naturally extends to symbolic terms when we add the following two rules:

\begin{prooftree}
	\AxiomC{}
	\UnaryInfC{$\Gamma \Vdash \alpha_j : \typeReal $}
\end{prooftree}

\begin{prooftree}
	\AxiomC{$\Gamma \Vdash \symM_1 : \typeReal$}
	\AxiomC{$\cdots$}
	\AxiomC{$\Gamma \Vdash \symM_{|f|} : \typeReal$}
	\TrinaryInfC{$\Gamma \Vdash \boxed{f}(\symM_1, \cdots, \symM_{|f|}) : \typeReal $}
\end{prooftree}

Let $\LambdaSymbolic$ be the set of all typable symbolic terms.
Note that any $M \in \Lambda$ directly corresponds to a symbolic term in the canonical way.

\begin{figure*}[!t]
	\begin{tcolorbox}[colback=white, colframe=black, arc=0mm, boxrule=1pt]
		\begin{minipage}{0.45\textwidth}
			\begin{prooftree}
				\AxiomC{}
				\UnaryInfC{$\symConf{(\lambda x.  \symM) \symN, \kappa, n}{\Delta} \toSym \symConf{\symM[\symN/x], \kappa, n}{\Delta}$}
			\end{prooftree}
		\end{minipage}
		\begin{minipage}{0.55\textwidth}
			\begin{prooftree}
				\AxiomC{}
				\UnaryInfC{$\symConf{\score(\symV), \kappa, n}{\Delta} \toSym \symConf{\symV, \kappa, n}{\Delta \cup \{\symV \geq 0 \}}$}
			\end{prooftree}
		\end{minipage}
		
		\begin{minipage}{0.6\textwidth}
			\begin{prooftree}
				\AxiomC{}
				\UnaryInfC{$\symConf{(\mu^\varphi_x. \symM) \symN, \kappa, n}{\Delta} \toSym \symConf{\symM[\symN/x, (\mu^\varphi_x. \symM)/\varphi], \kappa, n}{\Delta}$}
			\end{prooftree}
		\end{minipage}
		\begin{minipage}{0.4\textwidth}
			\begin{prooftree}
				\AxiomC{}
				\UnaryInfC{$\symConf{\sample, \kappa, n}{\Delta} \toSym \symConf{\alpha_{n}, \kappa, n+1}{\Delta}$}
			\end{prooftree}
		\end{minipage}
		
		\begin{minipage}{0.5\textwidth}
			\begin{prooftree}
				\AxiomC{}
				\UnaryInfC{$\symConf{\myif(\symV, \symN, \symP), \leftP::\kappa, n}{\Delta} \toSym \symConf{\symN, \kappa, n}{\Delta \cup \{\symV \leq 0\}}$}
			\end{prooftree}
		\end{minipage}
		\begin{minipage}{0.5\textwidth}
			\begin{prooftree}
				\AxiomC{}
				\UnaryInfC{$\symConf{\myif(\symV, \symN, \symP), \rightP::\kappa, n}{\Delta} \toSym \symConf{\symP, \kappa, n}{\Delta \cup \{\symV > 0\} }$}
			\end{prooftree}
		\end{minipage}
		
		\begin{minipage}{0.55\textwidth}
			\vspace{8.5mm}
			\begin{prooftree}
				\AxiomC{}
				\UnaryInfC{$\symConf{f(\symV_1, \cdots, \symV_{|f|}), \kappa, n} {\Delta} \toSym \symConf{\boxed{f}(\symV_1, \cdots, \symV_{|f|}), \kappa, n}{\Delta}$}
			\end{prooftree}
		\end{minipage}
		\begin{minipage}{0.45\textwidth}
			\begin{prooftree}
				\AxiomC{$\symConf{\symR, \kappa, n}{\Delta} \toSym \symConf{\symM, \kappa', n'}{\Delta'}$}
				\UnaryInfC{$\symConf{\symE[\symR], \kappa, n}{\Delta} \toSym \symConf{\symE[\symM], \kappa', n'}{\Delta'} $}
			\end{prooftree}
		\end{minipage}
		
	\end{tcolorbox}
	\caption{Small-step reduction for symbolic terms (symbolic execution).} \label{fig:symbolicSemantics}
\end{figure*}

\subsubsection{Symbolic Execution}

We now give an operational small-step semantics to symbolic terms.
This symbolic execution closely corresponds to reduction in the standard (CbN) semantics with the exception that every \sample-statement is resolved by a sample variable.
Symbolic redexes and evaluation contexts are defined as expected:
\begin{align*}
		\symR &\triangleq (\lambda x. \symM) \symN \mid (\mu^\varphi_x. \symM) \symN \mid \myif(\symV, \symN, \symP) \\
		&\quad\quad\mid f(\symV_1, \cdots, \symV_{|f|}) \mid \sample \mid \score(\symV)\\
		\symE &\triangleq [\cdot] \mid \symE \symM  \mid \myif(\symE, \symN, \symP) \mid \score(\symE) \\
		&\quad\quad\mid  f(\symV_1, \cdots, \symV_{k-1}, \symE, \symM_{k+1}, \cdots, \symM_{|f|}) 
\end{align*}

\subsubsection*{Symbolic Values}

As sample variables are taken in for real-valued numerals, whenever we resolve a sample statement, we can no longer evaluate primitive functions as some of the arguments may be variables.
A function symbol $f$ applied to arguments, therefore, does not evaluate to the function value but instead we postpone the evaluation and use the symbolic construct $\boxed{f}$. 
In particular, a (closed) symbolic value of type $\typeReal$ is no longer always a numeral. 
We can view $\boxed{f}$ as a function evaluation that is postponed. 
If we fix the value of the sample variables, a symbolic value, therefore, does again denotes a real number:
Let $\symV$ be a symbolic value of type $\typeReal$ (no $\lambda$ or $\mu$-abstraction) with sample-variables within $\{\alpha_0, \cdots, \alpha_{m-1}\}$. 
We can view a vector $\sigma \in \intreal^m$ as a substitution and define $\symV[\sigma] \in \real$ in the obvious way by substituting in values and evaluating primitive functions. 
Given $A \subseteq \real$ we define $\symV^{-1}(A) \triangleq \{\sigma \in \intreal^m \mid \symV[\sigma] \in A\}$.

\subsubsection*{Symbolic Inequality}

We define a symbolic inequality as a pair of the from $(\symV \bowtie r)$ where $\symV$ is a symbolic value, ${\bowtie} \in \{\leq, < , \geq, >\}$ and $r \in \real$.
A symbolic constraint $\Delta$ is a set of symbolic inequalities.
Given a symbolic constraint $\Delta = \{(\symV_i \bowtie_i r_i)\}_{i \in [n]}$ with sample variables contained within $\alpha_0, \cdots, \alpha_{m-1}$ we can define
$$\mathsf{Sat}_m(\Delta) \triangleq \{ \sigma \in \intreal^m \mid \forall i \in [n]: \symV_i[\sigma] \bowtie_i r_i \}$$
We can see every $\sigma \in \mathsf{Sat}_m(\Delta)$ also as an element in $\stdtrset^m$, i.e., a standard trace of length $m$.\index{$\mathsf{Sat}_m(\Delta)$, the set of traces of length $m$ that satisfy symbolic constraint $\Delta$}\index{$\Delta$, symbolic constraint}

\subsubsection*{Symbolic Configuration and Symbolic Execution}

A symbolic configuration has the from {\scriptsize$\symConf{\symM, \kappa, n} {\Delta}$} where $\symM$ is a symbolic term, $\kappa \in D^*$ a sequence of directions $n \in \natnum$ a natural number and $\Delta$  a symbolic constraint. \index{$\symConf{\symM, \kappa, n} {\Delta}$, symbolic configuration}
The conditional oracle $\kappa$ is used to resolve branching. During execution the constraints that a trace needs to satisfy to actually follow $\kappa$ are recorded in the constraint $\Delta$.
The natural number in each configuration references the number of sample variables that have already been substituted. 
We define the symbolic small-step reduction relation $\toSym$ via the rules in \refFig{symbolicSemantics}.\index{$\toSym$, symbolic reduction relation}

The symbolic execution we present here differs (especially on first glance) from the one used in \citep{MakOPW20}.
In their semantics, a symbolic configuration at all times contains a set of traces that can take this path.
In contrast, we annotate a symbolic configuration with an explicit set of symbolic inequalities.
As we fixed the outcomes of conditionals beforehand our reductions is deterministic.

\subsubsection*{Correspondence}

We can show the following correspondence theorem:

\begin{proposition}\label{prop:corr}
	For any term $M$. 
	If $\kappa \in D^*$ and there exist $\symV, n, \Delta$ (If they exists, they are unique) such that
	{\footnotesize$$\symConf{M, \kappa, 0} {\emptyset} \toSym^* \symConf{\symV, \epsilon, n} {\Delta}$$}
	then $\mathsf{Sat}_n(\Delta) = \termTrCO{M}{\kappa}$ otherwise $\termTrCO{M}{\kappa} = \emptyset$. 
\end{proposition}
\begin{proof}
	The proof is analogous to the proof in \citep[Thm.~13]{MakOPW20}.
	Every symbolic configuration {\scriptsize$\symConf{\symM, \kappa, n} {\Delta}$} can be seen as the pair $\llangle \symM, \_, \mathsf{Sat}_n(\Delta) \rrangle$ in the setting of \cite{MakOPW20} when we omit the weight parameter (denoted by $\_$). 
\end{proof}

\begin{example}
	Consider the term
	\begin{align*}
		\symM \triangleq \myif~&\sample + \sample - \num{1} \mythen x \myelse \\
		&\big(\myif~\num{0} \mythen \num{3} \myelse \num{4}\big)
	\end{align*}%
	For the conditional oracle $\rightP\leftP$ we get
	\begin{align*}
		\symConf{\symM, \rightP\leftP, 0} {\emptyset} \toSym^* \symConf{\num{3},\epsilon, 2}{\{\alpha_0 \boxed{+} \alpha_1 \boxed{-} \num{1} > 0, \num{0} \leq 0\}}
	\end{align*}%
	And the solution $\mathsf{Sat}_2$ of the symbolic constraint 
	$$\Delta \triangleq \{\alpha_0 \boxed{+} \alpha_1 \boxed{-} \num{1} > 0, \num{0} \leq 0\}$$ 
	is the set $\{s_0s_1 \in \stdtrset^2 \mid s_0 + s_1 > 1 \}$ which is exactly the set $\termTrCO{M}{\rightP\leftP}$ as stated in \refProp{corr}.
\end{example}

\subsubsection{Completeness Proof}

\begin{lemma}\label{lem:symPreimage}
	If $\symV$ is a symbolic value of type $\typeReal$ with sample variables among $\{\alpha_0, \cdots, \alpha_{m-1}\}$ where each variable occurs at most once and $\myint{a, b} \in \interval$ an interval then there exists a countable family of boxes $\{B_i\}_{i \in \indx}$ $(B_i \subseteq \intreal^m)$ such that $\bigcup_i B_i \almostSub \symV^{-1}(\myint{a, b})$.
\end{lemma}
\begin{proof}
	We do induction on the structure of $\symV$. The case of $\symV = \num{r}$ and $\symV = \alpha_j$ is trivial. So let $\symV = \boxed{f}(\symV_1, \cdots, \symV_{|f|})$.
	
	As $f$ is by assumption interval-separable there exist countable boxes $(B_i)_{i \in \indx}$ s.t., $\cup_i B_i \almostSub f^{-1}(\myint{a, b})$ for some countable set $\indx$. 
	Each $B_i$ is a box and can thus be written as  $B_i = \myint{a_i^1, b_i^1} \times \cdots \times \myint{a_i^{|f|}, b_i^{|f|}}$.
	Now define $C_i \triangleq \bigcap_{1\leq j \leq |f|} \symV_j^{-1}(\myint{a_i^j, b_i^j}) \subseteq \intreal^m$.
	These are all assignments such that each $\symV_j$ takes on a value in $\myint{a_i^j, b_i^j}$.
	As the countable union of Lebesgue null sets is a null we get $\bigcup_{i \in \indx} C_i \almostSub \symV^{-1}(\myint{a, b})$. Call this fact \textbf{(1)}.
	
	Now by induction for each $1 \leq j \leq |f|$ there exists a family of boxes $(B^{i, j}_k)_{k \in \indx_i^j}$ for some countable index set $\indx_i^j$, such that $\bigcup_{k \in \indx_i^j} B^{i, j}_k \almostSub \symV_j^{-1}(\myint{a_i^j, b_i^j})$.
	Now  $\bigcap_{1\leq j \leq |f|} \bigcup_{k \in \indx_i^j} B^{i, j}_k = \bigcup_{(k_1, \cdots, k_{|f|}) \in \indx_i^1 \times \cdots \times \indx_i^{|f|}} B^{i, 1}_{k_1} \cap \cdots B^{i, |f|}_{k_{|f|}}$	by distributing the intersection over the union. 
	We can put this together and get the following by again using the fact that the countable union of null sets is a null set:
 	\begin{align*}
			&\bigcup_{(k_1, \cdots, k_{|f|}) \in \indx_i^1 \times \cdots \times \indx_i^{|f|}} B^{i, 1}_{k_1} \cap \cdots B^{i, |f|}_{k_{|f|}} \\
			&\quad\quad= \bigcap_{1\leq j \leq |f|} \bigcup_{k \in \indx_i^j} B^{i, j}_k \almostSub \bigcap_{1\leq j \leq |f|} \symV_j^{-1}(\myint{a_i^j, b_i^j}) = C_i
	\end{align*}%
	Note that the index set $\indx_i^1 \times \cdots \times \indx_i^{|f|}$ is countable. 
	Combined with \textbf{(1)} we get
	{\small\begin{align*}
			\bigcup_{i \in \indx} \bigcup_{(k_1, \cdots, k_{|f|}) \in \indx_i^1 \times \cdots \times \indx_i^{|f|}} B^{i, 1}_{k_1} \cap \cdots B^{i, |f|}_{k_{|f|}} \almostSub \symV^{-1}(\myint{a, b})
	\end{align*}}%
	Note that the set $\{(i, k_1, \cdots, k_{|f|}) \mid i \in \indx, (k_1, \cdots, k_{|f|}) \in \indx_i^1 \times \cdots \times \indx_i^{|f|}\}$ is also countable as the countable product of countable sets.
	Also note that the finite intersection of boxes in the equation above is again a box. 
	We are thus done.
\end{proof}

\begin{lemma}\label{lem:RemDuplicates}
	If $A, B$ are two boxes in $\real^m$ then there exist finite boxes $\{C_i\}_{i \in [n]}$ that are pairwise almost disjoint and satisfy $A \cup B = \bigcup_i C_i$.
\end{lemma}

\begin{theoremRE}{\ref{theo:intComp}}
	If every $f \in \mathbb{F}$ is interval separable, then for every $M \in \Lambda_0$ there exists a countable set of pairwise-compatible interval traces $A \subseteq \termInterval{\toIntervalTerm{M}}$ such that $\omega(A) = \termProb{M}$; and if $M$ is AST then $\expVal(\toIntervalTerm{M}, A) = \expectedTermSteps{M}$.
\end{theoremRE}
\begin{proof}
	We can naturally identify traces in $\stdtrset^m$ with elements in $\intreal^m$.
	The definition of almost-surely fully contained, $\almostSub$, naturally extends to traces. 
	Fix any $\kappa \in D^*$.
	In the first step, we show that there exists a countable family of boxes $\{B_l\}_{l \in \indx}$ ($B_l \subseteq \intreal^m$) such that $\bigcup_{l \in \indx} B_l \almostSub \termTrCO{M}{\kappa}$.
	
	\textbf{First:}
	There either exists a value $\symV$ a natural number $m$ and constraint $\Delta$ (all of them unique) such that {\scriptsize$\symConf{\symM, \kappa, 0} {\emptyset} \toSym^* \symConf{\symV, \epsilon, m} {\Delta}$}
	or there exists none.
	In either case, we apply \refProp{corr}.
	In the latter case, we are done as $\termTrCO{M}{\kappa} = \emptyset$.
	In the former case, we get $\mathsf{Sat}_m(\Delta) = \termTrCO{M}{\kappa}$.
	Note that $\termTrCO{M}{\kappa} \subseteq \stdtrset^{m}$.
	Let $\Delta = \{(\symV_i \bowtie_i r_i)\}_{i \in [n]}$.
	Now by definition of $\mathsf{Sat}_m$, $\mathsf{Sat}_m(\Delta) = \{ \sigma \in \intreal^m \mid \forall i \in [n]: \symV_i[\sigma] \bowtie_i r_i \}$.
	We can write this as $\bigcap_{i \in [n]} \symV_i^{-1}(I_i)$ where $I_i$ is one of $(r_i, \infty)$, $[r_i, \infty)$, $(-\infty, r_i)$ or $(-\infty, r_i]$ depending on $\bowtie_i$. 
	Due to the CbN evaluation, each symbolic value contains each sample variable at at most one position.  
	As all of these sets can be given as a countable union of closed bounded intervals, we can apply \refLemma{symPreimage} and get a family $(B^i_k)_{k \in \indx_i}$ such that $\bigcup_{k \in \indx_i} B^i_k \almostSub \symV_i^{-1}(I_i)$.
	Now the finite intersection of countable unions of boxes is itself a countable union of boxes (refer to the proof of \refLemma{symPreimage}).
	There thus exists a family $(B_l)_{l \in \indx}$ with $\bigcup_{l \in \indx} B_l \almostSub \cap_{i \in [n]} \symV_i^{-1}(I_i) = \termTrCO{M}{\kappa}$.

	\textbf{Second:}
	So $\bigcup_{l \in \indx} B_l \almostSub \termTrCO{M}{\kappa}$.
	By \refLemma{RemDuplicates} we can assume that this family is pairwise almost disjoint. 
	Now each box $B_l \subseteq \intreal^{m}$ can naturally be seen as an interval trace within $\inttrset^{m}$.
	Let $A^{(\kappa)}$ be this set of interval traces. 
	As the boxes are pairwise almost disjoint the traces are pairwise compatible. 
	For each $\wpi \in A^{(\kappa)}$ we have $\containedTraces{\wpi} \in \termTrCO{M}{\kappa}$ so by \refLemma{ensureTerm} we get that $\wpi \in \termInterval{\toIntervalTerm{M}}$. 
	So $A^{(\kappa)} \subseteq \termInterval{\toIntervalTerm{M}}$.
	As the set of conditional oracles $D^*$ is countable we can take the union of all interval traces $A^{(\kappa)}$ for all $\kappa \in D^*$. 
	There thus exists a countable set of interval traces $A \subseteq \termInterval{\toIntervalTerm{M}}$ such that $\bigcup_{\wpi \in A} \containedTraces{\wpi} \almostSub \termTr{M}$. 
	This already implies that $\omega(A) = \mu_{\stdtrset}\big(\termTr{M}\big)$. 
	For the expected time to termination recall that for all $\tr \in \containedTraces{\wpi}$, $\numberSteps{\wpi}{\toIntervalTerm{M}} = \numberSteps{\tr}{M}$.
\end{proof}

\section{Additional Material - Section~\ref{sec:intersection}}

To state properties of the type system it is actually easiest to decompose this reduction relation.
A relation $\to_\det$, handling deterministic steps, and a relation $\to_{\myint{a, b}}$ for $\myint{a, b} \in \interval_{0, 1}$ performing probabilistic steps.
Those relations are defined in \refFig{decomposedSemantics}. 

\begin{figure*}[!t]
	\begin{tcolorbox}[colback=white, colframe=black, arc=0mm, boxrule=1pt, left=0pt, right=0pt]
		\small
		\begin{minipage}{0.22\textwidth}
			\begin{prooftree}
				\AxiomC{}
				\UnaryInfC{$(\lambda x. \calM) \calN \to_\det \calM[\calN/x] $}
			\end{prooftree}
		\end{minipage}
		\begin{minipage}{0.25\textwidth}
			\begin{prooftree}
				\AxiomC{}
				\UnaryInfC{$(\mu^\varphi_x. \calM) \calN \to_\det \calM[\calN/x, (\mu^\varphi_x. \calM)/\varphi] $}
			\end{prooftree}
		\end{minipage}
		\begin{minipage}{0.52\textwidth}
			\vspace{2.7mm}
			\begin{prooftree}
				\AxiomC{}
				\UnaryInfC{$f\big(\num{\myint{a_1, b_1}}, \cdots, \num{\myint{a_{|f|}, b_{|f|}}} \big) \to_\det \num{\hat{f}(a_1, b_1, \cdots, a_{|f|}, b_{|f|})}$}
			\end{prooftree}
		\end{minipage}
	
		\begin{minipage}{0.25\textwidth}
			\begin{prooftree}
				\AxiomC{$a \geq 0$}
				\UnaryInfC{$\score(\num{\myint{a, b}}) \to_\det \num{[a,b]}$}
			\end{prooftree}
		\end{minipage}
		\begin{minipage}{0.25\textwidth}
			\vspace{0.5mm}
			\begin{prooftree}
				\AxiomC{$b \leq 0$}
				\UnaryInfC{$\myif(\num{\myint{a, b}},\calN, \calP)\to_\det \calN$}
			\end{prooftree}
		\end{minipage}
		\begin{minipage}{0.25\textwidth}
			\vspace{0.9mm}
			\begin{prooftree}
				\AxiomC{$a > 0$}
				\UnaryInfC{$\myif(\num{\myint{a, b}}, \calN, \calP) \to_\det \calP$}
			\end{prooftree}
		\end{minipage}
		\begin{minipage}{0.24\textwidth}
			\begin{prooftree}
				\AxiomC{$\calR \to_\det \calM$}
				\UnaryInfC{$\calE[\calR] \to_\det \calE[\calM]$}
			\end{prooftree}
		\end{minipage}
		
		\tcblower
		\small
		\begin{minipage}{0.5\textwidth}
			\vspace{3.6mm}
			\begin{prooftree}
				\AxiomC{}
				\UnaryInfC{$ \sample \to_{\myint{a, b}} \num{\myint{a, b}}$}
			\end{prooftree}
		\end{minipage}
		\begin{minipage}{0.5\textwidth}
			\begin{prooftree}
				\AxiomC{$\calR \to_{\myint{a, b}} \calM$}
				\UnaryInfC{$\calE[\calR] \to_{\myint{a, b}} \calE[\calM]$}
			\end{prooftree}
		\end{minipage}
		
	\end{tcolorbox}

	\caption{Decomposed reduction into deterministic steps $\to_\det$ and probabilistic steps $\to_{\myint{a, b}}$} \label{fig:decomposedSemantics}
\end{figure*}

While our type system is designed such that the least upper bound over all derivation equals the probability of termination and thus looks very similar to the monadic system in \citet{DBLP:conf/ppdp/BreuvartL18}, we have to approach on an entirely different way. 
The system by \citeauthor{DBLP:conf/ppdp/BreuvartL18} relies on the countable nature of the execution tree and can state subject reduction by taking the weighted (finite) sum over the reduction relation. 
Due to the uncountable nature of SPCF, we cannot follow this approach. 
Instead, in our system we allow for enumeration of terminating interval traces and make use of the soundness and completeness of the interval-based semantics shown in \refSection{ibs}.
We write $[n]$ for the set $\{0, \cdots, n-1\}$, i.e., the first $n$ integers.

\begin{lemma}\label{lem:subsettyp}
	If $\vdash \calM : \calA$ and $\calB \subseteq \calA$ then $\vdash \calM : \calB$ 
\end{lemma}
\begin{proof}
	Easy induction on $\vdash \calM : \calA$.
\end{proof}

\subsection{Subject Reduction and Soundness}

\subsubsection{Subject Reduction}

We begin by showing that our system does enjoy subject reduction.
In our setting, the $\tau$ component gives the number of steps to termination.
Matching this intuition, the $\tau$ decrease by $1$ in each step.
Furthermore as each $\wpi$ is a terminating trace, each probabilistic reduction consumes the first element (c.f.~\cite{DBLP:conf/ppdp/BreuvartL18}). 

\begin{lemma}[Substitution]\label{lem:MOsubst}
	If $\Gamma; \{x_i:\sigma_i\}_{i \in [n]} \vdash \calM : \calA$ for distinct $x_i$ and for all $i \in [n]$ and $\calB \in \sigma_i$, $\Gamma \vdash \calN_i : \calB$ then $\Gamma \vdash \calM[\calN_i/x_i]_{i \in [n]} : \calA$
\end{lemma}
\begin{proof}
	An easy induction on $\calM$.
\end{proof}

\begin{lemma}[Deterministic Subject Reduction]\label{lem:detSR}
	If $\vdash \calM : \calA$, $\calA \neq \blist{}$ and $\calM$ has a deterministic redex and then $\calM \to_\det \calM'$ and $\vdash \calM' : \calA^{(\uparrow \epsilon, -1)}$. 
\end{lemma}
\begin{proof}
	Induction on $\calM \to_\det \calM'$. Case analysis on $\calM$.
	\begin{itemize}[leftmargin=*]
		\item $\calM = (\lambda x. \calN) \calP \to_\det \calN[\calP/x]$: Then the last step must have been:
		{\small\begin{prooftree}
				\AxiomC{$\{x:\sigma\} \vdash \calN : \calB$}
				\RightLabel{\rulename{abs}}
				\UnaryInfC{$\vdash \lambda x. \calN : \blist{(\sigma \to \calB, \epsilon, 0)}$}
				\AxiomC{$  \{\vdash \calP : \calC \mid \forall \calC \in \sigma \}$}
				\RightLabel{\rulename{app}}
				\BinaryInfC{$\vdash (\lambda x. \calN) \calP : \calB^{(\uparrow \epsilon, 1)} = \calA $}
		\end{prooftree}}%
		By substitution (\refLemma{MOsubst}) we can type $\vdash \calN[\calP/x] : \calB = \calA^{(\uparrow \epsilon, -1)}$ as required.
		
		\item $\calM = (\mu^\varphi_x. \calN) \calP \to_\det \calN[\calP/x, (\mu^\varphi_x. \calN)/\varphi]$: Then the last step must have been via \rulename{app} and \rulename{fix}, similar to above.
		We conclude via (\refLemma{MOsubst}). 

		\item $\calM = \myif(\num{\myint{a, b}}, \calN, \calP) \to_\det \calN$ and $b \leq 0$: Then the last step must have been:
		{\small\begin{prooftree}
				\AxiomC{}
				\UnaryInfC{$\vdash \num{\myint{a, b}} : \blist{(\myint{a, b}, \epsilon, 0)}$}
				\AxiomC{$\vdash \calN : \calB_{(\myint{a, b}, \epsilon, 0)} $}
				\RightLabel{\rulename{if}}
				\BinaryInfC{$\vdash \myif(\num{\myint{a, b}}, \calN, \calP) : \calB_{(\myint{a, b}, \epsilon, 0)}^{(\uparrow \epsilon, 1)} $}
		\end{prooftree}}%
		So $\vdash \calN : \calB_{(\myint{a, b}, \epsilon, 0)} = \calA^{(\uparrow \epsilon, -1)}$.
		
		\item $\calM = \myif(\num{\myint{a, b}}, \calN, \calP) \to_\det \calP$ and $a > 0$. Similar to the previous case.
		
		\item $\calM = \myif(\num{\myint{a, b}}, \calN, \calP)$ and $a < 0$ and $b \geq 0$. Note possible as by assumption $\calA \neq \blist{}$.
		
		\item $\calM = f(\num{\myint{a, b}},\num{[c, d]}) \to_\det \num{\hat{f}(a, b, c, d)}$. Then the last step must have been via \rulename{$f$} and \rulename{num} and $\calA = \blist{(\hat{f}(a, b, c, d), \epsilon, 1)}$ we can type 
		$$\vdash  \num{\hat{f}(a, b, c, d)} : \blist{(\hat{f}(a, b, c, d), \epsilon, 0)}$$
		via \rulename{num}.
		
		\item $\calM = \score(\myint{a, b}) \to_\det \myint{a, b}$ and $a \geq 0$. Then the last step must have been via \rulename{score} and \rulename{num} to $\calA = \blist{(\myint{a, b}, \epsilon, 1)}$ and we can type $\vdash  \num{\myint{a, b}} : \blist{(\myint{a, b}, \epsilon, 0)}$ via \rulename{num} as required.
		
		\item $\calM = \score(\myint{a, b})$ and $a < 0$. Not possible as by assumption $\calA \neq \blist{}$.
		
		\item $\calM = \calN \calP \to_\det \calN' \calP$ and $\calN \to_\det \calN'$. Then the last step must have been:
		{\small\begin{prooftree}
				\AxiomC{$\vdash \calN : \calB$}
				\AxiomC{$  \{\vdash \calP : \calD \mid \forall (\sigma \to \calC, \wpi, \tau) \in \calB, \calD \in \sigma \}$}
				\RightLabel{\rulename{app}}
				\BinaryInfC{$\vdash \calN \calP : \bigcup\limits_{(\sigma \to \calC, \wpi, \tau) \in \calB} \calC^{(\uparrow \wpi, \tau+1)}$}
		\end{prooftree}}%
		By induction we get $\vdash \calN' : \calB^{(\uparrow \epsilon, -1)}$.
		We can conclude using \rulename{app}, by choosing the same type derivations for each element in $\calB^{(\uparrow \epsilon, -1)}$ as in the original derivation.. 
		
		\item $\calM = \myif(\calN, \calP, \calQ) \to_\det \myif(\calN', \calP, \calQ)$ and $\calN \to_\det \calN'$: Then the last step must have been via \rulename{if}.
		We can use the IH on $N$ and conclude via \rulename{if} by choosing the same derivations.

		\item $\calM = \score(\calN), \calM = f(\calN, \calP), \calM = f(\num{\myint{a, b}}, \calN)$ are trivial. 
	\end{itemize}
\end{proof}

\begin{lemma}[Probabilistic Subject Reduction]\label{lem:probSR}
	If $\vdash \calM : \blist{(\alpha, \wpi, \tau)}$ and $\calM$ has a probabilistic redex then $\wpi = \myint{a, b}\wpi'$ and we have $\calM \to_{\myint{a, b}} \calM'$ and $\vdash \calM' : \blist{(\alpha, \wpi', \tau-1)}$
\end{lemma}
\begin{proof}
	Case analysis on $\calM$.
	\begin{itemize}[leftmargin=*]
		\item $\calM = \sample$. Then the last step is:
		{\small\begin{prooftree}
				\AxiomC{}
				\RightLabel{\rulename{sample}}
				\UnaryInfC{$\vdash \sample : \Blist{ (\myint{a, b}, \myint{a, b}, 1 )}$}
		\end{prooftree}}%
		for some $a, b$.
		We get $\sample \to_{\myint{a, b}} \calM' \triangleq \num{\myint{a, b}}$ and can obviously type: $\vdash \calM' : \blist{(\myint{a, b}, \epsilon, 0)}$ using \rulename{num}.
		
		\item $\calM = \calN \calP$ and $\calN$ has a probabilistic redex. The last step must have been:
		{\scriptsize\begin{prooftree}
				\AxiomC{$\vdash \calN : \Blist{(\sigma \to \blist{(\alpha, \wpi_3, \tau_3)}, \wpi_1, \tau_1)}$}
				\AxiomC{$  \{\vdash \calP : \calC \mid \forall \calC \in \sigma \}$}
				\RightLabel{\rulename{app}}
				\BinaryInfC{$\vdash \calN \calP: \blist{(\alpha, \wpi_1\wpi_3, \tau_1+\tau_3+1)}$}
		\end{prooftree}}%
		By induction we get that $\wpi_1 = \myint{a, b} \wpi_2$, $\calN \to_{\myint{a, b}} \calN'$ and $\vdash \calN : \blist{(\sigma \to \blist{(\alpha, \wpi_3, \tau_3)}, \wpi_2, \tau_1-1)}$.
		So $\wpi_1\wpi_3 = \myint{a, b} \wpi_2\wpi_3$.
		Now $\calN \calP \to_{\myint{a, b}} \calN' \calP$ and we can conclude $\vdash \calN' \calP : \blist{(\alpha, \wpi_2\wpi_3, \tau_1+\tau_3)}$ via \rulename{app}.

		\item All the other closre cases, i.e., $\calM = \myif(\calN, \calP, \calQ)$,  $\calM = f(\calN, \calP)$, $\calM = f(\num{\myint{a, b}}, \calN)$ and $\calM = \score(\calN)$ where $\calN$ has a probabilistic redex follow in the same fashion as above.

	\end{itemize}
\end{proof}

\begin{lemma}[Subject Reduction]\label{lem:monSubjectReduction}
	If $\vdash \calM : \blist{(\alpha, \wpi, \tau)}$ and $\calM$ is not a value, then either
	\begin{itemize}
		\item $\calM$ has a deterministic redex and $\calM \to_\det \calM'$ and $\vdash \calM' : \blist{(\alpha, \wpi, \tau-1)}$, or
		\item $\calM$ has a probabilistic redex then $\wpi = \myint{a, b}\wpi'$ and we have $\calM \to_{\myint{a, b}} \calM'$ and $\vdash \calM' : \blist{(\alpha, \wpi', \tau-1)}$
	\end{itemize}
\end{lemma}
\begin{proof}
	Follows from \refLemma{probSR} and \refLemma{detSR}.
\end{proof}

\begin{lemma}\label{lem:towardsSoundness}
	If $\vdash \calM : \blist{(\alpha_i, \wpi_i, \tau_i) \mid i \in [n]}$ then $\wpi_i \in \termInterval{\calM}$ and $\numberSteps{\wpi_i}{\calM} = \tau_i$ for all $i \in [n]$
\end{lemma}
\begin{proof}
	We show the easier observation that if $\vdash \calM : \blist{\alpha, \wpi, \tau}$, then $\wpi \in \termInterval{\wpi}$ and $\numberSteps{\wpi}{\calM} = \tau$.
	The result then follows by \refLemma{subsettyp} as we get $\vdash \calM : \blist{\alpha_i, \wpi_i, \tau_i}$ for every $i \in [n]$.
	
	As an easy corollary from Subject reduction (\refLemma{monSubjectReduction}) combined with the obvious properties of the decomposed semantics, we get that if $\vdash \calM : \blist{(\alpha, \wpi, \tau)}$ and $\langle \calM, \wpi \rangle \too \langle \calM', \wpi' \rangle$ we have $\vdash \calM' : \blist{(\alpha, \wpi', \tau-1)}$. 
	Call this observation \textbf{(1)}.
	
	We first show $\wpi \in \termInterval{\calM}$. 
	Let $\langle M, \wpi \rangle \triangleq \langle \calM_0, \wpi_0\rangle  \too \langle \calM_1, \wpi_1\rangle \too \langle \calM_2, \wpi_2\rangle \too \cdots$
	be the possibly infinite reduction sequence.
	From \textbf{(1)} we get $\vdash \calM_i : \blist{(\alpha, \wpi_i, \tau-i)}$. 
	The sequence can thus make \emph{at most} $\tau$-steps and is hence finite. 
	Let $\langle \calM, \wpi\rangle = \langle \calM_0, \wpi_0\rangle  \too \langle \calM_1, \wpi_1\rangle \too \langle \calM_2, \wpi_2\rangle \too \cdots \too \langle \calM_n, \wpi_n\rangle$
	be this finite, maximal sequence. 
	We assume for contraction that $\calM_n$ is not a value. 
	As  $\vdash \calM_n : \blist{(\alpha, \wpi_n, \tau-n)}$ we can use subject reduction (\refLemma{monSubjectReduction}) and get that $\langle \calM_n, \wpi_n\rangle$ can make a further step which contradicts the maximality.
	Now as $M_n$ is a value, we can inspect the typing rules and get that $\wpi_n = \epsilon$ as values can only be typed with an empty interval trace. 
	This already shows that $\wpi \in \termInterval{\calM}$. 
	
	Now by definition of the number of steps $\numberSteps{\wpi}{\calM} = n$.
	As $\calM_n$ is a value and $\vdash \calM_n : \blist{(\alpha, \wpi_n, \tau-n)}$  we get by inspection that $\tau-n = 0$, so $\tau = n = \numberSteps{\wpi}{\calM}$. 
\end{proof}

\subsubsection{Pairwise Compatibility}

We can easily see:

{Might want to include the proof}

\begin{lemma}[Pairwise Compatibility]\label{lem:pairwiseComp}
	If $\vdash \calM :  \blist{(\alpha_i, \wpi_i, \tau_i) \mid i \in [n]}$ then $\{\wpi_i\}_i$ are pairwise compatible.
\end{lemma}

\subsubsection{Soundness}

\begin{proposition}[Soundness]\label{prop:MonSound}
	For every interval term $\calM$ and $M \triangleleft \calM$
	\begin{tasks}[style=itemize](2)
		\task {$\bigvee\limits_{\vdash \calM : \calA} \omega(\calA) \leq \termProb{M}$}
		\task {$\bigvee\limits_{\vdash \calM : \calA} \expVal(\calA) \leq \expectedTermSteps{M}$}
	\end{tasks}
\end{proposition}
\begin{proof}
	Assume $\vdash \calM : \calA$ and $\calA = \blist{(\alpha_i, \wpi_i, \tau_i) \mid i \in [n]}$.
	By \refLemma{towardsSoundness} each $\wpi_i \in \termInterval{\calM}$.
	Furthermore, by \refLemma{pairwiseComp} the interval traces are pairwise compatible. 
	By the Soundness of the interval-based semantics (\refTheo{intSound}) we, therefore, conclude that
	$$\omega(\calA) \leq \termProb{M}$$
	For the second claim we can again use \refTheo{intSound} and the fact that $\tau_i = \numberSteps{\wpi_i}{\calM}$ (shown in \refLemma{towardsSoundness}) and get 
	$$\expVal(\calA) \leq \expectedTermSteps{M}$$
	As this holds for all $\vdash \calM : \calA$ it also holds for the least upper bound.
\end{proof}

\subsection{Subject Expansion and Completeness}

\subsubsection{Subject Expansion}

\begin{lemma}[Reverse Substitution]\label{lem:monRevSubst}
	If $\vdash \calM[\calN_i/x_i]_{i \in [n]} : \calA$ for distinct $x_i$ then there exist a $\{a_i\}_{i \in [n]}$, s.t., $\{x_i:a_i\}_{i \in [n]} \vdash \calM : \calA$ and for all $i \in [n]$ and $\calB \in a_i$, $\vdash \calN_i : \calB$
\end{lemma}
\begin{proof}
	Standard. By induction on $\calM$.
\end{proof}

\begin{lemma}[Deterministic Subject Expansion]\label{lem:detSE}
	If $\vdash \calM : \calA$ and $\calM' \to_\det \calM$ then $\vdash \calM' : \calA^{(\uparrow \epsilon, 1)}$
\end{lemma}
\begin{proof}
	We assume w.l.o.g.~that $\calA \neq \emptyset$.
	Induction on $\vdash \calM : \calA$. 
	Case analysis on $\calM'$.
	\begin{itemize}[leftmargin=*]
		\item $\calM' = (\lambda x. \calN) \calP \to_\det \calN[\calP/x]$: So $\vdash \calN[\calP/x] :\calA$. By \refLemma{monRevSubst} we get an $\sigma$, s.t., $\{x:\sigma\} \vdash \calN : \calA$ and for all $\calB \in \sigma$, $\vdash \calP : \calB$.
		We can conclude $\vdash \lambda x. \calN : \blist{(\sigma \to \calA, \epsilon, 1)}$ using \rulename{abs} and can the derive $\vdash (\lambda x. \calN) \calP : \calA^{(\uparrow \epsilon, 1)}$ via \rulename{app}. 
		
		\item $\calM' = (\mu^\varphi_x. \calN) \calP \to_\det \calN[\calP/x, (\mu^\varphi_x. \calN)/\varphi]$. \\
		So $\vdash \calN[\calP/x, (\mu^\varphi_x. \calN)/ \varphi] : \calA$. By \refLemma{monRevSubst} we get $a_x, a_\varphi$ such that $\{x:a_x, \varphi:a_\varphi\} \vdash \calN : \calA$ and for all $\calB \in a_x$, $\vdash \calP : \calB$ and all $\calB \in a_\varphi$, $\vdash \mu^\varphi_x. \calN : \calB$.
		We can thus type $\vdash \mu^\varphi_x. \calN : \Blist{(a_x \to \calA, \epsilon, 0)}$ using \rulename{fix} and conclude $\vdash (\mu^\varphi_x. \calN) \calP : \calA^{(\uparrow \epsilon, 1)}$ via \rulename{app}.
		
		\item $\calM' = \myif(\num{\myint{a, b}}, \calN, \calP) \to_\det \calN$ and $b \leq 0$ and $\vdash \calN : \calA$. 
		We can type $\vdash \num{\myint{a, b}} : \blist{(\myint{a, b}, \epsilon, 0)}$ via \rulename{num} and as $b \leq 0$ we can derive $\vdash \myif(\num{\myint{a, b}}, \calN, \calP) : \calA^{(\uparrow \epsilon, 1)}$ using \rulename{if}.
		
		\item $\calM' = \myif(\num{a, b}, \calN, \calP) \to_\det \calP$ and $a > 0$. Similar as the case above. 
		
		\item $\calM ' = f(\num{\myint{a, b}}, \num{\myint{c, d}}) \to_\det \num{\hat{f}(a, b, c, d)}$: So $\vdash \num{\hat{f}(a, b, c, d)} : \calA$, so we get that $\calA = \blist{(\hat{f}(a, b, c, d), \epsilon, 0)}$ as only \rulename{num} is applicable. 
		We can type $\vdash \num{\myint{a, b}} : \blist{(\myint{a, b}, \epsilon, 0)}$ via \rulename{num} and similar for $\num{\myint{c, d}}$ and can conclude using \rulename{$f_2$}.
		
		\item $\calM' = \calN' \calP \to_\det \calN \calP$ and $\calN' \to_\det \calN$.
		As $\vdash \calN \calP : \calA$ we get that the last step must have been:
		{\small\begin{prooftree}
				\AxiomC{$\vdash \calN : \calB$}
				\AxiomC{$  \{\vdash \calP : \calD \mid \forall (\sigma \to \calC, \wpi, \tau) \in \calB, \calD \in \sigma \}$}
				\RightLabel{\rulename{app}}
				\BinaryInfC{$\vdash \calN \calP : \bigcup\limits_{(\sigma \to \calC, \wpi, \tau) \in \calB} \calC^{(\uparrow \wpi, \tau+1)} = \calA $}
		\end{prooftree}}%
		By IH we get $\vdash \calN' : \calB^{(\uparrow \epsilon, 1)}$ and conclude using \rulename{app} by choosing the same derivations as in the original derivation. 
		
		\item $\calM' = \myif(\calN', \calP, \calQ) \to_\det \myif(\calN, \calP, \calQ)$ and $\calN' \to_\det \calN$. As $\vdash \myif(\calN, \calP, \calQ)$ we get that the last step must have been via \rulename{if}.
		As in the previous case we can apply induction choose the same derivations for $\calP$ and $\calQ$ and conclude back via \rulename{if}.

		\item $\calM' = \score(\calN'), \calM' = f(\calN', \calP) , \calM' = f(\num{\myint{a, b}}, \calN')$. Trivial.

	\end{itemize}
\end{proof}

\begin{lemma}[Probabilistic Subject Expansion]\label{lem:probSE}
	If $\vdash \calM_i : \calA_i$ and $\calM \to_{\myint{a_i, b_i}} \calM_i$ where $\{\myint{a_i, b_i}\}_i$ are almost disjoint then 
	$$\vdash \calM : \bigcup_i \calA_i^{(\uparrow\myint{a_i, b_i}, 1)}$$
\end{lemma}
\begin{proof}
	We can assume that $\calA_i \neq \blist{}$ as this case is trivial. 
	By induction on $\calM$.
	\begin{itemize}[leftmargin=*]
		\item $\calM = \sample$: Then $\calM_i = \num{\myint{a_i, b_i}}$ and as $\vdash \calM_i : \calA_i$ we get that $\calA_i = \blist{(\myint{a_i, b_i}, \epsilon, 0)}$ as the last step must be via \rulename{num}.
		As $\myint{a_i, b_i}$ are almost disjoint we can use the \rulename{sample}-rule to type \sample as required. 
		
		\item $\calM = \calN\calP$ and $\calN$ does a reduction step, i.e.,  $\calN \to_{\myint{a_i, b_i}} \calN_i$ and $\calM_i = \calN_i\calP$. As $\vdash \calN_i\calP : \calA_i$ we get that the last step must have been:
		{\small\begin{prooftree}
				\AxiomC{$\vdash \calN_i : \calB_i$}
				\AxiomC{$\{\vdash \calP : \calD \mid \forall (\sigma \to \calC, \wpi, \tau) \in \calB_i, \calD \in \sigma \}$}
				\RightLabel{\rulename{app}}
				\BinaryInfC{$\vdash \calN_i \calP : \bigcup\limits_{(\sigma \to \calC, \wpi, \tau) \in \calB_i} \calC^{( \uparrow \wpi, \tau+1)} = \calA_i$}
		\end{prooftree}}%
		By induction we get $\vdash \calN : \bigcup_i \calB_i^{(\uparrow \myint{a_i, b_i}, 1)}$. Define $\calB \triangleq \bigcup_i (\calB_i)^{(\uparrow \myint{a_i, b_i}, 1)}$.
		We get $\{\vdash \calP : \calD \mid \forall (\sigma \to \calC, \wpi, \tau) \in \calB, \calD \in c \}$ as $\calB$ is just the concatenation of all $\calB_i$, i.e., every type in $\calB$ is in at least on $\calB_i$.
		By using \rulename{app} we can thus type $\vdash \calN \calP : \bigcup_{(\sigma \to \calC, \wpi, \tau) \in \calB} \calC^{(\uparrow \wpi, \tau+1)}$.
		{\small\begin{align*}
			\textstyle\bigcup\limits_{(\sigma \to \calC, \wpi, \tau) \in \calB} \calC^{(\uparrow \wpi, \tau+1)} &= \bigcup\limits\limits_i \bigcup\limits_{(\sigma \to \calC, \wpi, \tau) \in \calB_i} \calC^{\uparrow(\myint{a_i, b_i}\wpi, \tau+1+1)}  \\
			&= \bigcup\limits_i \calA_i^{(\uparrow\myint{a_i, b_i}, 1)}
		\end{align*}}%
		as required.
		
		\item $\calM = \myif(\calN, \calP, \calQ)$ and $\calN$ does a reduction step, i.e., $\calN \to_{\myint{a_i, b_i}} \calN_i$ and $\calM_i = \myif(\calN_i, \calP, \calQ)$. The last step in each derivation must have been via \rulename{if} so 
		$\vdash \calN_i : \calB_i$, 
		$$\{\vdash \calP : \calC_{(\myint{a, b}, \wpi, \tau), i} \mid (\myint{a, b}, \wpi, \tau) \in \calB_i, b \leq 0 \}$$
		and 
		$$\{\vdash \calQ : \calD_{(\myint{a, b}, \wpi, \tau), i} \mid (\myint{a, b}, \wpi, \tau) \in \calB_i, a > 0 \}$$
		and 
		{\scriptsize$$\calA_i = \bigcup\limits_{(\myint{a, b}, \wpi, \tau) \in \calB_i \mid b \leq 0} \calC_{(\myint{a, b}, \wpi, \tau), i}^{(\uparrow \wpi, \tau)} \cup \bigcup\limits_{(\myint{a, b}, \wpi, \tau) \in \calB_i \mid a > 0} \calD_{(\myint{a, b}, \wpi, \tau)}^{(\uparrow \wpi, \tau), i}$$}%
		By induction we get $\vdash \calN : \bigcup_i \calB_i^{(\uparrow \myint{a_i, b_i}, 1)} \triangleq \calB$.
		Now each element $(\myint{a, b}, \wpi, \tau) \in \calB$ stems from exactly one category $i$ (from one $\calB_i$).
		So elements in $\calB$ can be seen as having the from $(\myint{a, b}, \wpi, \tau), i \in \calB$. 
		(This is just needed to take care of the indices).
		Using \rulename{if} we can type
		{\scriptsize\begin{align*}
				\textstyle\vdash &\textstyle\myif(\calN, \calP, \calQ) : \bigcup\limits_{(\myint{a, b}, \wpi, \tau), i \in \calB \mid b \leq 0} \calC_{(\myint{a, b}, \wpi, \tau), i}^{(\uparrow \wpi, \tau)} \cup \bigcup\limits_{(\myint{a, b}, \wpi, \tau), i \in \calB \mid a > 0} \calD_{(\myint{a, b}, \wpi, \tau), i}^{(\uparrow \wpi, \tau)} \\
				&\textstyle= \bigcup\limits_i \Big(\bigcup\limits_{(\myint{a, b}, \wpi, \tau) \in \calB_i \mid b \geq 0} \calC_{(\myint{a, b}, \wpi, \tau), i}^{(\uparrow \wpi, \tau)} \cup \bigcup\limits_{(\myint{a, b}, \wpi, \tau) \in \calB_i \mid a < 0} \calD_{(\myint{a, b}, \wpi, \tau), i}^{(\uparrow \wpi, \tau)}\Big) \\
				&= \bigcup\limits_i \calA_i^{(\uparrow\myint{a_i, b_i}, 1)}
		\end{align*}}%
		as required.

		\item $\calM = f(\calN, \calP)$ and $\calN$ does a reduction step, i.e., $\calN \to_{\myint{a_i, b_i}} \calN_i$ and $\calM_i = f(\calN_i, \calP)$. As $\vdash f(\calN_i, \calP) : \calA_i$ we get that the last step must have been via \rulename{$f_2$}, i.e., $\vdash \calN_i : \calB_i$, $\{\vdash \calP : \calC_{(\myint{a, b}, \wpi, \tau), i} \mid (\myint{a, b}, \wpi, \tau) \in \calB_i \}$ and 
		{\scriptsize$$\calA_i = \bigcup\limits_{(\myint{a, b}, \wpi, \tau) \in \calB_i} \quad \bigcup\limits_{(\myint{c, d}, \wpi', \tau') \in \calC_{(\myint{a, b}, \wpi, \tau), i} } \blist{(\hat{f}(a, b, c, d), \wpi\wpi', \tau+\tau'+1)}$$}%
		By induction we get  $\vdash \calN : \bigcup_i \calB_i^{(\uparrow \myint{a_i, b_i}, 1)}  \triangleq \calB$.
		We can now conclude using \rulename{$f_2$} as in the previous cases.
		
		\item $\calM = f(\num{\myint{a, b}}, \calN)$ and $\calN$ does a reduction step, i.e., $\calN \to_{\myint{a_i, b_i}} \calN_i$ and $\calM_i = f(\num{\myint{a, b}}, \calN_i)$.
		The last steps must thus have been:
		{\small\begin{prooftree}
				\AxiomC{}
				\RightLabel{\rulename{num}}
				\UnaryInfC{$\vdash \num{\myint{a, b}} : \blist{(\myint{a, b}, \epsilon, 0)}$}
				\AxiomC{$\vdash \calN_i : \calB_{(\myint{a, b}, \epsilon, 0), i}$}
				\RightLabel{\rulename{$f_2$}}
				\BinaryInfC{$\vdash f(\num{\myint{a, b}}, \calN_i) :  \bigcup\limits_{(\myint{c, d}, \wpi, \tau) \in \calB_{(\myint{a, b}, \epsilon, 0), i} } \blist{(\hat{f}(a, b, c, d), \wpi, \tau+1)}$}
		\end{prooftree}}%
		By induction we get  $\vdash \calN : \bigcup_i \calB_i^{(\uparrow \myint{a_i, b_i}, 1)}  \triangleq \calB$.
		We can trivially conclude via \rulename{$f_2$} and \rulename{num}.
		
		\item $\calM = \score(\calN)$ and $\calN$ does a reduction step, i.e., $\calN \to_{\myint{a_i, b_i}} \calN_i$ and $\calM_i = \score(\calN_i)$.
		The last step must have been via \rulename{score}. We can apply induction and trivially conclude via \rulename{score}.
	\end{itemize}
\end{proof}

\begin{lemma}[Subject Expansion]\label{lem:monSubjectExpansion}
	It holds that:
	\begin{itemize}
		\item If $\vdash \calM : \calA$ and $\calN \to_\det \calM$ then $\vdash \calN : \calA^{(\uparrow \epsilon, 1)}$
		\item If $\vdash \calM_i : \calA_i$ and $\calN \to_{\myint{a_i, b_i}} \calM_i$ where $\{\myint{a_i, b_i}\}_i$ are almost disjoint then 
		$$\textstyle\vdash \calN : \bigcup_i \calA_i^{(\uparrow\myint{a_i, b_i}, 1)}$$
	\end{itemize}
\end{lemma}
\begin{proof}
	Follows from \refLemma{detSE} and \refLemma{probSE}.
\end{proof}

\subsubsection{Completeness}

\subsubsection*{Na\"ive Attempt on Completeness:}
We have seen in the soundness proof that if  $\vdash \calM : \blist{(\alpha_i, \wpi_i, \tau_i) \mid i \in [n]}$ each $\wpi_i$ is a terminating trace.
For completeness we would like to reverse that process and show that any pairwise compatible set of traces can be achieved via a type derivations: 
That is, if $\{\wpi_i \mid i \in [n]\} \subseteq \termInterval{\calM}$ are pairwise compatible then 
$$\vdash \calM : \blist{(\alpha_i, \wpi_i, \numberSteps{\wpi_i}{\calM}) \mid i \in [n]}$$
for some types $\{\alpha_i\}_{i \in [n]}$.
This would immediately give us a completeness theorem as the interval-based semantics is itself complete.
However, the above does \textbf{not} hold.

\begin{example}\label{ex:strongComp}
	As an example consider the following simple term: $\calM \triangleq \toIntervalTerm{\myif \big(\sample - \tfrac{1}{2}, \sample, \num{0}\big)}$
	Then the two interval traces $\wpi_1 \triangleq \myint{0, \tfrac{1}{2}}\myint{0, \tfrac{1}{2}}$,  $\wpi_2 \triangleq \myint{0, \tfrac{1}{3}}\myint{\tfrac{1}{2}, 1}$ are clearly compatible but cannot be typed with the above system. 
	The interested reader is advised to try find a typing derivation. 
\end{example}

\subsubsection*{Strong Pairwise Compatibility}

To show completeness, we need to introduce the new concept of \emph{strong compatibility}. 
We call $\wpi_1, \wp_2$ \emph{strongly compatible} if $\wpi_1 \leftrightarrowtriangle \wpi_2$ is derivable by the following rules

\begin{minipage}{0.15\textwidth}
	\begin{prooftree}
		\AxiomC{}
		\UnaryInfC{$\epsilon \leftrightarrowtriangle \myint{a, b}\wpi$}
	\end{prooftree}
\end{minipage}%
\begin{minipage}{0.349\textwidth}
	\begin{prooftree}
		\AxiomC{$\myint{a, b}, \myint{c, d} \text{ are almost disjoint}$}
		\UnaryInfC{$\myint{a, b}\wpi_1 \leftrightarrowtriangle \myint{c, d}\wpi_2$}
	\end{prooftree}
\end{minipage}

\begin{minipage}{0.15\textwidth}
	\begin{prooftree}
		\AxiomC{}
		\UnaryInfC{$\myint{a, b}\wpi \leftrightarrowtriangle \epsilon$}
	\end{prooftree}
\end{minipage}%
\begin{minipage}{0.34\textwidth}
	\begin{prooftree}
		\AxiomC{$\wpi_1 \leftrightarrowtriangle \wpi_2$}
		\UnaryInfC{$\myint{a, b}\wpi_1 \leftrightarrowtriangle \myint{a, b}\wpi_2$}
	\end{prooftree}
\end{minipage}

Strongly compatible traces are either pairwise almost disjoint in the first position or agree on the first position and the remainder is also strongly compatible. 
If two traces are strongly compatible they can thus share a common, identical prefix but must be pairwise almost disjoint at the first position where they differ. 
Clearly every strongly compatible pair of interval traces is also compatible but not the other way around. 
As an example the two traces in \refExample{strongComp} are compatible but not strongly compatible. 
We can show the following:

\begin{lemma}\label{lem:spc}
	If $\{\wpi_i \mid i \in [n]\} \subseteq \inttrset$ then there exists interval traces $\{\wpi_j' \mid j \in [m]\} \subseteq \inttrset$ that are pairwise \emph{strongly} compatible with $\bigcup_{i\in[n]} \containedTraces{\wpi_i} = \bigcup_{j \in [m]} \containedTraces{\wpi_j'}$ and for each $j \in [m]$,  $\containedTraces{\wpi_j'} \subseteq \containedTraces{\wpi_i}$ for some $i \in [n]$. 
\end{lemma}
\begin{proof}
	We can give a constructive proof:
	We first analyse $\{\wpi_i(0)\}_{i \in [n]}$, i.e., the interval at the first position.
	Clearly there exists intervals $\{\myint{a_k, b_k}\}_{k \in \mathcal{K}}$ for a finite $\mathcal{K}$ that are all pairwise almost disjoint such that for each $i$ there is a set $\mathcal{K}_i \subseteq \mathcal{K}$ with $\wpi_i(0) = \bigcup_{k \in \mathcal{K}_i} \myint{a_k, b_k}$.
	This holds as we can always partition at overlapping position.
	We can thus replace every $\wpi_i$ with $|\mathcal{K}_i|$ many interval traces by replacing the first interval with the intervals in $[a_k, b_k]$ from $k \in \mathcal{K}_i$. 
	The resulting set of standard traces agrees with the one we started from.
	The first position of those traces are either pairwise identical or almost disjoint as required by the definition of strong compatibility. 
	For all traces that are identical on the first position we can proceed inductively.
\end{proof}

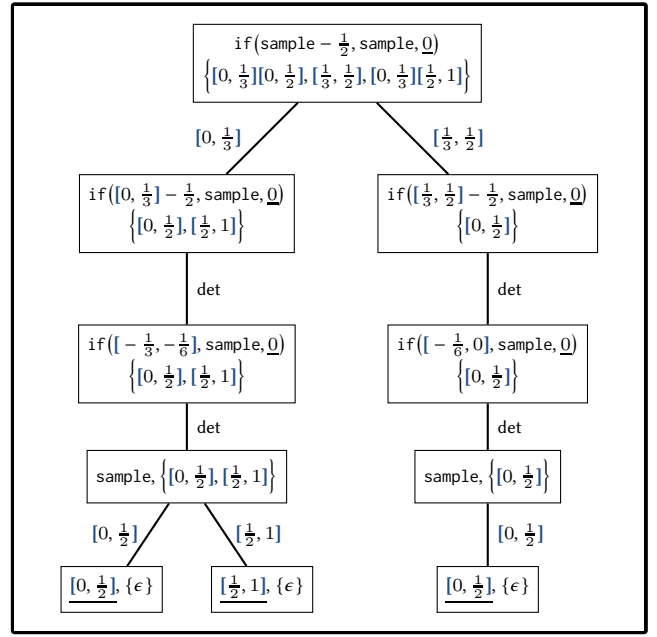
\begin{figure}	
	\begin{tcolorbox}[colback=white, colframe=black, arc=0mm, boxsep=1pt]
		\begin{center}
			\begin{tikzpicture}[scale=1]
				\node[rectangle,draw=black, align=center] at (-1.5,0) (n1) {\scriptsize$\myif \big(\sample - \tfrac{1}{2}, \sample, \num{0}\big)$\\\scriptsize $\Big\{\myint{0, \tfrac{1}{3}} \myint{0, \tfrac{1}{2}}, \myint{\tfrac{1}{3}, \tfrac{1}{2}}, \myint{0, \tfrac{1}{3}} \myint{\tfrac{1}{2}, 1}\Big\}$};

				\node[rectangle,draw=black, align=center] at (-3.5, -2) (n2) {\scriptsize$\myif \big(\myint{0, \tfrac{1}{3}} - \tfrac{1}{2}, \sample, \num{0}\big)$\\\scriptsize$\Big\{\myint{0, \tfrac{1}{2}}, \myint{\tfrac{1}{2}, 1}\Big\}$};
				
				\node[rectangle,draw=black, align=center] at (0.5, -2) (n3) {\scriptsize$\myif \big(\myint{\tfrac{1}{3}, \tfrac{1}{2}} - \tfrac{1}{2}, \sample, \num{0}\big)$\\\scriptsize$\Big\{\myint{0, \tfrac{1}{2}}\Big\}$};

				\node[rectangle,draw=black, align=center] at (-3.5, -4) (n4) {\scriptsize$\myif \big(\myint{-\tfrac{1}{3}, -\tfrac{1}{6}}, \sample, \num{0}\big)$\\\scriptsize$\Big\{\myint{0, \tfrac{1}{2}}, \myint{\tfrac{1}{2}, 1}\Big\}$};
				
				\node[rectangle,draw=black, align=center] at (0.5, -4) (n5) {\scriptsize$\myif \big(\myint{-\tfrac{1}{6}, 0}, \sample, \num{0}\big)$\\\scriptsize$\Big\{\myint{0, \tfrac{1}{2}}\Big\}$};

				\node[rectangle,draw=black, align=center] at (-3.5, -5.5) (n6) {\scriptsize$\sample, \Big\{\myint{0, \tfrac{1}{2}}, \myint{\tfrac{1}{2}, 1}\Big\}$};
				
				\node[rectangle,draw=black, align=center] at (0.5, -5.5) (n7) {\scriptsize$\sample, \Big\{\myint{0, \tfrac{1}{2}}\Big\}$};

				\node[rectangle,draw=black] at (-4.5, -7) (n8) {\scriptsize$\num{\myint{0, \tfrac{1}{2}}}, \{\epsilon\}$};
				
				\node[rectangle,draw=black] at (-2.5, -7) (n9) {\scriptsize$\num{\myint{\tfrac{1}{2}, 1}}, \{\epsilon\}$};
				
				\node[rectangle,draw=black] at (0.5, -7) (n10) {\scriptsize$\num{\myint{0, \tfrac{1}{2}}}, \{\epsilon\}$};

				\draw[-, thick] (n1) -- node[left,xshift=-4pt] {\scriptsize$\myint{0, \tfrac{1}{3}}$} (n2);
				
				\draw[-, thick] (n1) -- node[right,xshift=4pt] {\scriptsize$\myint{\tfrac{1}{3}, \tfrac{1}{2}}$} (n3);

				\draw[-, thick] (n2) -- node[right, pos=.5] {\scriptsize$\det$} (n4);
				
				\draw[-, thick] (n3) -- node[right, pos=.5] {\scriptsize$\det$} (n5);

				\draw[-, thick] (n4) -- node[right, pos=.5] {\scriptsize$\det$} (n6);
				
				\draw[-, thick] (n5) -- node[right, pos=.5] {\scriptsize$\det$} (n7);

				\draw[-, thick] (n6) -- node[left] {\scriptsize$\myint{0, \tfrac{1}{2}}$} (n8);
				\draw[-, thick] (n6) -- node[right] {\scriptsize$\myint{\tfrac{1}{2}, 1}$} (n9);
				
				\draw[-, thick] (n7) -- node[right] {\scriptsize$\myint{0, \tfrac{1}{2}}$} (n10);
				
			\end{tikzpicture}
		\end{center}
	\end{tcolorbox}

	\caption{Example reduction for the term from \refExample{strongComp} on a set of pairwise strongly compatible traces. Probabilistic and deterministic reduction steps are arranged as a tree.   }\label{fig:treeEx}
\end{figure}

\noindent The two traces in \refExample{strongComp} are not strongly compatible but can be replaced by the $3$ traces $\wpi'_1 \triangleq \myint{0, \tfrac{1}{3}} \myint{0, \tfrac{1}{2}}$, $\wpi'_2 \triangleq \myint{\tfrac{1}{3}, \tfrac{1}{2}} \myint{0, \tfrac{1}{2}}$ and $\wpi'_3 \triangleq \myint{0, \tfrac{1}{3}} \myint{\tfrac{1}{2}, 1}$ that denote the same set of standard traces but are pairwise strongly compatible.

\subsubsection*{Subject Expansion For Strongly Compatible Traces}

The crucial observation is that in the the statement of subject expansion (\refLemma{monSubjectExpansion}), the intervals in a probabilistic step should be pairwise almost disjoint. 
As we have seen in the example above pairwise compatible traces must not necessarily be almost disjoint in the first position.
But pairwise strongly compatible traces are: the first position is either almost disjoint or identical.
To make use of this idea we represent the reduction of a term given a set of interval traces as a tree. 
Nodes in the tree are of the from $(\calM, A)$ we $\calM$ is an interval term and $\emptyset \neq A \subseteq \termInterval{\calM}$ a set of strongly compatible interval traces.
The successors of a node are given by a relation $\leadsto$ where each transition is either labelled by $\det$, to represents a deterministic reduction or by an interval $\myint{a, b} \in \interval_{0, 1}$:

\begin{prooftree}
	\AxiomC{$\calM \to_\det \calN$}
	\UnaryInfC{$(\calM, A) \leadsto_\det (\calN, A)$}
\end{prooftree}

\begin{prooftree}
	\AxiomC{$\calM \to_{\myint{a, b}} \calN$}
	\AxiomC{$B = \{ \wpi \mid \myint{a, b}\wpi \in A \} \neq \emptyset$}
	\BinaryInfC{$(\calM, A) \leadsto_{\myint{a, b}} (\calN, B)$}
\end{prooftree}

\noindent If we again consider the example term fro \refExample{strongComp} and the pairwise strongly compatible traces $\wpi'_1, \wpi'_2, \wpi'_3$ from before we get the tree depicted in \refFig{treeEx}.
Every $\det$ step corresponds to a deterministic reduction. For every probabilistic reduction the set of interval traces is stripped by its first position.
As the set of traces is strongly compatible, the outgoing edges of every node are labelled by almost disjoint intervals. 

\begin{proposition}[Completeness]\label{prop:MonComp}
	If $\{\wpi_i \mid i \in [n]\} \subseteq \termInterval{\calM}$ are pairwise \textbf{strongly} compatible then 
	$$\vdash \calM : \blist{(\alpha_i, \wpi_i, \numberSteps{\wpi_i}{\calM}) \mid i \in [n]}$$
	for some types $\{\alpha_i \}_{i \in [n]}$
\end{proposition}
\begin{proof}
	We first make the following easy observation that follows immediately by the definition of strong compatibility:
	If $A \subseteq \termInterval{\calM}$ is pairwise strongly compatible and $(\calM, A) \leadsto^* (\calN, B)$ and $(\calN, B) \leadsto_{\myint{a_i, b_i}} (\calN_i, B_i)$ for $i \in [n]$ then $\myint{a_i, b_i}$ are almost disjoint. Call this observation \textbf{(1)}.
	For our proof we consider the tree that is generated by $(\calM, \{\wpi_i \mid i \in [n]\})$.
	Note that this tree is finite.
	We claim that for every node $(\calN, A)$ where $A = \{\dot{\wpi}_i \mid i \in [k]\}$ in this tree we can type $\vdash \calN : \blist{(\alpha_i, \dot{\wpi}_i, \numberSteps{\dot{\wpi}_i}{\calN}) \mid i \in [k]}$.
	We show this inductively by traversing the tree from the leafs up. 
	Formally, we do induction on the shortest path to a leaf.
	In the base case, the node in question is a leaf:
	as by assumption each $\wpi_i \in \termInterval{\calM}$ we get that each leaf of this tree has the from $(\calV, \{\epsilon\})$ for some closed value $\calV$.
	It is easy to check that for every value we can type $\vdash \calV : \blist{(\alpha, \epsilon, 0)}$ for some $\alpha$, by either using \rulename{num} (in case of a numeral) or \rulename{abs} or \rulename{fix} followed by \rulename{$\blist{}$} (in case of $\lambda$-or $\mu$-abstraction).
	Now consider the case where $(\calN, A)$ is a inner node.
	There are again two cases:
	\begin{itemize}[leftmargin=*]
		\item $(\calN, A) \leadsto_\det (\calP, A)$, so $\calN \to_\det \calP$. Write $A = \{\dot{\wpi}_i \mid i \in [k]\}$. By induction we can type $\vdash \calP : \blist{(\alpha_i, \dot{\wpi}_i, \numberSteps{\dot{\wpi}_i}{\calP}) \mid i \in [k]}$.
		Now by Subject Expansion (\refLemma{monSubjectExpansion}) we can type $\vdash \calN : \blist{(\alpha_i, \dot{\wpi}_i, \numberSteps{\dot{\wpi}_i}{\calP} + 1) \mid i \in [k]}$ as required,
		
		\item In the other case, $(\calN, A) \leadsto_{\myint{a_i, b_i}} (\calP_i, B_i)$ for $i \in [m]$. 
		Lets write $B_i = \{ \dot{\wpi}_i^j \mid j \in [k_i]\}$. 
		We have $A = \bigcup_{i \in [m]} \{ \myint{a_i, b_i}\dot{\wpi}_i^j \mid j \in [k_i]\}$.
		By induction we $\vdash \calP_i : \blist{(\alpha^i_j, \dot{\wpi}^i_j, \numberSteps{\dot{\wpi}^i_j}{\calP}) \mid j \in [k_i]} \triangleq \calA_i$.
		By \textbf{(1)} we get that the $\myint{a_i, b_i}$ are pairwise almost disjoint.
		By \refLemma{monSubjectExpansion} we can thus type  $\vdash \calN : \bigcup_i \calA_i^{(\uparrow\myint{a_i, b_i}, 1)}$ as required.
	\end{itemize}
\end{proof}

\subsection{Soundness and Completeness}

We can finally combine everything for a proof of \refTheo{soundCOmInt}.\\

\begin{theoremRE}{\ref{theo:soundCOmInt}}
	For every term $M \in \Lambda_0$,
	\begin{enumerate}
		\item $\bigvee\limits_{\vdash \toIntervalTerm{M} : \calA} \omega(\calA) = \termProb{M}$, and
		\item If $M$ is AST, $\bigvee\limits_{\vdash \toIntervalTerm{M} : \calA} \expVal(\calA) = \expectedTermSteps{M}$
	\end{enumerate}
\end{theoremRE}
\begin{proof}
	We first show the first part:\\
	We already showed $\bigvee_{\vdash \toIntervalTerm{M} : \calA} \omega(\calA) \leq \termProb{M}$ in \refProp{MonSound} as $M \triangleleft \toIntervalTerm{M}$.
	It remains to show that they are actually equal. 
	Let $\epsilon > 0$. We show that there exist a $\vdash \toIntervalTerm{M} : \calA$ such that $\omega(\calA) \geq \termProb{M} - \epsilon$.
	Using the completeness of the interval-based semantics (\refTheo{intComp}), we get a \emph{finite} set of pairwise compatible interval traces $\{\wpi'_j \mid j \in [n]\} \subseteq \termInterval{\toIntervalTerm{M}}$ such that $\sum_{j \in [n]} \omega(\wpi'_j) \geq \termProb{M} - \epsilon$.
	Now from \refLemma{spc} there exists a finite set of interval traces with the same weight that is furthermore pairwise strongly compatible. 
	Let $\{\wpi_i \mid i \in [m]\}$ be this set. 
	Note that $\{\wpi_i \mid i \in [m]\} \subseteq \termInterval{\toIntervalTerm{M}}$ as by\refLemma{spc} for every $i \in [m]$, $\containedTraces{\wpi_i} \subseteq \containedTraces{\wpi'_j}$ for some $j \in [n]$.
	By \refProp{MonComp} we get that  $\vdash \toIntervalTerm{M} : \blist{(\alpha_i, \wpi_i, \numberSteps{\wpi_i}{M}) \mid i \in [m]} \triangleq \calA$ for some types $\alpha_i$.
	Now obviously $\omega(\calA) = \sum_i \omega(\wpi'_i) \geq \termProb{M} - \epsilon$, so we are done as we can let $\epsilon$ tend to $0$.
	
	For the second part we can proceed as before by using the second part of \refTheo{intComp}. 
\end{proof}

\section{Additional Material - Section~\ref{sec:nonaffine}}

\begin{theoremRE}{\ref{theo:conditionForAST}}
	A finite {step distribution} $s$ is AST if and only if all of the following hold
	\begin{tasks}(3)
		\task \label{item:linear1}$\sum\limits_{i \in \mathbb{Z}} s(i) = 1$
		\task \label{item:linear2}$s \neq \delta_0$
		\task \label{item:linear3}$\sum\limits_{i \in \mathbb{Z}} i \cdot s(i) \leq 0$
	\end{tasks}
\end{theoremRE}
\begin{proof}
	$\boxed{\Leftarrow}$: We begin with the (arguably more interesting direction) that the three conditions together imply AST.
	We first note that due to condition \ref{item:linear1} the error state, $\bot$, is never reachable. We can thus concentrate on paths consisting of natural numbers and can neglect the possibly of moving to the error state. 
	Instead of considering the random walk on the half line we, we consider the more general wok on the integers, i.e., we remove the truncation at $0$.
	That is the Markov chain $\mathfrak{M} = (\intnum, \mathfrak{P})$ where the transition matrix $\mathfrak{P}$ is defined by $\mathfrak{P}(x, y) = s(y-x)$.
	It is easy to see that $s$ is AST if and only if $\mathfrak{M}$ eventually visits the non-positive numbers a.s.

	We then begin by checking the third condition (\ref{item:linear3}) for equality of strict inequality:

	\begin{itemize}[leftmargin=*]
		\item In the case of strict inequality, we have $\sum_{i \in \mathbb{Z}} i \cdot s(i) < 0$:
		
		Fix any starting state $m$ as in the definition of AST.
		We define integer valued random variables $X_0, X_1, \cdots$ by $X_i = m + \sum_{k=1}^{i} Y_i$ where $Y_i$ are independent random variables that are distributed according to $s$.
		It is easy to see that by the construction of the Markov chain $\mathfrak{M}$ we have $\mathfrak{P}^n(x, y) = \mathbb{P}(X_n = y)$, i.e., for the random variable $X_i$ the probability of $X_i = y$ is the probability of being in state $y$ after $n$ steps.  
		Here $\mathbb{P}$ is the probability distribution on the underlying (not specified) measurable space on which the $X_i$s are defined.

		With $\mathbb{E}(X_i)$ we denote the expectation of $X_i$ and with $\mathit{Var}(X_i)$ the variance defined in the standard way. 
		With $\mathbb{E}(s)$ we denote the expectation of $s$.
		We obviously have $\mathbb{E}(X_i) = m + i\cdot \mathbb{E}(s)$ and as each of the $Y_i$ are independent also $\mathit{Var}(X_i) = i\cdot\mathit{Var}(s)$.
		By assumption $\mathbb{E}(s) = \sum_{i \in \mathbb{Z}} i \cdot s(i) < 0$.
		Let $\epsilon = - \sum_{i \in \mathbb{Z}} i \cdot s(i) > 0$.
		So $\mathbb{E}(X_i) = m - i\cdot \epsilon$.
		
		All that remains now is to apply an appropriate concentration bound.
		Let $N$ be such that for every $i > N$ we have $\mathbb{E}(X_i) < 0$, which exists as $\mathbb{E}(X_i) = m + i\cdot \epsilon$.
		For each $i > N$ we have:
		\begin{align*}
			\mathbb{P}(X_i > 0) &\leq \mathbb{P}\Big( |X_i - \mathbb{E}(X_{i})| > - \mathbb{E}(X_{i})  \Big) \\
			&\myleq{\textbf{(1)}} \frac{\mathit{Var}(X_{i})}{(- \mathbb{E}(X_{i}))^2} = \frac{i \cdot \mathit{Var}(s) }{(m-\epsilon \cdot i)^2} 
		\end{align*}
		where \textbf{(1)} follows from Chebyshev's inequality. 
		If we let $i \to \infty$ we thus get that $\mathbb{P}(X_i > 0)$ converges to $0$.

		\item In the case of equality, we have $\sum_{i \in \mathbb{Z}} i \cdot s(i) = 0$: 
		First note that in this case the above reposing does not work. It does not even hold that $\mathbb{P}(X_i > 0)$ tends to $0$ as it has in the previous case.
		We again use the same construction of the RV $X_i$ as before. 
		We will show that $X_i$ does eventually become negative at least once. 
		
		From condition \ref{item:linear3} together with \ref{item:linear2} we get that there exists an $i^* < 0$ with $s(i^*) > 0$. 
		From any state $k$ we can now reach a non-positive number in $\lceil k / i^* \rceil$ steps with probability at least $s(i^*)^{\lceil k / i^* \rceil} > 0$ (just take the relative change $i^*$ so many times).
		Our proof now hinges on a famous theorem proved by George Pólya that states that any random walk on $\intnum^1$ or $\intnum^2$ with zero mean is recurrent (For a modern proof see e.g.~\citep{DBLP:journals/tamm/Novak14}). 
		As our random walk starts in $m$, i.e., $X_0 = m$ we thus get that the process $(X_i)_i$ does return to $m$ with probability $1$.
		We can then use the strong Markov property that states that if we have any stopping time $\tau$ (in our case the first time we revisit $m$) the process after $\tau$ is identical to the original one.
		We thus get that as $(X_i)_i$ is recurrent, i.e., visits $m$ again almost-surely, it also visits $m$ infinity many times a.s.
		As we have just shown, every time we visit $m$ there is a (lower bounded) positive probability (of $s(i^*)^{\lceil k / i^* \rceil} > 0$) of visiting a negative numbers.  
		So we eventually visit a negative number with certainty (see the zero-one law in \citep{DBLP:series/mcs/McIverM05} ).

	\end{itemize}	
	
	\noindent $\boxed{\Rightarrow}$: We now show the other direction.
	We prove this by contraposition. 
	It is easy to see that if $\sum_i s(i) < 1$ the walk is not AST as we have a positive probability of moving to the error state from any state. 
	Similar if $s = \delta_0$ we are obviously not terminating.
	Lastly if $\sum_{i \in \mathbb{Z}} i \cdot s(i) > 0$ we can follow similar reasoning as in the case of strictly negative expectation and show that the expectation \emph{increases} in each step.
\end{proof}

\begin{lemmaRE}{\ref{lem:uniformAST}}
	If $\{s_i\}_{i \in \indx}$ is a \emph{finite} family of {step distributions} and each $s_i$ is AST then $\{s_i\}_{i \in \indx}$ is uniform AST.
\end{lemmaRE}
\begin{proof}
	Fix any $m$.
	Fix any $\epsilon > 0$.
	By assumption 
	$$\lim_{n \to \infty} \mathfrak{P}_{s_i}^n(m, 0) = 1$$
	for every $i$. 
	So for any $i$ there exist a $N_i \in \natnum$ such that for every $n \geq N_i$, $\mathfrak{P}_{s_i}^n(m, 0) \geq 1 - \epsilon$ (By definition of the limit).
	Now define $N = \sum_i N_i$ which is finite as $\indx$ is finite.
	
	Now choose any $n \geq N$.
	We claim 
	$$\inf_{i_1, \cdots, i_n} \mathfrak{P}_{s_{i_1}} \cdots  \mathfrak{P}_{s_{i_n}} (m, 0) \geq 1 - \epsilon$$
	which would immediately give us the result.
	Choose arbitrary indices $i_1, \cdots, i_n$. We show $\mathfrak{P}_{s_{i_1}} \cdots  \mathfrak{P}_{s_{i_n}} (m, 0) \geq 1 - \epsilon$.
	As $ n \geq N$ there must exists a $i_* \in \indx$ that occurs at least $N_{i_*}$-many times among $i_1, \cdots, i_n$ by the pigeon hole principle. 

	As $0$ is an absorbing state we can see that $\mathfrak{P}_{s_i} \mathfrak{P}(m, 0) \geq \mathfrak{P}(m, 0)$ for any stochastic matrix $\mathfrak{P}$ \textbf{(1)}.
	Note that multiplication is commutative, i.e., $\mathfrak{P}_{s_i} \mathfrak{P}_{s_j}(m, 0) = \mathfrak{P}_{s_j} \mathfrak{P}_{s_i}(m, 0)$ (This does \emph{not} hold for general matrix multiplication but holds for $\mathfrak{P}_{s_i}$ as a random walk is invariant of the current state).
	We can thus reorder the indices $i_1, \cdots, i_n$ such that $i_*$ fills the last $N_{i_*}$ positions. 
	Together with \textbf{(1)} we thus get
	$$\mathfrak{P}_{s_{i_1}} \cdots  \mathfrak{P}_{s_{i_n}} (m, 0) \geq \mathfrak{P}^{N_{i_*}}_{s_{i_*}} (m, 0) \geq 1 - \epsilon$$
	as required.
\end{proof}

\subsection{Proof of Thm.~\ref{theo:ASTCount}}

This section is devoted to a proof of \refTheo{ASTCount}.
We begin by giving a rough outline of the proof for orientation.
The fundamental idea is to decompose the set of terminating traces according to the number of recursive calls made (not only on the first level).
We formalize this decomposition by a special kind of tree structure called number tree. 
We then show a direct correspondence between number trees and and terminating runs of the random walk generated by $\{\overline{\talloblong \mu^\varphi_x. M \mid r\talloblong}\}_{r \in \real}$.
Henceforth fix a term $\mu^\varphi_x. M$.

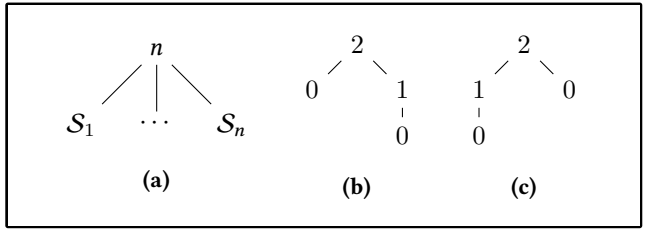
\begin{figure}
	
	\begin{tcolorbox}[colback=white, colframe=black, arc=0mm, boxrule=1pt]
		\begin{subfigure}{0.4\textwidth}
			\begin{center}
				\begin{tikzpicture}
					\node[] at (0,0) (n1) {$n$};
					
					\node[] at (-1,-1) (n2) {$\mathcal{S}_1$};
					
					\node[] at (0,-1) (n3) {$\cdots$};
					
					\node[] at (1,-1) (n4) {$\mathcal{S}_n$};

					\draw[-] (n1) -- (n2);
					\draw[-] (n1) -- (n3);
					\draw[-] (n1) -- (n4);
				\end{tikzpicture}
			\end{center}
		
			\subcaption{}\label{fig:numExamplea}
		\end{subfigure}
		\begin{subfigure}{0.3\textwidth}
			
			\begin{center}
				\begin{tikzpicture}[scale=0.6]
					\node[] at (0,0) (n1) {$2$};
					
					\node[] at (-1,-1) (n2) {$0$};

					\node[] at (1,-1) (n3) {$1$};
					
					\node[] at (1,-2) (n4) {$0$};

					\draw[-] (n1) -- (n2);
					\draw[-] (n1) -- (n3);
					\draw[-] (n3) -- (n4);
				\end{tikzpicture}
			\end{center}
		
			\subcaption{}\label{fig:numExampleb}
		\end{subfigure}
		\begin{subfigure}{0.28\textwidth}
			
			\begin{center}
				\begin{tikzpicture}[scale=0.6]
					\node[] at (0,0) (n1) {$2$};
					
					\node[] at (-1,-1) (n2) {$1$};

					\node[] at (1,-1) (n3) {$0$};
					
					\node[] at (-1,-2) (n4) {$0$};

					\draw[-] (n1) -- (n2);
					\draw[-] (n1) -- (n3);
					\draw[-] (n2) -- (n4);
				\end{tikzpicture}
			\end{center}
		
			\subcaption{}\label{fig:numExamplec}
		\end{subfigure}
	\end{tcolorbox}
	
	\caption{Example number trees.}\label{fig:numExample}
\end{figure}

\subsubsection*{Number Trees}

We define a \emph{number tree} by the following:
$$\mathcal{S} \triangleq n \triangleright [\mathcal{S}_1, \cdots, \mathcal{S}_n]$$
where $n \in \natnum$.\index{$\mathcal{S}$, a number tree}
We can depict each number tree by viewing $n$ as the label of the node and $\mathcal{S}_1, \cdots, \mathcal{S}_n$ as the children as depicted in \refFig{numExamplea}.
Note that the simplest tree is given by $0 \triangleright []$. 
Two (distinct) example trees are given in \refFig{numExampleb} and \refFig{numExamplec}.

\begin{figure*}
	\begin{tcolorbox}[colback=white, colframe=black, arc=0mm, boxrule=1pt]
		\begin{minipage}{0.4\textwidth}
			\begin{prooftree}
				\AxiomC{}
				\UnaryInfC{$\langle (\lambda x. M) V, \tr \rangle \placeto \langle M[V/x], \tr \rangle$ }
			\end{prooftree}
		\end{minipage}
		\begin{minipage}{0.3\textwidth}
			\vspace{0.4mm}
			\begin{prooftree}
				\AxiomC{}
				\UnaryInfC{$\langle (\boxed{\mu}) \num{r}, \square^r_{r'}::\tr \rangle \placeto \langle \num{r'}, \tr \rangle $}
			\end{prooftree}
		\end{minipage}
		\begin{minipage}{0.27\textwidth}
			\begin{prooftree}
				\AxiomC{}
				\UnaryInfC{$\langle \sample, r::\tr \rangle \placeto \langle \num{r}, \tr \rangle $}
			\end{prooftree}	
		\end{minipage}

		\begin{minipage}{0.33\textwidth}
			\begin{prooftree}
				\AxiomC{$r \leq 0$}
				\UnaryInfC{$\langle \myif(\num{r}, N, P), \tr \rangle \placeto \langle N, \tr \rangle$ }
			\end{prooftree}
		\end{minipage}
		\begin{minipage}{0.33\textwidth}
			\vspace{0.2mm}
			\begin{prooftree}
				\AxiomC{$r > 0$}
				\UnaryInfC{$\langle \myif(\num{r}, N, P), \tr \rangle \placeto \langle P, \tr \rangle $}
			\end{prooftree}
		\end{minipage}
		\begin{minipage}{0.32\textwidth}
			\begin{prooftree}
				\AxiomC{$r \geq 0$}
				\UnaryInfC{$\langle \score(\num{r}), \tr \rangle \placeto \langle \num{r}, \tr \rangle $}
			\end{prooftree}
		\end{minipage}

		\begin{minipage}{0.5\textwidth}
			\vspace{5.7mm}
			\begin{prooftree}
				\AxiomC{}
				\UnaryInfC{$\langle f(\num{r_1}, \cdots, \num{r_{|f|}}), \tr \rangle \placeto \langle \num{f(r_1, \cdots, r_{|f|})}, \tr \rangle $}
			\end{prooftree}
		\end{minipage}
		\begin{minipage}{0.5\textwidth}
			\begin{prooftree}
				\AxiomC{$\langle R, \tr \rangle \placeto \langle M, \tr' \rangle$}
				\UnaryInfC{$\langle E[R], \tr \rangle \placeto \langle E[M], \tr' \rangle $}
			\end{prooftree}
		\end{minipage}

	\end{tcolorbox}
	
	\caption{Small-step reduction rules for $\placeto$.}\label{fig:summarySem}
\end{figure*}

\subsubsection*{Summary Semantics}

We define a \emph{summary} as an element $\square^r_{r'}$ for $r, r' \in \real$. 
With $\mathfrak{O}$ we denote the set of all summaries and with $\mathbb{S}^\square \triangleq (\mathfrak{O} \cup \intreal)^*$ the set of summary traces.\index{$\mathfrak{O}$, the set of summary traces}
We can define a summary semantics working on summary traces in \refFig{summarySem}. 
For every recursive call, we substitute in a summary, i.e., an abbreviation for a trace, on the traces.
Note that this semantics closely corresponds to the semantics in \refFig{countRecCalls} used to count the recursive calls. 
The only difference is that we do not blindly substitute in a dummy value $\star$ but predefine the outcome via a summary\footnote{Note that, unlike in $\starto$ we do not need to count the number of calls, as we can simply count the number of summaries in a trace which equals the number of calls made.}. 
We set
$$\mathbb{T}_{r \mapsto r'}^{\square} = \{\tr \in \mathbb{S}^\square \mid \langle \repTerm{r}, \tr \rangle \placeto^* \langle \num{r'}, \epsilon \rangle \}$$
as all summary traces on which the term on argument $r$ evaluates to argument $r'$.

\subsubsection*{Number Trees as Traces}

The summary semantics explicitly lists recursive calls, as we can view the summary $\square^r_{r'}$ as a placeholder for a trace such that $(\mu^\varphi_x. M) \num{r}$ evaluates to $\num{r'}$.
The summaries allow us to partition the set of terminating traces according to the number of calls made on each level.
We can specify the number of calls by a number tree. For a tree $n \triangleright [\mathcal{S}_1, \cdots, \mathcal{S}_n]$ there should be $n$ direct recursive calls and inside those calls the number of calls is inductively specified by $\mathcal{S}_i$.
We consider the following example for some intuition:

\begin{example}\label{ex:ex1}
	Consider the term $\mu^\varphi_x. \varphi(\varphi x) \oplus \big( \num{0} \oplus \varphi x\big)$ and the number tree in \refFig{numExampleb}. 
	All traces that correspond to this tree should make $2$ recursive calls in the first level. In the first of those calls, no further call is made and on the second a single one is made and afterwards none.
	This corresponds to the following set of terminating traces: 
	\begin{align*}
		\{s \in \stdtrset^7 \mid &s_1 \in [0, \tfrac{1}{2}], s_2 \in (\tfrac{1}{2}, 1], s_3 \in [0, \tfrac{1}{2}], \\
		&s_4 \in (\tfrac{1}{2}, 1], s_5 \in (\tfrac{1}{2}, 1], s_6 \in (\tfrac{1}{2}, 1], s_7 \in [0, \tfrac{1}{2}] \}
	\end{align*}
\end{example}

\begin{definition}
	For each number tree $\mathcal{S}$ we can define a family of sets of traces $\{\mathbb{A}^\mathcal{S}_{r \mapsto r'}\}_{r, r' \in \real}$ by induction on $\mathcal{S}$ as follows:
	\begin{prooftree}
		\AxiomC{$\tr_1 \square^{r_1}_{r'_1} \tr_2 \cdots \square^{r_n}_{r'_n} \tr_{n+1} \in \mathbb{T}_{r \mapsto r'}^{\square}$}
		\AxiomC{$\big\{\dot{\tr}^{r_i}_{r'_i} \in \mathbb{A}^{\mathcal{S}_i}_{r_i \mapsto r'_i}\big\}_{i =1}^n$}
		\BinaryInfC{$\tr_1 \dot{\tr}^{r_1}_{r'_1} \tr_2 \cdots \dot{\tr}^{r_n}_{r'_n} \tr_{n+1} \in \mathbb{A}^{n \triangleright [\mathcal{S}_1, \cdots, \mathcal{S}_n]}_{r \mapsto r'}$}
	\end{prooftree}
\end{definition}

Elements in $\mathbb{A}^{n \triangleright [\mathcal{S}_1, \cdots, \mathcal{S}_n]}_{r \mapsto r'}$ are thus obtained by taking every summary trace with exactly $n$ summaries that takes $r$ to $r'$. For the $i$th summary (the $i$th recursive call) we then substitute in a trace from $\mathbb{A}^{\mathcal{S}_i}_{r_i \mapsto r'_i}$ that is recursively obtained from the $i$th child ($\mathcal{S}_i$).
Every number tree thus defines a specific set of traces.
By induction it is easy to see that for distinct trees the obtained sets of traces are disjoint. 
The interested reader is advised to match this definition with \refExample{ex1}.
We define $\mathbb{A}^\mathcal{S}_{r \mapsto \real} \triangleq \bigcup_{r' \in \real} \mathbb{A}^\mathcal{S}_{r \mapsto r'}$ as all terminating traces with recursion according to $\mathcal{S}$.
It is easy to see that $\mathbb{A}^\mathcal{S}_{r \mapsto \real}$ is measurable.

\subsubsection*{Probability Distributions on Number Trees}

We can view a number tree as being sampled from a counting distribution $t : \natnum \to \intreal$. 
For every node we sample a number $n$ according to $p$, add $n$ nodes and continue by sampling the child nodes.
Every counting distribution $t : \natnum \to \intreal$ thus gives a natural probability to a number tree $\mathcal{S}$ as just the product over all nodes in $\mathcal{S}$.
More generally we define:

\begin{definition}
	For a family of counting distributions $\{t_k\}_k : \natnum \to \intreal$ and a number tree $\mathcal{S}$ we define $\mathbb{P}^{\{t_k\}_k}_{\text{inf}}(\mathcal{S})$ by induction as
	$$\mathbb{P}^{\{t_k\}_k}_{\text{inf}}(n \triangleright  [\mathcal{S}_1, \cdots, \mathcal{S}_n]) \triangleq \big(\inf_k t_k(n)  \big) \cdot \prod_{i=1}^{n} \mathbb{P}^{\{t_k\}_k}_{\text{inf}}(\mathcal{S}_i) $$
	where we follow the usual convention that $\prod_{i=1}^{0} = 1$. 
\end{definition}

Note that if $\{t_k\}_k$ consist of a single element $\mathbb{P}^{\{t_k\}_k}_{\text{inf}}(\mathcal{S})$ for a tree $\mathcal{S}$ is the product of the probability of every node in that tree.
Taking the infimum follows the general scheme as e.g.~in the definition uniform AST (Def.~\ref{def:uniAST}).
We show that if we take the family $(\talloblong \mu^\varphi_x. M \mid r \talloblong)_{r \in \real}$ the probability of tree is a lower bound on the measure of $\mathbb{A}^{\mathcal{S}}_{r \mapsto \real}$.

\begin{example}
	For Ex.~\ref{ex:ex1} we get that the family $(\talloblong \mu^\varphi_x. M \mid r \talloblong)_{r \in \real}$ comprises a single element, namely the function $t$ defined by $t(2) = \tfrac{1}{2}$, $t(1) = \tfrac{1}{4}$ and $t(0) = \tfrac{1}{4}$.
	The probability of the tree $\mathcal{S}$ in \refFig{numExampleb}, as defined above then equals $\tfrac{1}{2}\tfrac{1}{4}\tfrac{1}{4}\tfrac{1}{4} = \tfrac{1}{128}$ which is less than or equal (in this case equal) to the measure of $\mathbb{A}^{\mathcal{S}}_{r \mapsto \real}$.
\end{example}

\begin{proposition}\label{prop:lowerBoundByNT}
	For every number tree $\mathcal{S}$ and any $r$ we have
	$$\mathbb{P}^{\{\talloblong \mu^\varphi_x. M \mid r' \talloblong\}_{r'}}_{\text{inf}}(\mathcal{S}) \leq \mu_{\stdtrset} \big( \mathbb{A}^{\mathcal{S}}_{r \mapsto \real} \big)$$
\end{proposition}
\begin{proof}
	By induction on $\mathcal{S}$ with $r$ universally quantified. 
	Let $\mathcal{S} = n \triangleright [\mathcal{S}_1, \cdots, \mathcal{S}_n]$.
	We first analyse $\mathbb{A}^{\mathcal{S}}_{r \mapsto \real}$:
	by definition every $\tr \in \mathbb{A}^{\mathcal{S}}_{r \mapsto r'}$ has the from $\tr = \tr_1 \dot{\tr}^{r_1}_{r'_1} \tr_2 \cdots \dot{\tr}^{r_n}_{r'_n} \tr_{n+1}$ for some $\tr_1 \square^{r_1}_{r'_1} \tr_2 \cdots \square^{r_n}_{r'_n} \tr_{n+1} \in \mathbb{T}_{r \mapsto r'}^{\square}$ and $\dot{\tr}^{r_i}_{r'_i} \in \mathbb{A}^{\mathcal{S}_i}_{r_i \mapsto r'_i}$.
	We can observe that $\{\tr_1 \cdots \tr_{n+1} \mid \tr_1 \square^{r_1}_{r'_1} \tr_2 \cdots \square^{r_n}_{r'_n} \tr_{n+1} \in \mathbb{T}_{r \mapsto r'}^{\square}, r' \in \real\} = \termTrCount{\repTerm{r}}{n}$ by comparing the relation $\starto$ and $\placeto$.
	If we concatenate two sets of traces the measure of the new set is the product of the individual measures.
	As we know that $n$ traces are substituted we can take the infimum over all possible arguments which gives us a trivial lower bound on $\mu_{\stdtrset} \big( \mathbb{A}^{\mathcal{S}}_{r \mapsto \real} \big)$:
	{\small\begin{align}\label{eq:e2}
			\small\mu_{\stdtrset} \big( \termTrCount{\repTerm{r}}{n} \big) \cdot \inf_{r_1, \cdots, r_n} \prod_{i=1}^{n} \mu_{\stdtrset} \big( \mathbb{A}^{\mathcal{S}_i}_{r_i \mapsto \real} \big) \leq  \mu_{\stdtrset} \big( \mathbb{A}^{\mathcal{S}}_{r \mapsto \real} \big) \tag*{\textbf{(i)}}
	\end{align}}%
	By induction we get that $\mathbb{P}^{\{\talloblong \mu^\varphi_x. M \mid r' \talloblong\}_{r'}}_{\text{inf}}(\mathcal{S}_i) \leq \mu_{\stdtrset} \big( \mathbb{A}^{\mathcal{S}_i}_{\dot{r} \mapsto \real} \big)$ for every $i$ and every $\dot{r}$.
	Note that the left hand side does not depend on $\dot{r}$ so in particular
	{\small\begin{align}\label{eq:e1}
			\small\prod_{i=1}^{n} \mathbb{P}^{\{\talloblong \mu^\varphi_x. M \mid r' \talloblong\}_{r'}}_{\text{inf}}(\mathcal{S}_i) \leq \inf_{r_1, \cdots, r_n} \prod_{i=1}^{n} \mu_{\stdtrset} \big( \mathbb{A}^{\mathcal{S}_i}_{r_i \mapsto \real} \big) \tag*{\textbf{(ii)}}
	\end{align}}%
	We can now put this all together and get:
	$$\small\begin{aligned}
		\mathbb{P}^{\{\talloblong \mu^\varphi_x. M \mid r' \talloblong\}_{r'}}_{\text{inf}}(\mathcal{S}) &\myeq{\textbf{(1)}} \inf_{r'} \talloblong \mu^\varphi_x. M \mid r' \talloblong(n) \cdot  \prod_{i=1}^{n} \mathbb{P}^{\{\talloblong \mu^\varphi_x. M \mid r' \talloblong\}_{r'}}_{\text{inf}}(\mathcal{S}_i)\\
		&\myleq{\textbf{(2)}} \talloblong \mu^\varphi_x. M \mid r \talloblong(n) \cdot  \inf_{r_1, \cdots, r_n} \prod_{i=1}^{n} \mu_{\stdtrset} \big( \mathbb{A}^{\mathcal{S}_i}_{r_i \mapsto \real} \big) \\
		&\myleq{\textbf{(3)}} \mu_{\stdtrset} \big( \termTrCount{\repTerm{r}}{n} \big) \cdot \inf_{r_1, \cdots, r_n} \prod_{i=1}^{n} \mu_{\stdtrset} \big( \mathbb{A}^{\mathcal{S}_i}_{r_i \mapsto \real} \big) \\
		&\myleq{\textbf{(4)}} \mu_{\stdtrset} \big( \mathbb{A}^{\mathcal{S}}_{r \mapsto \real} \big)
	\end{aligned}$$
	where \textbf{(1)} follows from the definition of $\mathbb{P}^{\{\talloblong \mu^\varphi_x. M \mid r' \talloblong\}_{r'}}_{\text{inf}}(\mathcal{S})$, \textbf{(2)} from the fact that $\inf_{r'} \talloblong \mu^\varphi_x. M \mid r' \talloblong(n) \leq \talloblong \mu^\varphi_x. M \mid r \talloblong(n)$ together with fact \ref{eq:e1}, \textbf{(3)} from the definition of $\talloblong \mu^\varphi_x. M \mid r' \talloblong$ and \textbf{(4)} from \ref{eq:e2}. 
\end{proof}

\subsubsection*{Number Trees as Terminating Runs}

It is easy to see that for every family of subprobability mass functions on the natural numbers $\{t_k\}_k$, $\sum_{\mathcal{S}} \mathbb{P}^{\{t_k\}_k}_{\text{inf}}(\mathcal{S}) \leq 1$.
Here the sum is taken over the countable set of (finite) number trees. 
What we can show is the following:

\begin{lemma}\label{lem:ASTforNT}
	If $\{t_k\}_k$ is a family of counting distributions and $\{\overline{t_k}\}_k$ is uniform AST then
	$$\sum_{\mathcal{S}} \mathbb{P}^{\{t_k\}_k}_{\text{inf}}(\mathcal{S}) = 1$$
\end{lemma}
\begin{proof}
	\newcommand{\toSeq}[1]{\mathfrak{F}(#1)}
	\newcommand{\toState}[1]{\mathfrak{H}(#1)}
	We define the following set of \emph{absolute} runs, i.e., sequences of states:
	{\small\begin{align*}
			\mathit{Runs}_A \triangleq \{ U \in \natnum^* &\mid U(i+1)-U(i) \geq -1, U(0)=1, \\
			&U(|u|-1)= 0, \forall 1 \leq i < |U|-1: \, U(i) \neq 0\}
	\end{align*}}%
	Think of elements in $\mathit{Runs}_A$ as terminating runs of the Markov chain that start in state $1$ and eventually reach state $0$. 
	The condition $U(i+1)-U(i) \geq -1$ is there to ensure that in each step the value never decrease by more than $1$ (Note that $\overline{t_k}$ never assign positive probability to values less than $-1$). 
	We associate a probability, $\mathbf{P}(U)$ to elements $U \in \mathit{Runs}_A$ by: 
	{\small$$\mathbf{P}(U) = \inf_{k_0, \cdots, k_{|U|-2}} \prod_{i=0}^{|U|-2}\mathfrak{P}_{\overline{t_{k_i}}}(U(i), U(i+1))$$}%
	which is just the probability of that run when taking the infimum over all possible choices of transition distribution. 
	What we observe now is that 
	\begin{align*}
		\lim_{n \to \infty} \Big(\inf_{k_1, \cdots, k_n} \mathfrak{P}_{\overline{t_{k_1}}} \cdots \mathfrak{P}_{\overline{t_{k_n}}} (1, 0)\Big) = \sum_{U \in \mathit{Runs}_A} \mathbf{P}(U)
	\end{align*}%
	, i.e., the probability of eventually reaching $0$ from $1$ is the same as the sum over the probability of each path that terminates starting in $1$. 
	By assumption $(\overline{t_k})_k$ is uniformly AST so the left hand side equals $1$. Call this \textbf{(1)}.
	Instead of analysis the absolute path we can also consider the relative change in each step.
	We define
	\begin{align*}
		\mathit{Runs}_R \triangleq \{ u \in (\natnum \cup \{-1\})^* &\mid \sum_{i=0}^{|u|-1} u(i) = -1, \\
		&\forall m < |u|-1 \sum_{i=0}^{m} u(i) > -1\}
	\end{align*}
	Each element $u \in \mathit{Runs}_R$ gives the relative change in each step such that starting from state $1$ we eventual terminate. 
	The sum of the relative change should thus be $-1$ but the sum of every strict prefix is at least $0$ (so that termination only occurs in the last step).
	There exists a bijective correspondence between elements in $\mathit{Runs}_A$ and $\mathit{Runs}_R$:
	For each $u \in \mathit{Runs}_R$, define $\toState{u} \in \mathit{Runs}_A$ as the sequence of length $|u|+1$ defined by $\toState{u}(i) \triangleq 1 + \sum_{j=0}^{i-1} u(j)$.
	It is easy to verify that $\toState{\cdot}$ is a bijection.
	
	Now lastly we observe that there is a bijection between the set of number trees and $\mathit{Runs}_R$.
	For each number tree $\mathit{S}$ we inductively define a sequence of integers $\toSeq{\mathcal{S}} \in \mathit{Runs}_R$ by
	\begin{align*}
		\toSeq{n \triangleright [\mathcal{S}_1, \cdots, \mathcal{S}_n] }= (n-1)::\toSeq{\mathcal{S}_1} \cdots \toSeq{\mathcal{S}_n}
	\end{align*}
	It is an easy proof to show that $\toSeq{\cdot}$ forms a bijection. 
	We thus have the bijective situation depicted below.
	
	\begin{center}
		\begin{tikzpicture}
			\node[] at (0,0) (n1) {$\mathit{Runs}_A$};
			
			\node[] at (2,0) (n2) {$\mathit{Runs}_R$};
			
			\node[] at (4,0) (n3) {$\mathit{NTree}$};
			
			\draw[->, thick] (n1) to[out=30,in=150] node[above] {$\toState{\cdot}^{-1}$} (n2);
			\draw[->, thick] (n2) to[out=210,in=330] node[below] {$\toState{\cdot}$} (n1);
			
			\draw[->, thick] (n2) to[out=30,in=150] node[above] {$\toSeq{\cdot}^{-1}$} (n3);
			\draw[->, thick] (n3) to[out=210,in=330] node[below] {$\toSeq{\cdot}$} (n2);
		\end{tikzpicture}
	\end{center}
	
	It is now easy to see that for every number tree $\mathcal{S}$, $\mathbb{P}^{\{t_k\}_k}_{\text{inf}}(\mathcal{S}) = \mathbf{P}(\toState{\toSeq{\mathcal{S}}})$ as $\mathbb{P}^{\{t_k\}_k}_{\text{inf}}$ gives probability of the relative change $\{t_k\}_k$ which is exactly the same as weighting the transition directly as in the definition of $\mathbf{P}$.

	As $\mathfrak{H} \circ \mathfrak{F}$ is a bijection and by \textbf{(1)} we thus get $\sum_{\mathcal{S}} \mathbb{P}^{\{t_k\}_k}_{\text{inf}}(\mathcal{S}) = 1$ as required. 
\end{proof}

\begin{theoremRE}{\ref{theo:ASTCount}}
	If $\{\overline{\talloblong \mu^\varphi_x. M \mid r\talloblong}\}_{r \in \real}$ is uniform AST then $\mu^\varphi_x. M$ terminates a.s.~on every argument.
\end{theoremRE}
\begin{proof}
	We obviously have $\termTr{(\mu^\varphi_x. M) \num{r}} \supseteq \biguplus_{\mathcal{S}} \mathbb{A}^{\mathcal{S}}_{r \mapsto \real}$ (in fact they are equal but we do not require this for the proof). We can thus deduce:
	\begin{align*}
		\mu_{\stdtrset} \big(\termTr{(\mu^\varphi_x. M) \num{r}}\big) &\geq \mu_{\stdtrset} \Big( \biguplus_{\mathcal{S}} \mathbb{A}^{\mathcal{S}}_{r \mapsto \real} \Big) \\
		&\myeq{\textbf{(1)}} \sum_{\mathcal{S}} \mu_{\stdtrset} \big(\mathbb{A}^{\mathcal{S}}_{r \mapsto \real}\big) \\
		&\mygeq{\textbf{(2)}} \sum_{\mathcal{S}} \mathbb{P}^{\{\talloblong \mu^\varphi_x. M \mid r' \talloblong\}_{r'}}_{\text{inf}}(\mathcal{S}) \\
		&\myeq{\textbf{(3)}} 1
	\end{align*}
	where \textbf{(1)} follows from the fact that $\mathbb{A}^{\mathcal{S}}_{r \mapsto \real}$ is disjoint for distinct number trees, \textbf{(2)} from \refProp{lowerBoundByNT} and \textbf{(3)} from \refLemma{ASTforNT}.
\end{proof}

\subsection{Partial Order on Counting Distributions}

We first show:

\begin{lemma}\label{lem:aproxStep}
	Let $p, q : \natnum \to \intreal$ be finite counting distributions with $p \sqsubseteq q$. 
	If $\overline{p}$ is an AST step distribution then $\overline{q}$ is an AST step distribution.
\end{lemma}
\begin{proof}
	For convenience lets write $\hat{p}(i) \triangleq \sum_{j \leq i} p(j)$ similarly for $\hat{q}$.
	As $p \sqsubseteq q$, we have $\hat{p}(i) \leq \hat{q}(i)$ for every $i$.
	We now use \refTheo{conditionForAST}: As $\overline{p}$ is an AST step distribution we have $\sum_i i \cdot p(i) = 1$, $\sum_i i \cdot p(i) \leq 1$ and $p \neq \delta_1$.
	We now show that $q$ satisfies all those conditions as well and can then use the other direction from \refTheo{conditionForAST}.

	As $p$ is finite we can view $p : \{0, \cdots, k\} \to \intreal$ for some $k$.
	As $\sum_i i \cdot p(i) = 1$ we get $\hat{p}(k) = 1$ and thus by assumption $\sum_i i \cdot q(i) = \hat{q}(k) = 1$ \textbf{(1)}.  
	We can hence also view $q$ as a function $q: \{0, \cdots, k\} \to \intreal$.
	As $\sum_i i \cdot p(i) \leq 1$ and $p \neq \delta_1$ we get that $\hat{p}(0) = p(0) > 0$, so $q(0)$ is also positive and thus $q \neq \delta_1$ \textbf{(2)}.
	
	In the following, it remains to show that $\sum_i i \cdot q(i) \leq 1$.
	We use the combinatorial fact that $\sum_{i \in \natnum} i \cdot p(i) = \sum_{i \in \natnum} \sum_{j > i} p(j)$ and show:
	\begin{align*}
		\sum_{i \in \natnum} i \cdot p(i) &= \sum_{i \in \natnum} \sum_{j > i} p(j) = \sum_i \big(1 - \sum_{j \leq i} p(j)\big) \\
		&= \sum_{i=0}^k \big(1 - \hat{p}(i)\big) \\
		&= (k+1) - \sum_{i=0}^k \hat{p}(i)
	\end{align*}
	Analogously $\sum_{i \in \natnum} i \cdot q(i) = (k+1) - \sum_{i=0}^k \hat{q}(i)$.
	As $\hat{p}(i) \leq \hat{q}(i)$ for all $i$ we get:
	\begin{align*}
		\sum_i i \cdot q(i) &= (k+1) - \sum_{i=0}^k \hat{q}(i) \\
		&\leq (k+1) - \sum_{i=0}^k \hat{p}(i) \\
		&= \sum_i i \cdot p(i) \\
		&\leq 1 \tag*{\textbf{(3)}}
	\end{align*}
	We are done as \textbf{(1)}, \textbf{(2)} and \textbf{(3)} imply that $\overline{q}$ is AST by \refTheo{conditionForAST}.
\end{proof}

Similarly we can show the following as we can easily extend $\sqsubseteq$ to distributions on $\natnum$ that result from runs of the Markov chain $\mathfrak{P}_{t_i}$. \\

\begin{lemmaRE}{\ref{lem:order}}
	If $s$, $\{t_i\}_{i \in \indx}$ are counting distributions and for all $i \in \indx$, $s \sqsubseteq t_i$ and $\overline{s}$ is AST then $\{\overline{t_i}\}_{i \in \indx}$ is uniform AST.
\end{lemmaRE}

\subsection{Ensure Progress}\label{subsection:progress}

\begin{figure*}[!t]
	\begin{tcolorbox}[colback=white, colframe=black, arc=0mm, boxrule=1pt]
		\begin{subfigure}[b]{0.3\textwidth}
			$$\alpha, \beta \triangleq \typeReal \mid \typeReal^\top \mid \alpha \to \beta$$
			\subcaption{}\label{fig:simplRecAvoida}
		\end{subfigure}
		\begin{subfigure}[b]{0.7\textwidth}
			\begin{minipage}{0.25\textwidth}
				\vspace{2.7mm}
				\begin{prooftree}
					\AxiomC{}
					\UnaryInfC{$\alpha \sqsubseteq \alpha$}
				\end{prooftree}
			\end{minipage}
			\begin{minipage}{0.25\textwidth}
				\vspace{3.5mm}
				\begin{prooftree}
					\AxiomC{}
					\UnaryInfC{$\typeReal \sqsubseteq \typeReal^\top $}
				\end{prooftree}
			\end{minipage}
			\begin{minipage}{0.45\textwidth}
				\begin{prooftree}
					\AxiomC{$\alpha' \sqsubseteq \alpha$}
					\AxiomC{$\beta \sqsubseteq \beta'$}
					\BinaryInfC{$\alpha \to \beta \sqsubseteq \alpha' \to \beta' $}
				\end{prooftree}
			\end{minipage}
			\subcaption{}\label{fig:simplRecAvoidb}
		\end{subfigure}

		\tcblower
		\begin{subfigure}{1\textwidth}
			\begin{minipage}{0.2\textwidth}
				\begin{prooftree}
					\AxiomC{$x:\alpha \in \Gamma$}
					\UnaryInfC{$\Gamma \vdash x: \alpha $}
				\end{prooftree}
			\end{minipage}
			\begin{minipage}{0.2\textwidth}
				\vspace{0.5mm}
				\begin{prooftree}
					\AxiomC{$\Gamma; x:\alpha \vdash M : \beta$}
					\UnaryInfC{$\Gamma \vdash \lambda x. M : \alpha \to \beta $}
				\end{prooftree}
			\end{minipage}
			\begin{minipage}{0.2\textwidth}
				\vspace{4.5mm}
				\begin{prooftree}
					\AxiomC{}
					\UnaryInfC{$\Gamma \vdash \boxed{\mu} : \typeReal^\top \to \typeReal^\top $}
				\end{prooftree}
			\end{minipage}
			\begin{minipage}{0.19\textwidth}
				\vspace{3mm}
				\begin{prooftree}
					\AxiomC{}
					\UnaryInfC{$\Gamma \vdash \num{r} : \typeReal $}
				\end{prooftree}
			\end{minipage}
			\begin{minipage}{0.19\textwidth}
				\vspace{3mm}
				\begin{prooftree}
					\AxiomC{}
					\UnaryInfC{$\Gamma \vdash \sample : \typeReal $}
				\end{prooftree}
			\end{minipage}

			\begin{minipage}{0.3\textwidth}
				\begin{prooftree}
					\AxiomC{$\Gamma \vdash M : \alpha$}
					\AxiomC{$\alpha \sqsubseteq \beta$}
					\BinaryInfC{$\Gamma \vdash M : \beta$ }
				\end{prooftree}
			\end{minipage}
			\begin{minipage}{0.3\textwidth}
				\begin{prooftree}
					\AxiomC{$\Gamma\vdash M : \alpha \to \beta$}
					\AxiomC{$\Gamma \vdash N : \alpha$}
					\BinaryInfC{$\Gamma \vdash M N : \beta$}
				\end{prooftree}
			\end{minipage}
			\begin{minipage}{0.39\textwidth}
				\vspace{0.8mm}
				\begin{prooftree}
					\AxiomC{$\Gamma \vdash M : \typeReal$}
					\AxiomC{$\Gamma \vdash N : \alpha$}
					\AxiomC{$\Gamma \vdash P : \alpha$}
					\TrinaryInfC{$\Gamma \vdash \myif(M, N, P) : \alpha$ }
				\end{prooftree}
			\end{minipage}
		
			\begin{minipage}{0.39\textwidth}
				\vspace{2mm}
				\begin{prooftree}
					\AxiomC{$\Gamma \vdash M_1 : \typeReal$}
					\AxiomC{$\cdots$}
					\AxiomC{$\Gamma \vdash M_{|f|} : \typeReal$}
					\TrinaryInfC{$\Gamma \vdash f(M_1, \cdots, M_{|f|}) : \typeReal $}
				\end{prooftree}
			\end{minipage}
			\begin{minipage}{0.2\textwidth}
				\vspace{0.8mm}
				\begin{prooftree}
					\AxiomC{$\Gamma \vdash M : \typeReal$}
					\UnaryInfC{$\Gamma \vdash \score(M) : \typeReal $}
				\end{prooftree}
			\end{minipage}
			\begin{minipage}{0.39\textwidth}
				\vspace{2mm}
				\begin{prooftree}
					\AxiomC{$\Gamma \vdash M_1 : \typeReal^\top$}
					\AxiomC{$\cdots$}
					\AxiomC{$\Gamma \vdash M_{|f|} : \typeReal^\top$}
					\TrinaryInfC{$\Gamma \vdash f(M_1, \cdots, M_{|f|}) : \typeReal^\top$}
				\end{prooftree}
			\end{minipage}
			
			\subcaption{}\label{fig:simplRecAvoidc}
		\end{subfigure}
	\end{tcolorbox}
	
	\caption{Simple typing judgments that guarantee that recursive outcomes are never used inside inside conditionals. }\label{fig:simplRecAvoid}
\end{figure*}

The problem with formally counting the number of recursive calls is that the returned value of a prior call can influence not only what the next calls are but also how many calls are made.
As an example consider $\mu^\varphi_x. \myif f x \mythen f x \myelse f x + f x$ where the number of recursive calls is either $2$ or $3$ depending on the outcome of the first.
We present a type system that guarantees evaluation via $\to^\star$ to succeed, i.e., whenever $\repTerm{r}$ is typable, $\sum_n \talloblong \mu^\varphi_x. M \mid r\talloblong(n) = 1$ for all $r \in \real$.
There are two conceptually different reasons why $\sum_n \talloblong \mu^\varphi_x. M \mid r\talloblong(n) = 1$. Either we get stuck (on a non-null set of traces) on terms of the from $\myif(\star, N, P)$ or $\score(\star)$. The other case is to get stuck on terms of the from $\score(r)$ for $r < 0$. 
We focus on the first cause, which informally occurs whenever a recursive outcome is subsequently used in guards or scores and thereby influences the control flow. 
The second cause, on the other hand depends on the concrete denotation of a program, and at such can not be analysed statically. 

The crux of our approach is thus to disallow the outcome of recursive calls to influence branching in the programs, i.e., recursive outcomes may not be used inside guards of conditionals or \score-constructs.
Obviously, this cannot be characterised purely syntactically as the property we seek is semantic in nature. 
We enforce this by a more involved simple type system where we add a dedicated type $\typeReal^\top$ for recursive outcomes that cannot be used within guards. 
We define simple types in \refFig{simplRecAvoida}.
The idea of $\typeReal^\top$ being more restrictive than $\typeReal$ can be formalized via a subtyping relation given in \refFig{simplRecAvoidb}.
Typing judgments are of the form $\Gamma \vdash M : \alpha$ and given in \refFig{simplRecAvoidc}. 
The crucial step is the rule for conditionals combined with the fixpoint rule.
For conditionals we require that the term in the guard position has $\typeReal$ and at the same time the recursive abstraction $\boxed{\mu}$ has the more restrictive return type $\typeReal^\top$. 
Combined with subtyping this gives a semantic guarantee that the recursive abstraction cannot be used inside conditionals. 
Note that the type system works with the simplified fixpoint constructs, $\boxed{\mu}$.
For now on we assume that the fixed $\mu^\varphi_x. M$ satisfies $\vdash \repTerm{r} : \typeReal^\top$ for some $r$\footnote{We obviously have that $\repTerm{r}$ is typable for {some} $r$ iff it is typable for all $r$.}.

Whenever $\repTerm{r}$ is typable in the system in \refFig{simplRecAvoid} recursive outcomes cannot be used inside the conditionals or score constructs. 
If, in addition, no \score-constructs get stuck, this already ensures that $\starto$ enjoys progress\footnote{While we can statically ensure that a recursive outcome, $\star$, never occurs insider a guard or a \score-construct, we can not ensure that we only score on non-negative values. 
Checking if the argument of every score is non-negative requires the inspection of the denotion of a subprogram and is thus very involved. For most interesting program it is, however, easy to verify the concrete score value as it is e.g.~a constant.  }.

\begin{lemma}
	If $\repTerm{r}$ is typable in the system from \refFig{simplRecAvoid} and no subterm of the from $\score(r)$ for $r < 0$ is reachable, then $\sum_n \talloblong \mu^\varphi_x. M \mid r\talloblong(n) = 1$ for all $r$.
\end{lemma}
\begin{proof}
	We extend the system in \refFig{simplRecAvoid} by the axiom $\Gamma \vdash \star : \typeReal^\top$ and cn show subject reduction w.r.t.~$\starto$.
	As terms of the from $\myif(\star, N, P)$ or $\score(\star)$ are not typable in \refFig{simplRecAvoid} execution via $\starto$ can never reach terms that contain such subterms.
	As by assumption no \score-construct can fail, our reduction does enjoy progress.
	This directly implies that $\sum_n \talloblong \mu^\varphi_x. M \mid r\talloblong(n) = 1$ for all $r$ by the same argument as in \cite[Lem.~7]{MakOPW20}.
\end{proof}

\begin{figure*}
	\begin{tcolorbox}[colback=white, colframe=black, arc=0mm, boxrule=1pt]
		\begin{minipage}{0.25\textwidth}
			\vspace{3.8mm}
			\begin{prooftree}
				\AxiomC{}
				\UnaryInfC{$\{x:[\alpha]\} \vdash x: \alpha $}
			\end{prooftree}
		\end{minipage}
		\begin{minipage}{0.25\textwidth}
			\begin{prooftree}
				\AxiomC{$\Gamma; x:a \vdash M : \alpha$}
				\UnaryInfC{$\Gamma \vdash \lambda x. M : a \to \alpha $}
			\end{prooftree}
		\end{minipage}
		\begin{minipage}{0.5\textwidth}
			\vspace{0.2mm}
			\begin{prooftree}
				\AxiomC{$\Gamma \vdash M : [\alpha_i] \to \beta$}
				\AxiomC{$\{\Gamma_i \vdash N : \alpha_i\}$}
				\BinaryInfC{$\uplus_i \Gamma_i \uplus \Gamma \vdash M N : \beta $}
			\end{prooftree}
		\end{minipage}

		\begin{minipage}{0.33\textwidth}
			\begin{prooftree}
				\AxiomC{$\Gamma \vdash M : \typeReal$}
				\AxiomC{$\Delta \vdash N : \alpha$}
				\BinaryInfC{$\Gamma \uplus \Delta \vdash \myif(M, N, P) : \alpha $}
			\end{prooftree}
		\end{minipage}
		\begin{minipage}{0.33\textwidth}
			\begin{prooftree}
				\AxiomC{$\Gamma \vdash M : \typeReal$}
				\AxiomC{$\Delta \vdash P : \alpha$}
				\BinaryInfC{$\Gamma \uplus \Delta \vdash \myif(M, N, P) : \alpha $}
			\end{prooftree}
		\end{minipage}
		\begin{minipage}{0.32\textwidth}
			\vspace{2mm}
			\begin{prooftree}
				\AxiomC{}
				\UnaryInfC{$\emptyset \vdash \sample : \typeReal$}
			\end{prooftree}
		\end{minipage}
		
		\begin{minipage}{0.5\textwidth}
			\begin{prooftree}
				\AxiomC{$\Gamma_1 \vdash M_1 : \typeReal$}
				\AxiomC{$\cdots$}
				\AxiomC{$\Gamma_{|f|} \vdash M_{|f|} : \typeReal$}
				\TrinaryInfC{$\Gamma_1 \uplus \cdots \uplus \Gamma_{|f|}  \vdash f(M_1, \cdots, M_{|f|}) : \typeReal $}
			\end{prooftree}
		\end{minipage}
		\begin{minipage}{0.2\textwidth}
			\vspace{3mm}
			\begin{prooftree}
				\AxiomC{}
				\UnaryInfC{$\emptyset \vdash \num{r} : \typeReal$}
			\end{prooftree}
		\end{minipage}
		\begin{minipage}{0.27\textwidth}
			\vspace{0.5mm}
			\begin{prooftree}
				\AxiomC{$\Gamma \vdash M : \typeReal$}
				\UnaryInfC{$\Gamma \vdash \score(M) : \typeReal$}
			\end{prooftree}
		\end{minipage}
	\end{tcolorbox}
	
	\caption{Non-idempotent Intersection Type System that counts the number of semantic uses of a variable. Note that this system is not syntax-guided due to the two rules for conditionals.} \label{fig:NIIsystem}
\end{figure*}

\subsection{Intersection Counting System}\label{subsection:countingType}

To count the number of occurrences we employ a non-idempotent intersection (NII) type system.
Intersection types are defined by the mutually recursive grammar $\alpha, \beta \triangleq \typeReal \mid a \to \alpha$ and $a \triangleq [\alpha_1, \cdots, \alpha_n]$ where $[\alpha_1, \cdots, \alpha_n]$ denotes a \emph{multiset}.
A typing context $\Gamma$ is a partial map from variables to intersections. The disjoint union of two contexts $\Gamma, \Delta$ denoted by $\Gamma \uplus \Delta$ is the elementwise disjoint union of multiset. 
For an overview on non-idempotent intersection types see \citep{DBLP:journals/igpl/BucciarelliKV17}.
Typing judgments are of the from $\Gamma \vdash M : \alpha$ and given by the rules in \refFig{NIIsystem}.
Due to the non-idempotent nature, for each type derivation we can read of the number of semantic occurrences as the cardinality of the intersection type. 
E.g.~$\mu^\varphi_x. M$ is a first-order fixpoint and $\{\varphi:a, x:b\} \vdash M : \typeReal$ we get a path in which $\varphi$ is used exactly $|a|$-many times. 
Here $|a|$ denote the cardinality of an intersection type.

We can easily see that the type system gives a upper bound on the recursive rank, as each type derivation outlines a possible execution and the cardinality of a intersection type represents the semantic use cases of a variable.

\begin{lemma}\label{lem:upperOnRecRank}
	Let $\mu^\varphi_x. M$ be a first-order fixpoint term with recuive rank $m$ (as defined in \refSection{pra}).
	Then $m \leq \max\limits_{\{\varphi:a, x:b\} \vdash M : \typeReal} |a|$
\end{lemma}

This lemma justifies the use of the NII-type system to upper bound the recursive rank and thus make use of Cor.~\ref{corr:era} without computing the recursive rank directly. 
A direct computation would involve probabilistic reasoning, whereas the type system gives a ``easy to compute'' upper bound. Note that for a term $\mu^\varphi_x. M$, the quantity $\max\limits_{\{\varphi:a, x:b\} \vdash M : \typeReal} |a|$ used in \refLemma{upperOnRecRank} is effectively computable.

\subsection{Further example}\label{sec:furtherEx}

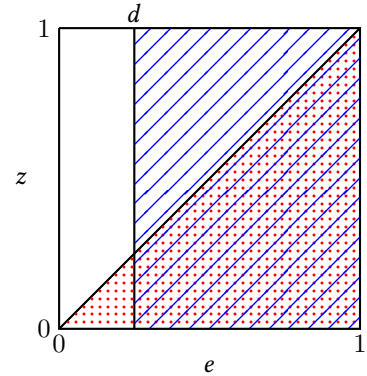
\begin{figure}
	\begin{center}
		
		\makeatletter
		\tikzset{%
			hatch distance/.store in=\hatchdistance,
			hatch distance=5pt,
			hatch thickness/.store in=\hatchthickness,
			hatch thickness=5pt
		}
		\pgfdeclarepatternformonly[\hatchdistance,\hatchthickness]{north east hatch}%
		{\pgfqpoint{-1pt}{-1pt}}%
		{\pgfqpoint{\hatchdistance}{\hatchdistance}}%
		{\pgfpoint{\hatchdistance-1pt}{\hatchdistance-1pt}}%
		{
			\pgfsetcolor{\tikz@pattern@color}
			\pgfsetlinewidth{\hatchthickness}
			\pgfpathmoveto{\pgfqpoint{0pt}{0pt}}
			\pgfpathlineto{\pgfqpoint{\hatchdistance}{\hatchdistance}}
			\pgfusepath{stroke}
		}
		\makeatother
		\begin{tikzpicture}

			\coordinate[] (A) at (0, 0);
			\coordinate[] (B) at (4, 0);
			\coordinate[] (C) at (4, -4);
			\coordinate[] (D) at (0, -4);
			
			\draw[-, thick] (A) -- (B) -- (C) -- (D) -- (A);
			
			\coordinate[] (X) at (1, 0);
			\coordinate[] (Y) at (1, -4);

			\draw[-, thick] (X) -- (Y);

			\draw[-, thick] (D) -- (B);
			
			\draw[pattern=dots, pattern color=red] (D) -- (B) -- (C) -- (D);

			\draw[pattern=north east hatch, hatch distance=3mm, hatch thickness=.5pt, pattern color=blue] (X) -- (Y) -- (C) -- (B) -- (X);
			
			\node[] at (-0.5, -2) () {$z$};
			\node[] at (-0.2, 0) () {$1$};
			\node[] at (-0.2, -4) () {$0$};

			\node[] at (2, -4.5) () {$e$};
			\node[] at (0, -4.2) () {$0$};
			\node[] at (4, -4.2) () {$1$};
			
			\node[] at (1, 0.2) () {$d$};
			
		\end{tikzpicture}
	\end{center}

	\caption{A geometric interpretation of the probability in \refSection{furtherEx}. Each point in the square correspond to a value of $e$ (the error value sampled in the let) and $z$ the sampled value in the binary choice. 
		The blue (striped) area are all value pairs such that $e > d$, i.e., all sampled values for $e$ such that the first conditional takes the right branch. 
		The red (dotted) area contains all value pairs such that $z \leq e$, i.e., the left branching in the binary sample is taken. }
	\label{fig:geometricProb}
\end{figure}

We consider the \refExample{complexExample}, i.e., the addition of \refExample{runningExample} where we use probabilistic outcomes as first class citizens. 
Recall that $p$ is the acceptance probability of a print. 
For each print we first sample a value $e$ uniform on $[0, 1]$ which represents how broken the product is, i.e., $e = 0$ is a completely fine product and $e = 1$ would correspond to a total failure. 
Whenever the quality is less than $p$ we accept the print.
In the other case, as before in \refExample{runningExample}, there is a chance of $\mathit{sig}(x)$ of the staff being tired and making mistakes.
In case a mistake is made, we do however not have a fair binary choice between printing $2$ or $3$ copies, but instead this depends on the quality of the most recent print $q$. 
With probability $e$ we reprint $3$ copies and otherwise only $2$. With increasing $e$, i.e., the more damaged the last print was, the more likely it is to reprint $3$ instead of $2$. 
The term we analyse in \refExample{complexExample} is the following (parameterised by $p$): 
\begin{align*}
	\mu^\varphi_x. &\mylet e = \sample \myin \myif \, e \leq p \mythen x \myelse \\
	&\,\Big(\big( \varphi^3(x+1) \oplus_e \varphi^2(x+1) \big) \oplus_{\mathit{sig}(x)} \varphi^2(x+1) \Big)
\end{align*}%
In particular, note the use of a probabilistic sample ($e$) as a first class value and the subsequent use as a probability $\oplus_e$.
Such behavior cannot be modelled via discrete distributions, as the quality $e$ is an intrinsic continuous value. 

We want to check for which instantiation of $p$ this term is AST on every input. 
To make use of \refTheo{ASTCount} we extract the counting pattern $\talloblong \mu^\varphi_x. M \mid r \talloblong$ for the program above.
If we fix $p$ this would become easier. 
However, for our demonstration, we treat $p$ as a variable. 
As we want the $p$ to stay flexible, this can be done via some basic geometric reasoning. 
It is easy to see that $\talloblong \mu^\varphi_x. M \mid r \talloblong(0) = p$, $\talloblong \mu^\varphi_x. M \mid r \talloblong(1) = 0$ and $\talloblong \mu^\varphi_x. M \mid r \talloblong(n) = 0$ for all $n > 3$, so it remains to compute $\talloblong \mu^\varphi_x. M \mid r \talloblong(2)$ and $\talloblong \mu^\varphi_x. M \mid r \talloblong(3)$.
Lets start with $\talloblong \mu^\varphi_x. M \mid r \talloblong(3)$. In order to make $3$ recursive calls we must have that the sampled value $e$ satisfies $e > p$ and in the later binary choice ($\oplus_e$) the sampled value (lets cal it $z$) must satisfy $z \leq e$.
If we let $e$ and $z$ be sampled iid from a uniform distribution on $[0, 1]$ we can interpret the desired probability as volume of the intersection of the red (dotted) and blue (striped) area in \refFig{geometricProb}.
We can compute this volume which is $\tfrac{1 - p^2}{2}$.
We thus get $\talloblong \mu^\varphi_x. M \mid r \talloblong(3) = \mathit{sig}(r) * \tfrac{1 - p^2}{2}$.
Similarly we can compute $\talloblong \mu^\varphi_x. M \mid r \talloblong(2) = (1-p)(1- \tfrac{1+p}{2}\mathit{sig}(r))$.

We now want to use \refTheo{ASTCount} to find values of $p$ such that the term is AST on every input.
We again make use of \refLemma{order}.
We define
$$s \defined p \delta_0 + \tfrac{(1-p)^2}{2}\delta_2 + \tfrac{1-p^2}{2}\delta_3$$
It is routine to check that $s \sqsubseteq \talloblong \mu^\varphi_x. M \mid r \talloblong$ for every $r$.
To analyse for which $p$ $\overline{s}$ is AST we use \refTheo{conditionForAST}; so we need to find values for $p$ such that the expectation of $p$ is less than or equal $1$\footnote{Note that this is equivalent to the fact that the exception of $\overline{s}$ (which is shifted by $-1$) is less than or equal $0$ }.
We can compute:
\begin{align*}
	& \quad\quad p*0 + \tfrac{(1-p)^2}{2} * 2 + \tfrac{1-p^2}{2} * 3 \leq 1 \\
	&\Leftrightarrow (1-p)^2 + \tfrac{3(1-p^2)}{2} \leq 1\\
	&\Leftrightarrow -\tfrac{1}{2}p^2 - 2p + \tfrac{3}{2} \leq 0 \\
	&\Leftrightarrow (p \leq -2 - \sqrt{7}) r (p \geq \sqrt{7} - 2)
\end{align*} 
As we assumed $p \in [0, 1]$, $\overline{s}$ is AST iff $p \geq \sqrt{7} - 2$.
We have $s \sqsubseteq \talloblong \mu^\varphi_x. M \mid r \talloblong$ for every $r$ so we can appeal to \refLemma{order} and get that $\{\talloblong \mu^\varphi_x. M \mid r \talloblong_r\}_{r \in \real}$ is uniform AST whenever $p \geq \sqrt{7} - 2$.
By \refTheo{ASTCount} we can thus conclude that when $p \geq \sqrt{7} - 2$ the program is AST on every input.

This example demonstrated well, that when we use (continuous) random outcomes as first class values, the analysis becomes very intricate.
Such examples can not be expressed in PHORS \cite{DBLP:conf/lics/KobayashiLG19} or with binary probabilistic choice \cite{DBLP:journals/toplas/LagoG19,DBLP:journals/jacm/KaminskiKMO18,DBLP:conf/lics/OlmedoKKM16}.
Our framework can analyse such example efficiently.

As we will see in the next section, we can automate this process entirely. I.e., the probability computation and the derivation of $s$ can be done fully automatically (for a fixed $p \geq \sqrt{7} - 2$).

\section{Additional Material - Section~\ref{sec:proofsystemFull}}

\subsection{Detailed Algorithm Description}

In this section we give a detailed (and formal) description of our algorithm.

\begin{figure*}
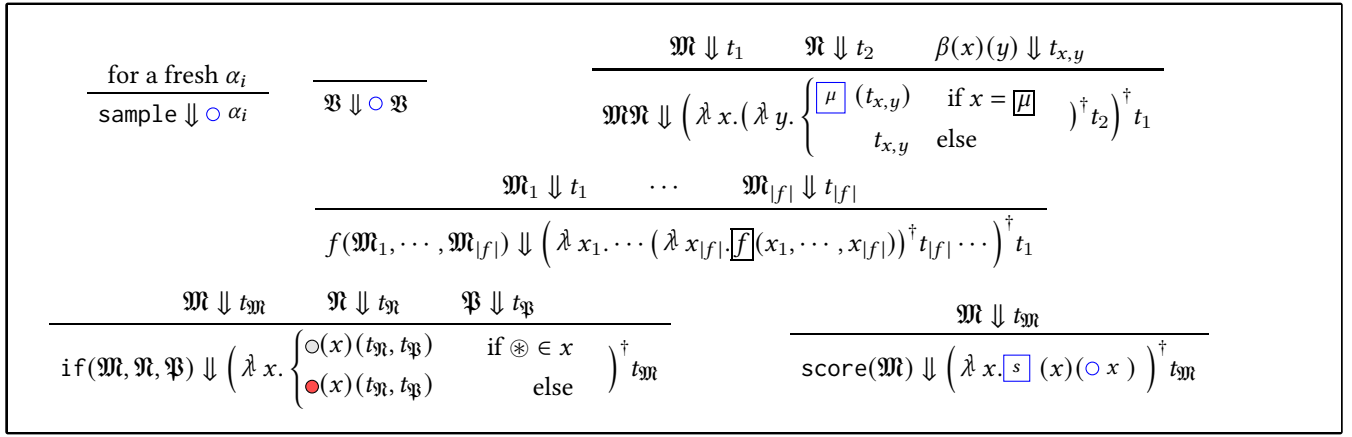

	\begin{tcolorbox}[colback=white, colframe=black, arc=0mm, boxrule=1pt]
		\begin{minipage}{0.2\textwidth}
			\begin{prooftree}
				\AxiomC{$\text{for a fresh } \alpha_i$}
				\UnaryInfC{$\sample \Downarrow \oLeafT{\alpha_i}$}
			\end{prooftree}
		\end{minipage}
		\begin{minipage}{0.1\textwidth}
			\begin{prooftree}
				\small
				\AxiomC{}
				\UnaryInfC{$\symV \Downarrow \oLeafT{\symV}$}
			\end{prooftree}
		\end{minipage}
		\begin{minipage}{0.7\textwidth}
			\begin{prooftree}
				\AxiomC{$\symM \Downarrow t_1$}
				\AxiomC{$\symN \Downarrow t_2$}
				\AxiomC{$\beta(x)(y) \Downarrow t_{x, y}$}
				\TrinaryInfC{$\symM \symN \Downarrow \Big(\metalambda x. \big(\metalambda y. \begin{cases}
						\begin{matrix}
							\oFixT{t_{x, y}} \quad \text{if } x = \boxed{\mu}\\
							t_{x, y} \quad \text{else}
						\end{matrix}
					\end{cases} \big)^\dagger t_2\Big)^\dagger t_1$}
			\end{prooftree}
		\end{minipage}

		\begin{prooftree}
			\AxiomC{$\symM_1 \Downarrow t_1$}
			\AxiomC{$\cdots$}
			\AxiomC{$\symM_{|f|} \Downarrow t_{|f|}$}
			\TrinaryInfC{$f(\symM_1, \cdots, \symM_{|f|}) \Downarrow \Big(\metalambda x_1. \cdots \big(\metalambda x_{|f|}. \boxed{f}(x_1, \cdots, x_{|f|})  \big)^\dagger t_{|f|} \cdots \Big)^\dagger t_1$}
		\end{prooftree}

		\begin{minipage}{0.5\textwidth}
			\begin{prooftree}
				\AxiomC{$\symM \Downarrow t_\symM$}
				\AxiomC{$\symN \Downarrow t_\symN$}
				\AxiomC{$\symP \Downarrow t_\symP$}
				\TrinaryInfC{$\myif(\symM, \symN, \symP) \Downarrow \Big(\metalambda x.
					\begin{cases}
						\begin{aligned}
							&\oAndT{t_\symN}{t_\symP}{x} \quad &\text{if } \circledast \in x \\
							&\oAndcT{t_\symN}{t_\symP}{x}\quad &\text{else}
						\end{aligned}
					\end{cases} \Big)^\dagger t_\symM$}
			\end{prooftree}
		\end{minipage}
		\begin{minipage}{0.5\textwidth}
			\begin{prooftree}
				\AxiomC{$\symM \Downarrow t_\symM$}
				\UnaryInfC{$\score(\symM) \Downarrow \Big( \metalambda x. \oScoreT{\oLeafT{x}}{x}  \Big)^\dagger t_\symM$}
			\end{prooftree}
		\end{minipage}
		
	\end{tcolorbox}
	
	\caption{Big-step symbolic execution where symbolic terms denote execution trees.}\label{fig:ExecTree}
\end{figure*}

\subsubsection*{Symbolic Execution Trees}

The key step is to evaluate a term symbolically and use sample variables to postpone sample decision (c.f. symbolic terms in \refSection{symbolicTerms}).
We extend the syntax of symbolic terms by a new symbol, $\circledast$, that will be used as an unknown argument. 
We now present symbolic execution as a big-step semantics, where branching on the term level is represented as branching of a tree. 
As we are, in particular, interested in recursive calls we annotate each call made in the semantics.
We define (symbolic) executions trees by:
\begin{align*}
	\mathit{ETree} \ni \treex \triangleq \oLeafT{\symV} &\mid \oFixT{\treex} \mid \oScoreT{\treex}{\symV} \\
	&\mid \oAndT{\treex_1}{\treex_2}{\symV} \mid \oAndcT{\treex_1}{\treex_2}{\symV}
\end{align*}
where $\symV \in \mathit{Val}$ is a symbolic value\footnote{As we have done in \refSection{extraction}, we extend symbolic values by a special symbol $\star$. }. \index{$\treex$, a (symbolic) executions tree}
We choose a more space-economical way to present trees. 
We occasionally depict execution trees as tree of degree $2$ where $\oAndT{\treex_1}{\treex_2}{\symV}$ represents a binary branch. 
Note that an execution tree condenses all of the information we are interested in.
For branching, it records the symbolic value as the condition; for score, it records the symbolic value that is scored, and finally every recursive call is recorded. 
To construct an execution tree from a program it remains to fold\footnote{Tree folding is standard in functional programming. In our case, fold traverses the tree and replaces every leaf with a tree given the folded function.} execution trees:

\begin{definition}
	Given a function $H: \mathit{Val} \to \mathit{ETree}$ we can define the lifted tree fold $H^\dagger : \mathit{ETree} \to \mathit{ETree}$ by induction as follows:
	\begin{align*}
		H^\dagger \big(\oLeafT{\symV}\big) &= H(\symV)\\
		H^\dagger \big(\oFixT{\treex}\big) &= \oFixT{H^\dagger \treex}\\
		H^\dagger \Big(\oScoreT{\treex}{\symV}\Big) &= \oScoreT{H^\dagger \treex}{\symV}\\
		H^\dagger \left(\oAndT{\treex_1}{\treex_2}{\symV}\right) &= \oAndT{H^\dagger \treex_1}{H^\dagger \treex_2}{\symV}\\
		H^\dagger \left(\oAndcT{\treex_1}{\treex_2}{\symV}\right) &= \oAndcT{H^\dagger \treex_1}{H^\dagger \treex_2}{\symV}
	\end{align*}
	
\end{definition}

We can now define a big-step semantics by giving a symbolic execution tree for each program,  denoted $M \Downarrow \mathfrak{T}$.
The big-step rules are given in \refFig{ExecTree} where $\beta(x)(y)$ performs a $\beta$-step, i.e., $\beta(\lambda x. M) (V) \triangleq M[V/x]$ and $\beta(\boxed{\mu})(V) \triangleq \star$.
$\metalambda$ binds the argument of an anonymous function. To avoid confusion, we use the special symbol to distinguish it from abstractions within our language. 
Note that this system inherits the structure of a standard big-step semantic (see e.g.~\citep{DBLP:conf/icfp/BorgstromLGS16}). 
As we execute symbolically and cannot resolve branching, we operate on trees and fold each reduction step. 
For each resolved conditional we introduce a binary branch at every conditional, a $\oScoreT{\treex}{V}$ for every score construct, and a $\oFixT{\treex}$ for every recursive call. 
For every term $M$ there exist a $M \Downarrow \treex$ and $\treex$ is unique up to reordering of the sample variables. 
The term we analyse is $\repTerm{\circledast}$, i.e., the body with the argument replaced by the distinguished symbol $\circledast$.

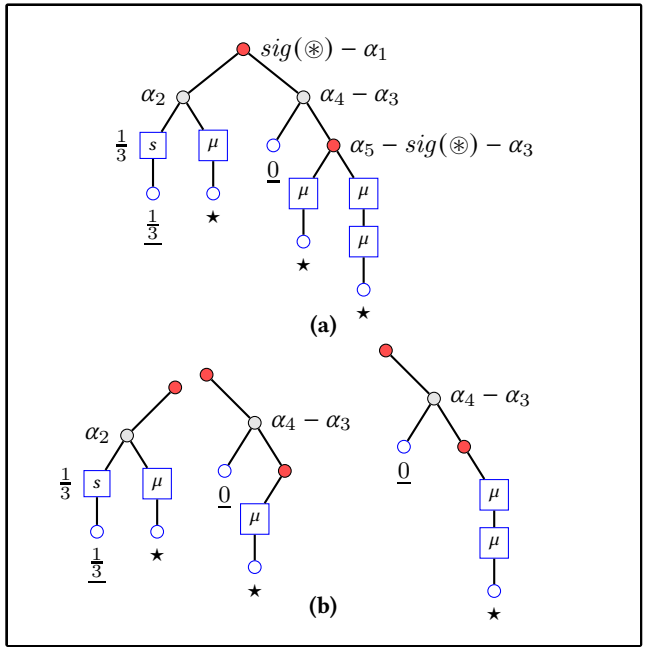
\begin{figure}
	\begin{tcolorbox}[colback=white, colframe=black, arc=0mm, boxrule=1pt]
		\begin{subfigure}{1\textwidth}
			\begin{center}
				\begin{tikzpicture}[scale=0.8]
					\node[circle, inner sep=1.7pt, draw, fill=red!70, label={[]right:$\mathit{sig}(\circledast) - \alpha_1$}] at (0,0) (n1){};

					\node[circle, inner sep=1.7pt, draw, fill=black!10, label=left:{$\alpha_2$}] at (-1,-0.8) (n11){};

					\node[rectangle, draw=blue, label=left:{\ensuremath{\tfrac{1}{3}}}] at (-1.5,-1.6) (n12){\ensuremath{\scriptstyle s}};
					
					\node[circle, draw=blue, fill=white, inner sep=1.7pt, label=below:{\small\ensuremath{\num{\tfrac{1}{3}}}}] at (-1.5,-2.4) (n13){};

					\node[rectangle, draw=blue] at (-0.5,-1.6) (n14){\ensuremath{\scriptstyle \mu}};
					
					\node[circle, draw=blue, fill=white, inner sep=1.7pt, label=below:{\small$\star$}] at (-0.5,-2.4) (n15){};

					\draw[-, thick] (n11) to (n12);
					\draw[-, thick] (n12) to (n13);
					
					\draw[-, thick] (n11) to (n14);
					\draw[-, thick] (n14) to (n15);

					\node[circle, inner sep=1.7pt, draw, fill=black!10, label=right:{$\alpha_4 - \alpha_3$}] at (1,-0.8) (n21){};
					
					\node[circle, draw=blue, fill=white, inner sep=1.7pt, label=below:{\small$\num{0}$}] at (0.5,-1.6) (n22){};

					\node[circle, inner sep=1.7pt, draw, fill=red!70, label={[]right:$\alpha_5 - \mathit{sig}(\circledast) - \alpha_3$}] at (1.5,-1.6) (n23){};

					\node[rectangle, draw=blue] at (1,-2.4) (n24){\ensuremath{\scriptstyle \mu}};
					
					\node[circle, draw=blue, fill=white, inner sep=1.7pt, label=below:{\small$\star$}] at (1,-3.2) (n25){};
					
					\node[rectangle, draw=blue] at (2,-2.4) (n26){\ensuremath{\scriptstyle \mu}};
					\node[rectangle, draw=blue] at (2,-3.2) (n27){\ensuremath{\scriptstyle \mu}};
					
					\node[circle, draw=blue, fill=white, inner sep=1.7pt, label=below:{\small$\star$}] at (2,-4) (n28){};

					\draw[-, thick] (n21) to (n22);
					
					\draw[-, thick] (n21) to (n23);
					
					\draw[-, thick] (n23) to (n24);
					\draw[-, thick] (n24) to (n25);
					
					\draw[-, thick] (n23) to (n26);
					\draw[-, thick] (n26) to (n27);
					\draw[-, thick] (n27) to (n28);

					\draw[-, thick] (n1) to (n11);
					\draw[-, thick] (n1) to (n21);
				\end{tikzpicture}
			\end{center}
			\vspace{-0.5cm}
			\subcaption{}\label{fig:ex2Treea}
		\end{subfigure}\\
		\begin{subfigure}{1\textwidth}
			
			\begin{minipage}{0.25\textwidth}
				\begin{center}
					\begin{tikzpicture}[scale=0.8]
						\node[circle, inner sep=1.7pt, draw, fill=red!70] at (-0.2,0) (n1){};

						\node[circle, inner sep=1.7pt, draw, fill=black!10, label=left:{$\alpha_2$}] at (-1,-0.8) (n11){};

						\node[rectangle, draw=blue, label=left:{\ensuremath{\tfrac{1}{3}}}] at (-1.5,-1.6) (n12){\ensuremath{\scriptstyle s}};
						
						\node[circle, draw=blue, fill=white, inner sep=1.7pt, label=below:{\small\ensuremath{\num{\tfrac{1}{3}}}}] at (-1.5,-2.4) (n13){};

						\node[rectangle, draw=blue] at (-0.5,-1.6) (n14){\ensuremath{\scriptstyle \mu}};
						
						\node[circle, draw=blue, fill=white, inner sep=1.7pt, label=below:{\small$\star$}] at (-0.5,-2.4) (n15){};

						\draw[-, thick] (n11) to (n12);
						\draw[-, thick] (n12) to (n13);
						
						\draw[-, thick] (n11) to (n14);
						\draw[-, thick] (n14) to (n15);

						\draw[-, thick] (n1) to (n11);
					\end{tikzpicture}
				\end{center}
			\end{minipage}
			\begin{minipage}{0.32\textwidth}
				\begin{center}
					\begin{tikzpicture}[scale=0.8]
						\node[circle, inner sep=1.7pt, draw, fill=red!70] at (0.2,0) (n1){};

						\node[circle, inner sep=1.7pt, draw, fill=black!10, label=right:{$\alpha_4 - \alpha_3$}] at (1,-0.8) (n21){};
						
						\node[circle, draw=blue, fill=white, inner sep=1.7pt, label=below:{\small$\num{0}$}] at (0.5,-1.6) (n22){};

						\node[circle, inner sep=1.7pt, draw, fill=red!70] at (1.5,-1.6) (n23){};

						\node[rectangle, draw=blue] at (1,-2.4) (n24){\ensuremath{\scriptstyle \mu}};
						
						\node[circle, draw=blue, fill=white, inner sep=1.7pt, label=below:{\small$\star$}] at (1,-3.2) (n25){};

						\draw[-, thick] (n21) to (n22);
						
						\draw[-, thick] (n21) to (n23);
						
						\draw[-, thick] (n23) to (n24);
						\draw[-, thick] (n24) to (n25);

						\draw[-, thick] (n1) to (n21);
					\end{tikzpicture}
				\end{center}
			\end{minipage}
			\begin{minipage}{0.3\textwidth}
				\begin{center}
					\begin{tikzpicture}[scale=0.8]
						\node[circle, inner sep=1.7pt, draw, fill=red!70] at (0.2,0) (n1){};

						\node[circle, inner sep=1.7pt, draw, fill=black!10, label=right:{$\alpha_4 - \alpha_3$}] at (1,-0.8) (n21){};
						
						\node[circle, draw=blue, fill=white, inner sep=1.7pt, label=below:{\small$\num{0}$}] at (0.5,-1.6) (n22){};

						\node[circle, inner sep=1.7pt, draw, fill=red!70] at (1.5,-1.6) (n23){};

						\node[rectangle, draw=blue] at (2,-2.4) (n26){\ensuremath{\scriptstyle \mu}};
						\node[rectangle, draw=blue] at (2,-3.2) (n27){\ensuremath{\scriptstyle \mu}};
						
						\node[circle, draw=blue, fill=white, inner sep=1.7pt, label=below:{\small$\star$}] at (2,-4) (n28){};

						\draw[-, thick] (n21) to (n22);
						
						\draw[-, thick] (n21) to (n23);
						
						\draw[-, thick] (n23) to (n26);
						\draw[-, thick] (n26) to (n27);
						\draw[-, thick] (n27) to (n28);

						\draw[-, thick] (n1) to (n21);
					\end{tikzpicture}
				\end{center}
			\end{minipage}
			\vspace{-0.7cm}
			\subcaption{}\label{fig:ex2Treeb}
		\end{subfigure}
	\end{tcolorbox}
	
	\caption{Symbolic execution trees for the running example and all possible strategies \textbf{\small(b)}.}\label{fig:ex2Tree}
\end{figure}

\begin{example}\label{ex:complexRunning}
	As an running example to demonstrate our tool consider the following non-trivial term
	\begin{align*}
		\mu^\varphi_x. \big( \score(\num{\tfrac{1}{3}}) \oplus \varphi \, x \big) \oplus_{\mathit{sig}(x)} &\Big( \mylet p = \sample \myin \nonumber \\
		&\num{0} \oplus_p \big(\varphi \, x \oplus_{x + p} \varphi(\varphi \, x)\big) \Big)
	\end{align*}
	where $\mathit{sig}(x)$ is the sigmoid function that squashes the real line into $\myint{0, 1}$.
	Checking this program for AST is challenging as the analysis depends on a complex interplay between the {actual argument} $x$ and the probabilistic outcomes. 
	Note that for $M$ as above: $\talloblong \mu^\varphi_x. M \mid r\talloblong \neq \talloblong \mu^\varphi_x. M \mid r'\talloblong$, if $r \neq r'$.
	The term we analyse in our big-step system is $\repTerm{r}$ which in our case is:
	\begin{align*}
		\big( \score(\num{\tfrac{1}{3}}) &\oplus \boxed{\mu} \num{\circledast} \big) \oplus_{\mathit{sig}(\num{\circledast})} \\
		&\Big( \mylet p = \sample \myin \num{0} \oplus_p \big(\boxed{\mu} \num{\circledast} \oplus_{\mathit{sig}(\num{\circledast}) + p} \boxed{\mu}(\boxed{\mu} \num{\circledast})\big) \Big)
	\end{align*}
	The tree $\treex$ with $\repTerm{\circledast} \Downarrow \treex$ is depicted in \refFig{ex2Treea}.
	The interested reader is advised to check the construction herself. 
\end{example}

\subsubsection*{Strategies}

Informally, each red inner node does contain the unknown argument $\circledast$ so we cannot determine its probabilistic behaviour without knowing its concrete value.
The route we pursue here is to simply ignore every branching at red nodes and not treating it as a quantitative but non-deterministic branching.
Loosely speaking, we let the environment decide which branch to take.  
We define strategies by
\begin{align*}
	\trees \triangleq \oLeafT{\symV} &\mid \oFixT{\trees} \mid \oScoreT{\trees}{\symV} \mid \oAndT{\trees_1}{\trees_2}{\symV} \\
	&\mid \oAndLeftcT{\trees}{\symV} \mid \oAndRightcT{\trees}{\symV} 
\end{align*}%
So strategies almost agree with execution trees but can choose which path to follow for each red node. 
A strategy $\trees$ is compatible with an execution tree $\treex$ (written $\trees \prec \treex$) if it matches the structure.

\subsubsection*{Paths and Probability}

As we arranged execution in a tree, we effectively postponed branching decision. 
Each branch in a strategy (or execution) corresponds to a branching path of the problem. 
A path is a sequence in $\kappa \in \{\leftP, \rightP\}^*$ that resolves binary branching decision. 
For a strategy $\trees$ we denote with $\pathsTerm(\treex)$ the set of terminating paths, i.e., paths that lead to a leaf. 
In the first example strategy in \refFig{ex2Treeb} paths include $\leftP\leftP$ and $\leftP\rightP$. 

For any strategy $\trees$ and terminating path $\kappa \in \pathsTerm(\trees)$ we count the numbers of recursive calls on that path, i.e., the number of times that a fixpoint node, $\oFixT{\cdot}$, is traversed. 
We denote this number with $|\boxed{\mu}|(\trees, \kappa) \in \natnum$. 
For a set $C$ of natural numbers we abbreviate $\pathsTerm(\trees, C) \triangleq \{\kappa \in \pathsTerm(\trees) \mid |\boxed{\mu}|(\trees, \kappa) \in C\}$. \index{$\pathsTerm(\trees, C)$, terminating paths in strategy $\trees$ s.t.~the number of fixpoint nodes is contained in $C$}

\begin{figure}[!t]
	\begin{tcolorbox}[colback=white, colframe=black, arc=0mm, boxrule=1pt,left=0pt, right=0pt]
		\small
		\begin{align*}
			\const^\star(\oLeafT{\symV}, \epsilon) &\triangleq \intreal^m\\
			\const^\star(\oFixT{\treex}, \kappa) &\triangleq \const^\star(\trees, \kappa)\\
			\const^\star(\oAndLeftcT{\trees_1}{\trees_2}{\symV}, \leftP\kappa) &\triangleq \const^\star(\trees_1, \kappa)\\
			\const^\star(\oAndRightcT{\trees_1}{\trees_2}{\symV}, \rightP\kappa) &\triangleq \const^\star(\trees_2, \kappa)\\
			\const^\star(\oScoreT{\trees}{\symV}, \kappa) &\triangleq \const^\star(\trees, \kappa) \cap \symV^{-1}[0, \infty)\\
			\const^\star(\oAndT{\trees_1}{\trees_2}{\symV}, \leftP\kappa) &\triangleq \const^\star(\trees_1, \kappa) \cap \symV^{-1}(-\infty, 0]\\
			\const^\star(\oAndT{\trees_1}{\trees_2}{\symV}, \rightP\kappa) &\triangleq \const^\star(\trees_2, \kappa) \cap \symV^{-1}(0, \infty)	
		\end{align*}
	\end{tcolorbox}
	
	\caption{Inductive definition of $\const^\star(\trees, \kappa)$ for $\kappa \in \pathsTerm(\trees)$.}\label{fig:constDef}
\end{figure}

Assume that all sample variables occurring in a execution tree $\treex$ are within $\{\alpha_0, \cdots, \alpha_{m-1}\}$ and $\trees \prec \treex$ (so sample variables within $\trees$ are also within $\{\alpha_0, \cdots, \alpha_{m-1}\}$).
Then each path $\kappa \in \pathsTerm(\trees)$ denotes a measurable subset of $\intreal^m$ in the natural way as all assignment such that this path is followed. 
We denote this set with $\const^\star(\trees, \kappa) \subseteq \intreal^m$ and it is defined by induction in \refFig{constDef}.
$\const^\star(\trees, \kappa) \subseteq \intreal^m$ denotes the set of assignments for $\alpha_0, \cdots, \alpha_{m-1}$ such that the branching and \score-constructs are evaluated according to $\kappa$. 
Red nodes are ignored as we do not interpret them probabilistically.
It is easy to see that $\const^\star(\trees, \kappa)$ is measurable.
We abbreviate $\mathbb{P}^\star(\trees,\kappa) \triangleq \lambda_m \big(\const^\star(\trees, \kappa)\big)$, i.e., the Lebesgue measure of all those assignments. \index{$\mathbb{P}^\star(\trees,\kappa)$, the probability of path $\kappa$ in strategy $\trees$}

\subsubsection*{The Algorithm}

We are now in a position to present our algorithm. 
Given a term $\mu^\varphi_x. M$ we begin by computing $\repTerm{\circledast} \Downarrow \treex_\circledast$. Note that such a tree always exist and is, up to sample variables, unique. 
For a strategy $\trees$, we abbreviate $\mathbb{P}^\star(\trees, n) \defined \sum\limits_{\kappa \in \pathsTerm(\trees, \{0, \cdots, n\})}  \mathbb{P}^\star(\trees,\kappa)$, i.e., the probability that in $\trees$ at most $n$ call are made. 
\index{$\mathbb{P}^\star(\trees, n)$, the joint probability of all paths in $\trees$ making at most $n$ recursive calls}

We can now define:
{\small\begin{align*}
		\mathbb{P}_{\mathit{approx}}(0) &\defined \min_{\trees \in \mathit{Strat}(\treex_\circledast)} \mathbb{P}^\star(\trees, 0)\\
		\mathbb{P}_{\mathit{approx}}(n > 0) &\defined \\
		&\quad\quad\quad\Big(\min_{\trees \in \mathit{Strat}(\treex_\circledast)} \mathbb{P}^\star((\trees, n)\Big) - \Big(\min_{\trees \in \mathit{Strat}(\treex_\circledast)} \mathbb{P}^\star((\trees, n-1)\Big)
\end{align*}}%
We can understand $\mathbb{P}_{\mathit{approx}}(n)$ as the least probability that $n$ calls are made even if the environment chooses in the worst (worst here meaning more recursive calls) possible way. 

\begin{example}\label{ex:appex2}
	Consider all strategies for the running example listed in \refFig{ex2Treeb}.
	We can compute $\mathbb{P}_{\mathrm{approx}}(0) = \mathbb{P}_{\mathrm{approx}}(2)  = \tfrac{1}{2}$ and $\mathbb{P}_{\mathrm{approx}}(n) = 0$ for all other $n$.
\end{example}

As the same sampling outcome can be used within multiple branching, we must make sure that the non-deterministic interpretation of branching does not lose any information. 
We call a execution tree $\treex$ \emph{sufficiently independent} if every sample variable that is used in a red node is not used in the subtree rooted at that node. 
Informally speaking, this means that probabilistic outcomes that we over-approximated by switching to a non-deterministic view may not be used afterwards. 
They can, of course, be used prior to the non-deterministic node. 
The correctness of our approach is then stated as follows:\\

\begin{theoremRE}{\ref{theo:soundPS}}
	If $\treex_\circledast$ is sufficiently independent, then for every $r \in \real$, $\mathbb{P}_{\mathrm{approx}} \sqsubseteq \talloblong \mu^\varphi_x. M \mid r \talloblong$. 
\end{theoremRE}\\

Note that our approach still does not provide a straightforward way to implement it. 
While $\repTerm{\circledast} \Downarrow \treex_\circledast$ can be computed effectively and the (finitely many) strategies with $\trees \prec \treex$ can be enumerated we still to compute $\mathbb{P}(\trees, n)$ and therefore the Lebesgue measure of a certain set. 
However, our approach does a big leap towards automation as we no longer need to consider individual arguments. 
As we argue later (in the implementation section) the Lebesgue measure of a set can be computed or approximated efficiently for certain primitive functions.

\subsection{Correctness Proof}

It remains to show the correctness of our approach, by proving \refTheo{soundPS}.
For the proof it is actually easiest to ignore some of the previous work.
Instead of analysing $\repTerm{\circledast}$ we fix a {actual} argumengt $r$ and investigate $\repTerm{r}$.
Most notably, we get that $\repTerm{r} \Downarrow \treex_r$ for a (up to sample variables unique) $\treex_r$ and we know that $\treex_r$ does not contain a single red node (as it does not contain $\boxed{\circledast}$).

\subsubsection*{Paths in Trees}

Similar to the way we defined paths in strategies, we can also define paths in execution trees. 
For a execution tree $\treex$ we denote with $\pathsTerm(\treex)$ all terminating paths in $\treex$ and for a $\kappa \in \pathsTerm(\treex)$ with $|\boxed{\mu}|(\trees, \kappa) \in \natnum$ the number of times a fixpoint node is traversed.
The set of terminating traces for the execution tree in \refFig{ex2Treea} includes e.g.~$\rightP\rightP\rightP$ $\leftP\rightP$.  
As before, for a set $C$ of natural numbers we abbreviate $\pathsTerm(\treex, C) \triangleq \{\kappa \in \pathsTerm(\treex) \mid |\boxed{\mu}|(\treex, \kappa) \in C\}$.

\subsubsection*{Correspondence}

\begin{figure}
	\begin{tcolorbox}[colback=white, colframe=black, arc=0mm, boxrule=1pt]
		\small
		\begin{align*}
			\const(\oLeafT{\symV}, \epsilon) &\triangleq \intreal^m\\
			\const(\oFixT{\treex}, \kappa) &\triangleq \const(\treex, \kappa)\\
			\const(\oScoreT{\treex}{\symV}, \kappa) &\triangleq \const(\treex, \kappa) \cap \symV^{-1}[0, \infty)\\
			\const(\oAndT{\treex_1}{\treex_2}{\symV}, \leftP\kappa)  &\triangleq \const(\treex_1, \kappa) \cap \symV^{-1}(-\infty, 0]\\
			\const(\oAndT{\treex_1}{\treex_2}{\symV}, \rightP\kappa)  &\triangleq \const(\treex_2, \kappa) \cap \symV^{-1}(0, \infty)\\
			\const(\oAndcT{\treex_1}{\treex_2}{\symV}, \leftP\kappa)  &\triangleq \const(\treex_1, \kappa) \cap \symV^{-1}(-\infty, 0]\\
			\const(\oAndcT{\treex_1}{\treex_2}{\symV}, \rightP\kappa)  &\triangleq \const(\treex_2, \kappa) \cap \symV^{-1}(0, \infty)			
		\end{align*}
	\end{tcolorbox}
	
	\caption{Inductive definition of $\const(\treex, \kappa)$ for $\kappa \in \pathsTerm(\treex)$.}\label{fig:constDef1}
\end{figure}

For every execution tree $\treex$ that does not contain $\circledast$ and $\kappa \in \pathsTerm(\treex)$ we define a measurable set $\const(\treex, \kappa)$ by induction in \refFig{constDef1}.
Note that $\treex$ must not be obtained via $\Downarrow$.
This is similar to the definition in \refFig{constDef} with the exception that, as $\circledast$ is not contained, every branch (both red and white) restricts the set of assignments. 
Informally, $\const(\treex, \kappa)$ includes all assignments to the sample variables, such that the branching according to $\kappa$ is taken and all \score constructs do not fail. 
As before, we define $\mathbb{P}(\treex,\kappa) \triangleq \lambda_m \big(\const(\treex, \kappa)\big)$.\index{$\mathbb{P}(\treex,\kappa)$, the probability of path $\kappa$ in tree $\treex$}
We can now show a intuitive correspondence between the paths in $\repTerm{r} \Downarrow \treex_r$ and the small step semantics $\starto$ from \refFig{countRecCalls} (which is similar to \refProp{corr}).

\begin{proposition}\label{prop:SymbolicBig}
	If $r \in \real$ and $\repTerm{r} \Downarrow \treex_r$ and $n \in \natnum$ then,
	$$\sum_{\kappa \in \pathsTerm(\treex_r, \{n\})} \mathbb{P}(\treex_r,\kappa) = \mu_{\stdtrset} \big( \termTrCount{\repTerm{r}}{n} \big)$$
\end{proposition}

This proposition states, that if we are interested in the number of recursive calls, say $n$.
Then the set of paths $\kappa \in \pathsTerm(\treex_r, \{n\})$ are all paths on which $n$ calls are made and the constraints along those paths characterize exactly the traces on which $n$ calls are made in the $\starto$ semantics (\refFig{countRecCalls}).
Note that not all traces in $\termTrCount{\repTerm{r}}{n}$ are of length at most $m$.

\subsubsection*{Replacing Probabilistic by Nondeterministic Choice}

We can show the following (which does not depend on the fact that $\treex$ must be obtained via our big-step semantics $\Downarrow$).
In particular note, that all trees obtained via $\Downarrow$ and do not contain $\circledast$ also do not contain a red node. For general $\treex$ this does not hold, i.e., there can be trees containing red nodes but no $\circledast$.
We need the following simple fact:

\begin{lemma}\label{lem:simpleFact1}
	If $\dot{a} \leq a$ and $\dot{b} \leq b$ and $p \in \intreal$ then $\dot{a} \leq pa + (1-p)b$ or $\dot{b} \leq pa + (1-p)b$.
\end{lemma}
\begin{proof}
	Assume for contradiction $pa + (1-p)b < \dot{a}$ and $pa + (1-p)b < \dot{b}$ then
	$pa+(1-p)b < p\dot{a} + (1-p)\dot{b}$.
	But obviously also $p\dot{a}+(1-p)\dot{b} \leq pa + (1-p)b$, a contradiction. 
\end{proof}

\begin{proposition}\label{prop:lowerBoundStrat}
	If $\treex$ is sufficiently independent and does not contain $\circledast$ and $C \subseteq \natnum$ then there exists a strategy $\trees \prec \treex$, s.t., 
	$$\sum\limits_{\kappa \in \pathsTerm(\trees, C)}  \mathbb{P}^\star(\trees,\kappa) \leq \sum\limits_{\kappa \in \pathsTerm(\treex, C)} \mathbb{P}(\treex,\kappa)$$
\end{proposition}
\begin{proof}
	We generalize the statement.
	For a measurable set $A \subseteq \intreal^m$ we define $\mathbb{P}_A(\treex, \kappa) \triangleq \lambda_m (\const(\treex, \kappa) \cap A)$ and $\mathbb{P}^\star_A(\trees, p) \triangleq \lambda_m (\const^\star(\trees, p) \cap A)$.
	Note that $\mathbb{P}_{\intreal^m}(\treex, p) = \mathbb{P}(\treex, p)$ and $\mathbb{P}^\star_{\intreal^m}(\trees, p) = \mathbb{P}^\star(\trees, p)$.
	
	We now show that the statement holds with $\mathbb{P}_A$ instead of $\mathbb{P}$ and $\mathbb{P}^\star_A$ instead of $\mathbb{P}^\star$ for any measurable $A \subseteq \intreal^m$ which obviously subsumes our initial obligation.
	The proof goes by induction on $\treex$ with $A \subseteq \intreal^m$ universally quantified. 
	\begin{itemize}[leftmargin=*]
		\item If $\treex = \oLeafT{\symV}$ then define $\trees \triangleq \oLeafT{\symV}$.
		It is easy to check that this strategy does satisfy the condition.  
		
		\item If $\treex = \oFixT{\treex'}$. By induction there is a $\trees' \prec \treex'$ that satisfies the conditions. Define $\trees \triangleq \oFixT{\trees'}$ which trivial satisfies the condition. 
		
		\item If $\treex = \oScoreT{\treex'}{\symV}$. 
		Define $A' \triangleq \symV^{-1}[0, \infty) \cap A$ which is obviously measurable. 
		Now by induction there is a $\trees' \prec \treex'$ such that
		{\small$$\sum\limits_{\kappa \in \pathsTerm(\trees', C)}  \mathbb{P}^\star_{A'}(\trees',\kappa) \leq \sum\limits_{\kappa \in \pathsTerm(\treex', C)} \mathbb{P}_{A'}(\treex',\kappa)$$}%
		Define $\trees \triangleq \oScoreT{\trees'}{\symV}$.
		Now for every $\kappa \in \pathsTerm(\trees)$ we have 
		{\small\begin{align*}
				\mathbb{P}^\star_{A}(\trees,\kappa) &= \lambda_m \big( \const^\star(\trees, \kappa) \cap A \big) \\
				&= \lambda_m \big( \const^\star(\trees', \kappa) \cap \symV^{-1}[0, \infty) \cap A \big)\\
				&= \mathbb{P}^\star_{\symV^{-1}[0, \infty) \cap A}(\trees',\kappa) \\
				&= \mathbb{P}^\star_{A'}(\trees',\kappa)
		\end{align*}}%
		Analogously $\mathbb{P}_{A}(\treex,\kappa) = \mathbb{P}_{A'}(\treex',\kappa)$.
		So using the IH we get
		{\small\begin{align*}
				\sum\limits_{\kappa \in \pathsTerm(\trees, C)}  \mathbb{P}^\star_A(\trees,\kappa) &= \sum\limits_{\kappa \in \pathsTerm(\trees', C)}  \mathbb{P}^\star_{A'}(\trees',\kappa) \\
				&\leq \sum\limits_{\kappa \in \pathsTerm(\treex', C)}  \mathbb{P}_{A'}(\treex',\kappa) \\
				&= \sum\limits_{\kappa \in \pathsTerm(\treex, C)} \mathbb{P}_A(\treex,\kappa)
		\end{align*}}%
		
		\item If $\treex = \oAndT{\treex_1}{\treex_2}{\symV}$: 
		We define the set $A_1 \triangleq \symV^{-1}(-\infty, 0] \cap A$ and $A_2 \triangleq \symV^{-1}(0, \infty) \cap A$. 
		Both are measurable. 
		By induction there are strategies $\trees_1, \trees_2$ such that 
		{\small\begin{align}\label{eq:IH1}
				\sum\limits_{\kappa \in \pathsTerm(\trees_i, C)}  \mathbb{P}^\star_{A_i}(\trees_i,\kappa) \leq \sum\limits_{\kappa \in \pathsTerm(\treex_i, C)} \mathbb{P}_{A_i}(\treex_i,\kappa) \tag*{\textbf{(1)}}
		\end{align}}%
		for $i \in \{1, 2\}$. 
		We define $\trees \triangleq \oAndT{\trees_1}{\trees_2}{\symV}$ and claim that this fulfils the criterion. 
		We observe the following, for any $\kappa \in \pathsTerm(\trees_1)$ we have:
		{\small\begin{align*}
				\mathbb{P}^\star_A(\trees, \leftP\kappa) &=  \lambda_m\big( \const^\star(\trees, \leftP\kappa) \cap A \big) \\
				&= \lambda_m\big( \const^\star(\trees_1, \kappa) \cap \symV^{-1}(-\infty, 0] \cap A \big)\\
				&= \mathbb{P}^\star_{\symV^{-1}(-\infty, 0] \cap A}(\trees_1, \kappa)  = \mathbb{P}^\star_{A_1}(\trees_1, \kappa)
			\end{align*}
		}%
		and analogously for every $\kappa \in \pathsTerm(\trees_2)$, $\mathbb{P}^\star_A(\trees, \rightP\kappa) = \mathbb{P}^\star_{A_2}(\trees_2, \kappa)$.
		The same also holds for $\mathbb{P}$ instead of $\mathbb{P}^\star$.
		We can now check:
		{\small\begin{align*}
				&\sum\limits_{\kappa \in \pathsTerm(\trees, C)}  \mathbb{P}^\star_A(\trees,\kappa) \\
				&= \sum\limits_{\kappa \in \pathsTerm(\trees_1, C)}  \mathbb{P}^\star_A(\trees,L\kappa) + \sum\limits_{\kappa \in \pathsTerm(\trees_2, C)}  \mathbb{P}^\star_A(\trees,R\kappa)\\
				&= \sum\limits_{\kappa \in \pathsTerm(\trees_1, C)}  \mathbb{P}^\star_{A_1}(\trees_1, \kappa) + \sum\limits_{\kappa \in \pathsTerm(\trees_2, C)}  \mathbb{P}^\star_{A_2}(\trees_2, \kappa)
			\end{align*}
		}%
		And using the same reasoning we have
		{\small\begin{align*}
				&\sum\limits_{\kappa \in \pathsTerm(\treex, C)} \mathbb{P}_A(\treex,\kappa) \\
				&= \sum\limits_{\kappa \in \pathsTerm(\treex_1, C)}  \mathbb{P}_{A_1}(\treex_1, \kappa) + \sum\limits_{\kappa \in \pathsTerm(\treex_2, C)}  \mathbb{P}_{A_2}(\treex_2, \kappa)
			\end{align*}
		}%
		We can now conclude using the inequalities we obtained via induction \ref{eq:IH1}.
		
		\item If $\treex = \oAndcT{\treex_1}{\treex_2}{f}$:
		We can assume that $\lambda_m(A) > 0$ as otherwise the statement is obvious as any strategy would work since both sides are equal to zero. 
		
		Let $\kappa \in \pathsTerm(\treex_1)$: 
		We make use of the assumption of sufficient independence. 
		As by assumption $\symV$ does not contain sample variables occurring in $\treex_1$, we get that $\symV^{-1}(-\infty, 0]$ and $\const(\treex_1, \kappa)$ are conditionally independent w.r.t. to $\lambda_m$.
		In particular,
		{\small\begin{align*}
			\lambda_m &\big( \symV^{-1}(-\infty, 0] \cap \const(\treex_1, \kappa) \mid A   \big) \\
			&= \lambda_m \big( \symV^{-1}(-\infty, 0] \mid A   \big) \cdot \lambda_m \big( \const(\treex_1, \kappa) \mid A   \big)
		\end{align*}}%
		We can multiply both sides by $\lambda_m(A)$ and derive
		{\small\begin{align*}
			\lambda_m &\big( \symV^{-1}(-\infty, 0] \cap \const(\treex_1, \kappa) \cap A \big) \\
			&= \lambda_m \big( \const(\treex_1, \kappa) \cap A \big) \cdot \lambda_m \big( \symV^{-1}(-\infty, 0] \mid A \big)
		\end{align*}}%
		We can now derive:
		{\small\begin{align*}
				\mathbb{P}_A(\treex, \leftP\kappa) &=  \lambda_m\big( \const(\treex, \leftP\kappa) \cap A \big) \\
				&= \lambda_m\big( \const(\treex_1, \kappa) \cap \symV^{-1}(-\infty, 0] \cap A \big)\\
				&= \lambda_m(\const(\treex_1, \kappa) \cap A) \cdot \lambda_m(\symV^{-1}(-\infty, 0] \mid A) \\
				&= \mathbb{P}_{A}(\treex_1, \kappa) \cdot \lambda_m(\symV^{-1}(-\infty, 0] \mid A)
		\end{align*}}%
		Analogously $\mathbb{P}_A(\treex, \rightP\kappa) = \mathbb{P}_{A}(\treex_2, \kappa) \cdot \mu(\symV^{-1}(0, \infty) \mid A)$ for $\kappa \in \pathsTerm(\treex_2)$.
		Now:
		{\small\begin{align*}
				&\sum\limits_{\kappa \in \pathsTerm(\treex, C)} \mathbb{P}_A(\treex,\kappa) \\
				&= \sum\limits_{\kappa \in \pathsTerm(\treex_1, C)}  \mathbb{P}^\star_A(\treex,\leftP\kappa) + \sum\limits_{\kappa \in \pathsTerm(\treex_2, C)}  \mathbb{P}^\star_A(\treex,\rightP\kappa)\\
				&= \sum\limits_{\kappa \in \pathsTerm(\treex_1, C)}  \mathbb{P}_{A}(\treex_1, \kappa) \lambda_m(\symV^{-1}(-\infty, 0] \mid A) \\
				&\quad\quad+ \sum\limits_{\kappa \in \pathsTerm(\treex_2, C)}  \mathbb{P}_{A}(\treex_2, \kappa) \lambda_m(\symV^{-1}(0, \infty) \mid A)
			\end{align*}
		}%
		By the IH there are strategies $\trees_1, \trees_2$ such that 
		{\small$$\sum\limits_{\kappa \in \pathsTerm(\trees_i, C)}  \mathbb{P}^\star_{A}(\trees_i,\kappa) \leq \sum\limits_{\kappa \in \pathsTerm(\treex_i, C)} \mathbb{P}_{A}(\treex_i,\kappa)$$}%
		for $i \in \{1, 2\}$.
		Now as $\lambda_m(\symV^{-1}(-\infty, 0] \mid A) + \lambda_m(\symV^{-1}(0, \infty) \mid A) = 1$ we can apply \refLemma{simpleFact1}.
		So there exists $i^* \in \{1, 2\}$ such that 
		{\small$$\sum\limits_{\kappa \in \pathsTerm(\trees_{i^*}, C)}  \mathbb{P}^\star_{A}(\trees_{i^*},\kappa) \leq \sum\limits_{\kappa \in \pathsTerm(\treex, C)} \mathbb{P}_A(\treex,\kappa)$$}%
		In case where ${i^*} = 1$, we define $\trees = \oAndLeftcT{\trees_1}{\symV}$.
		We can observe that for all $\kappa \in \pathsTerm(\trees_1)$ we have $\mathbb{P}^\star_{A}(\trees_1,\kappa) = \mathbb{P}^\star_{A}(\trees,\leftP\kappa)$ as $\const^\star$ does not add any constraint. 
		So $\trees$ does satisfy the desired property.  
		In the case of ${i^*} = 2$, define $\trees = \oAndRightcT{\trees_2}{\symV}$.
	\end{itemize}
\end{proof}

\subsubsection*{Changing the Node Colour}

As $\repTerm{r}$ does not contain $\circledast$ we get that, when $\repTerm{r} \Downarrow \treex_r$, $\treex_r$ does not contain any red nodes. 
We do however want to colour $\treex_r$ similarly to what we did with $\treex_\circledast$ (recall $\repTerm{\circledast} \Downarrow \treex_\circledast$). 
The first step is to observe that $\treex_\circledast$ and $\treex_r$ do agree structurally if we ignore node colours and the values at nodes.  
In fact if we replace every occurrence of $\circledast$ in $\treex_\circledast$ with $r$, we get, up to the colouring (and reordering of sample variables), exactly $\treex_r$.
To fix the colouring we do the following: Denote with $\treex_r^\bullet$ the tree $\treex_r$ but with all nodes that depend on $r$ coloured in red. 
Formally that is $\treex_r^\bullet \triangleq \treex_\circledast[r/\circledast]$ where $\treex_\circledast[r/\circledast]$ denotes $\treex_\circledast$ with all occurrence of $\circledast$ replaced by $r$.
In particular $\treex_r^\bullet$ and $\treex_r$ agree up to reordering of sample variables and colouring of nodes. 
Now $\treex_r^\bullet$ does contain red nodes, but does not contains $\circledast$, in particular every symbolic value at branching nodes (both red and white) denotes a function and we can use \refProp{lowerBoundStrat}.
We can then finally show:\\

\begin{theoremRE}{\ref{theo:soundPS}}
	If $\treex_\circledast$ is sufficiently independent, then for every $r \in \real$, $\mathbb{P}_{\mathrm{approx}} \sqsubseteq \talloblong \mu^\varphi_x. M \mid r \talloblong$
\end{theoremRE}
\begin{proof}
	We have $\repTerm{\circledast} \Downarrow \treex_\circledast$.
	Choose any $r \in \real$ and any $n \in \natnum$.
	Let $\repTerm{r} \Downarrow \treex_r$.
	And $\treex_r^\bullet \triangleq \treex_\circledast[\num{r}/\circledast]$. 
	As we argued before $\treex_r^\bullet$ and $\treex_r$ are identical up to the colouring of nodes. 
	Furthermore the strategies for $\treex_r^\bullet$ and $\treex_\circledast$ are identical (up to different labels of red nodes). 
	By \refProp{lowerBoundStrat} there exists a strategy $\trees_r \prec \treex_r^\bullet$ such that 
	\begin{align}\label{eq:stratEx}
			\begin{split}
				\sum\limits_{\kappa \in \pathsTerm(\trees_r, \{0, \cdots, n\})}  &\mathbb{P}^\star(\trees_r,\kappa)\\
				&\leq \sum\limits_{\kappa \in \pathsTerm(\treex_r^\bullet, \{0, \cdots, n\})} \mathbb{P}(\treex_r^\bullet,\kappa) 
			\end{split}\tag*{\textbf{(i)}}
	\end{align}%
	As the strategies for $\treex_r^\bullet$ and $\treex_\circledast$ are identical (up to red values at red nodes) we get that $\trees_r$ is also a strategy for $\treex_\circledast$ (after changing the values at red nodes).
	Thus
	$$
		\begin{aligned}
			\sum_{m \leq n} \mathbb{P}_{\mathrm{approx}}(m) &\myeq{\textbf{(1)}} \min_{\trees' \prec \treex_\circledast}  \sum_{\kappa \in \pathsTerm(\trees', \{0, \cdots, n\})} \mathbb{P}^\star(\trees', \kappa) \\
			&\myleq{\textbf{(2)}} \sum\limits_{\kappa \in \pathsTerm(\trees_r, \{0, \cdots, n\})}  \mathbb{P}^\star(\trees_r,\kappa) \\
			&\myleq{\textbf{(3)}} \sum\limits_{\kappa \in \pathsTerm(\treex_r, \{0, \cdots, n\})} \mathbb{P}(\treex_r,\kappa) \\
			&\myeq{\textbf{(4)}} \sum_{m \leq n} \mu_{\stdtrset}\big( \termTrCount{\repTerm{r}}{m} \big) \\
			&\myeq{\textbf{(5)}} \sum_{m \leq n} \talloblong \mu^\varphi_x. M \mid r \talloblong(m)
		\end{aligned}
		$$
	where \textbf{(1)} is a simple telescoping sum (see the definition of $\mathbb{P}_\mathit{approx}$), \textbf{(2)} follows as $\trees_r \prec \treex_\circledast$ (as the strategies for $\treex_\circledast$ $\treex_r^\bullet$ are almost identical as we argued before), \textbf{(3)} is by the choice of $\trees_r$ (c.f. \ref{eq:stratEx}), \textbf{(4)} follows from \refProp{SymbolicBig} and \textbf{(5)} by definition of $\talloblong \mu^\varphi_x. M \mid r \talloblong$.
	Thus $\mathbb{P}_{\mathrm{approx}} \sqsubseteq \talloblong \mu^\varphi_x. M \mid r \talloblong$ as required.
\end{proof}

\section{Additional Material - Section~\ref{sec:implementation}: Implementation}

\subsection{Lower Bound Computation}

We can turn our interval-based semantics into an effective lower bound computation algorithm by iteratively searching for terminating interval traces. 

To do so effectively, our algorithm evaluates a given term symbolically (see \refSection{symbolicTerms}) in a breath-first manor. 
Once we identified a conditional oracle leading to termination, i.e., a probabilistic execution leading to a value, we collect the symbolic constraints along this path.
Let $\{\symV_i \bowtie_i r_i\}_{i \in [m]}$ be those constraints.

To approximate the probability of this path, i.e., the Lebesgue measure of sample-variable assignments that satisfy all constraints along this path, we use our interval approach.
Let $\alpha_1, \cdots, \alpha_n$ be the sample variables occurring in $\symV_1, \cdots, \symV_m$. 
We use a standard sweep algorithm to split $[0, 1]^n$ into smaller boxes. In each step we choose a variable among $\{\alpha_1, \cdots, \alpha_n\}$ and split the current box in half along the chosen dimension. 
For the resulting smaller boxes we check if the guards are satisfied (using the interval-based reasoning) and in case they are not, split the boxes again; If the box does satisfies all constraints we add the respective volume to the total count. 
We stop the computation once the analysed parts of the box exceed a user specified probability, i.e., the current branch is analysed  such that discovering new terminating interval traces would only contribute very little to the lower bound.

\subsubsection*{Optimization}

Our prototype implementation should be considered a proof of concept and as such is not optimized.
The only optimization we use is a dependency analysis that identifies symbolic contains that do not share sample variables and computes the probability individually. 

We conjecture, that our implementation can be optimized significantly, by optimizing the split routine. At the moment we split a box along its longest dimension to keep boxes as ``square'' as possible. 
Ideally one would have a heuristic, that identifies which dimension should be split and at which value to split to minimize the overall number on overall splits. 
This would decrease the number of computation steps significantly.

\subsubsection{Experimental Results:}\label{section:exTerms}

As lower bound computation is a iterative, possibly non-terminating, process we set a termination condition.
This can either be given as a time constraints, leading the termination to be stopped after a given time or as a depth constraints where terms are evaluated up to a given depth. 
We use the following example programs. 
Wherever possible we try to use examples used in the implementation of \cite{DBLP:conf/lics/KobayashiLG19}.
Our results are, however, only partially comparable to \cite{DBLP:conf/lics/KobayashiLG19}. 
On the one hand, they only consider discrete distributions, which is obviously easier to analyse than the interval-based reasoning we use for continuous distributions. 
On the other hand, the main contribution of \cite{DBLP:conf/lics/KobayashiLG19} is the insight that the termination probability can be defined as the least fixpoint of higher-order fixpoint equations. 
Their tool therefore works on manually extracted fixpoint equations. 
As they already noted in their paper, not every fixpoint equation corresponds to a program; so we can only apply very few of their examples to our framework.

\subsubsection*{Examples}

\begin{itemize}
	\item $\mathit{geo}_p \defined \Big( \mu^\varphi_x. x \oplus_p \varphi(x+1) \Big) \num{0}$
	The simple example \refExample{3dprinting} from \refSection{intro} .
	This term ``computes'' the geometric distribution, i.e., the output follows the mass function $n \mapsto (1-p)^np$. 
	It is AST for every $p > 0$.
	
	\item {\footnotesize$\mathit{1dRW}_{p, m} \defined \Big( \mu^\varphi_x. \myif x \mythen 0 \myelse \varphi(x-1) \oplus_p \varphi(x+1) \Big) \num{m}$}%
	
	The $p$-biased 1-dimensional random walk with a probability of $p$ moving towards $0$. 
	The walk is known to be AST if and only if $p \geq \tfrac{1}{2}$. In case of $p = \tfrac{1}{2}$ this program is not PAST.
	Due to the non PAST nature this program (for $p = \tfrac{1}{2}$) is intrinsically hard to analyse, as the termination probability decreases significantly with increasing evaluation depth, requiring to consider very long executions. 
	For a $p > \tfrac{1}{2}$ this program is PAST and, as a results, allow for better (faster) lower bound computation. 
	
	\item $\mathit{gr} \defined \Big(\mu^\varphi_x. x \oplus \varphi(\varphi(\varphi x))\Big) \num{0}$
	
	Program inspired by \cite{DBLP:conf/lics/OlmedoKKM16}.
	As we can infer from our counting based framework, this program is actually not AST and terminates with probability $\tfrac{\sqrt{5}-1}{2}$, the reciprocal of the golden ratio (see \cite{DBLP:conf/lics/OlmedoKKM16}).
	Note that due to the CbN nature of our analysis, the left branch of the probabilistic choice must be $x$. For example, the term $\Big(\mu^\varphi_x. \num{0} \oplus \varphi(\varphi(\varphi x))\Big) \num{0}$ is trivially AST as the CbN evaluation causes the argument (in this case $\varphi x$) to be ignored without prior evaluation.
	
	\item $\mathit{print}_p \defined \Big( \mu^\varphi_x. x \oplus_p \varphi(\varphi(x)) \Big) \num{0}$
	
	Essentially, the example \refExample{3dprinting} from \refSection{intro}.
	This program is AST iff $p \geq \tfrac{1}{2}$ and is case of $p = \tfrac{1}{2}$ it is not PAST. 
	For $p = \tfrac{1}{4}$ this is comparable to the term ``Ex2.3-1'' from the full version of \cite{DBLP:conf/lics/KobayashiLG19}.

	\item $\mathit{3print}_p \defined \Big( \mu^\varphi_x. x \oplus_p \varphi(\varphi(\varphi(x))) \Big) \num{0}$
	
	Similar to the previous case with three instead of $2$ recursive calls. 
	For $p = \tfrac{1}{4}$ this is comparable to the term ``Ex2.3-v2'' from the full version of \cite{DBLP:conf/lics/KobayashiLG19}.
	
	\item $\mathit{bin}_{p, m} \defined \Big(\mu^\varphi_x. \myif x \mythen 0 \myelse (f (x-1) \oplus_p f(x)) \Big) \num{m}$
	
	Inspired by \cite{DBLP:journals/pacmpl/McIverMKK18}.

	\item \begin{align*}
		&\mathit{pedestrian} \defined \\
		\Big(\mu^\varphi_x. \myif &x \mythen 0 \myelse \\
		&\mylet s = \sample \myin\\
		&\quad s + \varphi\big( (x - s) \oplus_{\num{0.}7} (x + s) \big)\Big) \sample
	\end{align*}

	The term describes a random walk on $\real_+$ that models the situation of a forgetful pedestrian.
	The example is taken from \cite{MakOPW20}.
	
\end{itemize}

\subsubsection*{Experimental Setup}

Our experiential results are listed in Table \ref{tab:resLBi}. Where $\termProb{M}$ gives the actual probability of termination, LB the lower bound computed by our tool\footnote{We emphasis again that our tool works with rational numbers and thus perfect precision. For readability we give the first 10 decimal digits of the rational output.  }, Depth gives the evaluation depth at which we abort the search\footnote{As mentioned previously the computation is a ongoing, possibly infinite computation that must be ended at some point. This can be done by either specifying a target depth of time. To keep the results as independent from the concrete machine as possible, we specify a target depth to increase reproducibility.    }, \#V gives the number of values up to that depth and \#Nodes the total number of terms explored.  
Finally $t$ gives the time in milliseconds.

\subsection{AST Verification}

Our proof method from \refSection{proofsystem} gives us a straightforward implementation as all operations are on a finite tree. 
Our tool first computes the execution tree and its strategies. 
The key difficultly is to compute $\mathbb{P}(\trees, \kappa)$ for strategy $\trees$ and a path $\kappa \in \pathsTerm(\trees)$, i.e., compute the weight associated with a path.
We restrict the primitive operations to addition and multiplication by a constant (and thus subtraction). 
Under this restriction, each symbolic value $\symV$ denotes a \emph{linear} function in the sample variables. 
The weight of a path is thus the Lebesgue measure of an intersection of half planes or equivalently the volume of a polyhedron (a subset of $\real^d$ of the from $\{\vec{x} \mid A \vec{x} \leq b \}$) \cite{DBLP:journals/siamcomp/DyerF88}. 
As shown in \cite{lasserre1983analytical} the volume of such a polyhedron although $\#P$ hard, can be computed via a simple recursive scheme.
We use the optimized implementation of this scheme in \cite{Bueler2000} to effective compute the volume.
Our tool thus perform all basic operations on trees and refers to the tool from \cite{Bueler2000} for the probabilistic computations. 

\subsubsection*{Experimental Results}

Our tool can verify AST for all examples in this paper (with the identified bounds on free variables like $p$ in \refExample{3dprinting} or \refExample{complexExample}).
Our results are given in Table \ref{tab:resLASTi}.
The distribution $\mathbb{P}_{\mathit{approx}}$ is the one \emph{automatically} inferred by our tool.

\begin{table*}
	\caption{Experimental Results for Lower Bound Computations. We give the actual probability of termination $\termProb{M}$ (if known), the Lower Bound computed (LB), the depth at which we stopped the exploration (Depth), the number of identified values (\#Values) and total nodes (\#Nodes) as well as the time in milliseconds ($t$). }
	\label{tab:resLBi}
	\begin{tabular}{|c||c|c|c|c|c|c|}
		\hline
		Term $M$ & $\termProb{M}$ & LB & Depth & \#Values & \#Nodes & t \\
		\hline
		\hline
		$\mathit{geo}_{\tfrac{1}{2}}$ & $1$ (see \refTheo{ASTCount}) & $0.9999990463$ & $100$ & $20$ & $356$ & $78$ \\
		\hline
		$\mathit{geo}_{\tfrac{1}{5}}$ & $1$ (see \refTheo{ASTCount}) & $0.9995620416$ & $200$ & $40$ & $1211$ & $192$ \\
		\hline
		$\mathit{1dRW}_{\tfrac{1}{2}, 1}$ & $1$ & $0.8036193847$ & $200$ & $65535$ & $1376252$ & $28223$ \\
		\hline
		$\mathit{1dRW}_{\tfrac{7}{10}, 1}$ & $1$ & $0.9720964250$ & $150$ & $8191$ & $204796$ & $10224$ \\
		\hline
		$\mathit{gr}$ & $\tfrac{\sqrt{5}-1}{2}$ & $0.6112594604$ & $80$ & $1773$ & $2046981$ & $4389$ \\
		\hline
		$\mathit{print}_{\tfrac{1}{2}}$ & $1$ (see \refTheo{ASTCount}) & $0.8318119049$ & $90$ & $23714$ & $5056590$ & $15749$ \\
		\hline
		$\mathit{print}_{\tfrac{1}{4}}$ & $? (< 1)$ & $0.3328795089$ & $90$ & $23714$ & $5056590$ & $15749$ \\
		\hline
		$\mathit{3print}_{\tfrac{3}{4}}$ & $1$ (see \refTheo{ASTCount}) & $0.9606655982$ & $80$ & $1773$ & $2046981$ & $4622$\\
		\hline
		$\mathit{bin}_{\tfrac{1}{2}, 2}$ & $1$ & $0.9998493194$ & $100$ & $9445$ & $118907$ & $2265$\\
		\hline
		$\mathit{pedestrian}$ & $1$ & $0.6002376673$ & $40$ & $7$ & $197$ & $4493$\\
		\hline
	\end{tabular}
\end{table*}

\begin{table*}
	\caption{Experimental Results for AST Verification. For each term (all of which our tool can verified to be AST) we give the counting distribution $\mathbb{P}_{\mathit{approx}}$ computed by our tool (which is analysed via \refTheo{conditionForAST}). We also give the time used by our internal computation $t_{\mathit{int}}$, by the volume computation via VINCI (\cite{Bueler2000}) $t_{\mathit{vol}}$ and the total time $t = t_{\mathit{int}} + t_{\mathit{vol}}$ in milliseconds. }
	\label{tab:resLASTi}
	\begin{tabular}{|c||c|c|c|c|}
		\hline
		$M$ & $\mathbb{P}_{\mathit{approx}}$ & $t_{\mathit{int}}$  & $t_{\mathit{vol}}$ & $t$\\
		\hline
		\hline
		$\mathit{geo}_{\tfrac{1}{2}}$ & $\tfrac{1}{2}\delta_0 + \tfrac{1}{2}\delta_1$ & 140 & 99 & 239\\
		\hline
		\refExample{3dprinting}, $p = \tfrac{1}{2}$ ($\mathit{print}_{\tfrac{1}{2}}$) & $\tfrac{1}{2}\delta_0 + \tfrac{1}{2}\delta_2$ & 138 & 99 & 237\\
		\hline
		$\mathit{3print}_{\tfrac{2}{3}}$ & $\tfrac{2}{3}\delta_0 + \tfrac{1}{3}\delta_3$ & 274 & 123 & 297\\
		\hline
		\refExample{runningExample}, $p = 0.6$ & $ 0.6\delta_0 + 0.2\delta_2 + 0.2 \delta_3$ & 154 & 242 & 396\\
		\hline
		\refExample{complexRunning} & $ 0.5\delta_0 + 0.5\delta_2$ & 150 & 255 & 405\\
		\hline
		\refExample{complexExample}, $p = 0.65$ & $ 0.65\delta_0 + 061250\delta_2 + 0.288750\delta_3$ & 158 & 215 & 373\\
		\hline
	\end{tabular}
\end{table*}

\onecolumn

\printindex

\end{document}

%% file: makrox.tex
\newcommand{\refLemma}[1]{Lem.~\ref{lem:#1}}
\newcommand{\refTheo}[1]{Thm.~\ref{theo:#1}}
\newcommand{\refProp}[1]{Prop.~\ref{prop:#1}}

\newcommand{\refFig}[1]{Fig.~\ref{fig:#1}}

\newcommand{\refExample}[1]{Ex.~\ref{ex:#1}}

\newcommand{\refSection}[1]{Sec.~\ref{sec:#1}}

\newcommand{\rulename}[1]{{\scriptsize \textbf{(#1)}} }
\newcommand{\sample}{\texttt{sample}}

\newcommand{\num}[1]{\underline{#1}}
\newcommand{\myif}{\texttt{if}}
\newcommand{\mylet}{\texttt{let} \, }
\newcommand{\myin}{\, \texttt{in} \,}
\newcommand{\mythen}{\, \texttt{then} \,}
\newcommand{\myelse}{\, \texttt{else} \,}
\newcommand{\score}{\texttt{score}}

\newcommand{\interval}{\mathfrak{I}}

\newcommand{\typeReal}{\textbf{\textsf{R}}}

\newcommand{\LambdaInterval}{\Lambda_{\interval}}
\newcommand{\LambdaSymbolic}{\Lambda_{\text{sym}}}

\newcommand{\toIntervalTerm}[1]{#1^{2\interval}}

\newcommand{\indx}{\mathcal{I}}

\newcommand{\cbvto}{\xrightarrow[]{\scriptscriptstyle \mathfrak{V}}}

\newcommand{\stdtrset}{\mathbb{S}}
\newcommand{\inttrset}{\mathbb{S}_{\mathfrak{I}}}
\newcommand{\tr}{\boldsymbol{s}}
\newcommand{\too}{\leadsto}

\newcommand{\termTr}[1]{\mathbb{T}_{#1, \text{term}}}

\newcommand{\termProb}[1]{\mathbb{P}_{\text{term}}(#1)}

\newcommand{\termInterval}[1]{\mathbb{T}_{#1, \text{term}}^\mathfrak{I}}

\newcommand{\termTrCO}[2]{\mathbb{T}_{#1, \text{term}}^{(#2)}}

\newcommand{\wpi}{\wp}

\newcommand{\numberSteps}[2]{\#_\downarrow^{#1}(#2)}

\newcommand{\expectedTermSteps}[1]{\mathbb{E}_{\text{term}}(#1)}

\newcommand{\containedTraces}[1]{\llparenthesis \,#1 \,\rrparenthesis}

\newcommand{\toConditional}{\xrightarrow[]{\scriptscriptstyle co}}

\newcommand{\toSym}{\xrightarrow[]{\scriptscriptstyle \text{sym}}}

\newcommand{\placeto}{\xrightarrow[]{\square}}
\newcommand{\starto}{\stackrel{\raisebox{-2pt}{$\scriptscriptstyle\star$}}{\to}}

\newcommand{\repTerm}[1]{\text{\textbf{body}}_{\mu^\varphi_x. M}(#1)}

\newcommand{\termTrCount}[2]{\mathbb{T}^\star_{#1; #2}}

\newcommand\defined{\coloneqq} 

\newcommand{\myeq}[1]{\stackrel{\mathclap{\normalfont\mbox{\tiny #1}}}{=}}
\newcommand{\myleq}[1]{\stackrel{\mathclap{\normalfont\mbox{\tiny #1}}}{\leq}}
\newcommand{\mygeq}[1]{\stackrel{\mathclap{\normalfont\mbox{\tiny #1}}}{\geq}}

\renewcommand{\det}{\textsf{det}}

\newcommand{\treex}{\mathfrak{T}}

\newcommand{\trees}{\mathfrak{S}}

\newcommand{\expVal}{\mathbb{E}}

\renewcommand{\boxed}[1]{\text{\fboxsep=.1em\fbox{$\displaystyle#1$}}}

\def\metalambda{\mathop{\scalerel*{\stackengine{1.9pt}{$\lambda$}{%
				\kern3.4pt\smash{\clipbox{2pt -.5pt 0pt -.5pt}{$\lambda$}}}{O}{l}{F}{F}{L}}{X}\mkern1mu}}

\newcommand{\real}{\mathbb{R}}
\newcommand{\ration}{\mathbb{Q}}
\newcommand{\natnum}{\mathbb{N}}
\newcommand{\intnum}{\mathbb{Z}}
\newcommand{\posreal}{\mathbb{R}_+}
\newcommand{\intreal}{\mathbb{R}_{[0,1]}}

\newcommand{\leftP}{\boldsymbol{L}}
\newcommand{\rightP}{\boldsymbol{R}}

\newcommand{\calM}{\mathcal{M}}
\newcommand{\calN}{\mathcal{N}}
\newcommand{\calP}{\mathcal{P}}
\newcommand{\calQ}{\mathcal{Q}}
\newcommand{\calV}{\mathcal{V}}

\newcommand{\calE}{\mathcal{E}}
\newcommand{\calR}{\mathcal{R}}

\newcommand{\calA}{\mathcal{A}}
\newcommand{\calB}{\mathcal{B}}
\newcommand{\calC}{\mathcal{C}}
\newcommand{\calD}{\mathcal{D}}

\newcommand{\blist}[1]{\big\lBrace #1 \big\rBrace}
\newcommand{\Blist}[1]{\Big\lBrace #1 \Big\rBrace}

\newcommand{\bset}[1]{\big\{ #1 \big\}}

\newcommand{\symM}{{\mathfrak{M}}}
\newcommand{\symN}{{\mathfrak{N}}}
\newcommand{\symP}{{\mathfrak{P}}}
\newcommand{\symR}{{\mathfrak{R}}}
\newcommand{\symV}{{\mathfrak{V}}}

\newcommand{\symE}{{\mathfrak{E}}}

\definecolor{intcolor}{RGB}{30,76,135}

\newcommand{\myint}[1]{ {\color{intcolor} \boldsymbol{[}} #1 {\color{intcolor} \boldsymbol{]}}}

\newcommand{\almostSub}{\Subset}

\makeatletter
\DeclareFontFamily{OMX}{MnSymbolE}{}
\DeclareSymbolFont{MnLargeSymbols}{OMX}{MnSymbolE}{m}{n}
\SetSymbolFont{MnLargeSymbols}{bold}{OMX}{MnSymbolE}{b}{n}
\DeclareFontShape{OMX}{MnSymbolE}{m}{n}{
	<-6>  MnSymbolE5
	<6-7>  MnSymbolE6
	<7-8>  MnSymbolE7
	<8-9>  MnSymbolE8
	<9-10> MnSymbolE9
	<10-12> MnSymbolE10
	<12->   MnSymbolE12
}{}
\DeclareFontShape{OMX}{MnSymbolE}{b}{n}{
	<-6>  MnSymbolE-Bold5
	<6-7>  MnSymbolE-Bold6
	<7-8>  MnSymbolE-Bold7
	<8-9>  MnSymbolE-Bold8
	<9-10> MnSymbolE-Bold9
	<10-12> MnSymbolE-Bold10
	<12->   MnSymbolE-Bold12
}{}

\let\llangle\@undefined
\let\rrangle\@undefined
\DeclareMathDelimiter{\llangle}{\mathopen}%
{MnLargeSymbols}{'164}{MnLargeSymbols}{'164}
\DeclareMathDelimiter{\rrangle}{\mathclose}%
{MnLargeSymbols}{'171}{MnLargeSymbols}{'171}
\makeatother

\makeatletter
\@ifpackageloaded{stix}{%
}{%
  \DeclareFontEncoding{LS2}{}{\noaccents@}
  \DeclareFontSubstitution{LS2}{stix}{m}{n}
  \DeclareSymbolFont{stix@largesymbols}{LS2}{stixex}{m}{n}
  \SetSymbolFont{stix@largesymbols}{bold}{LS2}{stixex}{b}{n}
  \DeclareMathDelimiter{\lBrace}{\mathopen} {stix@largesymbols}{"E8}%
                                            {stix@largesymbols}{"0E}
  \DeclareMathDelimiter{\rBrace}{\mathclose}{stix@largesymbols}{"E9}%
                                            {stix@largesymbols}{"0F}
}
\makeatother

\newcommand{\symConf}[2]{\left[\begin{matrix}
		#1\\
		#2
	\end{matrix} \right]}

\newcommand{\pathsTerm}{\mathsf{Paths}^{\blacktriangledown}}

\newcommand{\const}{\mathsf{Const}}

\newcommand{\oLeafT}[1]{\begin{tikzpicture}[baseline=-3pt, scale=0.8, every node/.style={outer sep=0}]
	\node[circle, draw=blue, fill=white, inner sep=1.7pt, label=right:{\small\ensuremath{#1}}] at (0,0) (){};
	\end{tikzpicture}}

\newcommand{\oScoreT}[2]{\begin{tikzpicture}[baseline=-3pt, scale=0.8, every node/.style={outer sep=0}]
	\node[rectangle, draw=blue, label=right:{\ensuremath{\small(#2)(#1)}}] at (0,0) (n1){\ensuremath{\scriptstyle s}};
	\end{tikzpicture}}

\newcommand{\oFixT}[1]{\begin{tikzpicture}[baseline=-3pt, scale=0.8, every node/.style={outer sep=0}]
	\node[rectangle, draw=blue, label=right:{\ensuremath{\small(#1)}}] at (0,0) (n1){\ensuremath{\scriptstyle \mu}};
	\end{tikzpicture}}

\newcommand{\oAndT}[3]{\begin{tikzpicture}[baseline=-3pt, scale=0.8, every node/.style={inner sep=0,outer sep=0}]
	\node[circle, inner sep=1.7pt, draw=black, fill=black!10, label=right:{\ensuremath{\small(#3) (#1, #2)}}] at (0,0) (n1){};
	\end{tikzpicture}}

\newcommand{\oAndcT}[3]{\begin{tikzpicture}[baseline=-3pt, scale=0.8, every node/.style={inner sep=0,outer sep=0}]
	\node[circle, inner sep=1.7pt, draw=black, fill=red!70, label=right:{\ensuremath{\small(#3) (#1, #2)}}] at (0,0) (n1){};
	\end{tikzpicture}}

\newcommand{\oAndLeftcT}[2]{\begin{tikzpicture}[baseline=-3pt, scale=0.8, every node/.style={inner sep=0,outer sep=0}]
	\node[circle, inner sep=1.7pt, draw=black, fill=red!70, label=right:{\ensuremath{\small(#2) (#1, \times)}}] at (0,0) (n1){};
	\end{tikzpicture}}

\newcommand{\oAndRightcT}[2]{\begin{tikzpicture}[baseline=-3pt, scale=0.8, every node/.style={inner sep=0,outer sep=0}]
	\node[circle, inner sep=1.7pt, draw=black, fill=red!70, label=right:{\ensuremath{\small(#2) (\times, #1)}}] at (0,0) (n1){};
	\end{tikzpicture}}

\tikzset{
diagonal fill/.style 2 args={fill=#2, path picture={
\fill[#1, sharp corners] (path picture bounding box.south west) -|
                         (path picture bounding box.north east) -- cycle;}},
reversed diagonal fill/.style 2 args={fill=#2, path picture={
\fill[#1, sharp corners] (path picture bounding box.north west) |- 
						 (path picture bounding box.south east) -- cycle;}}
}